\documentclass[twocolumn,showpacs,preprintnumbers,amsmath,amssymb]{revtex4}
\usepackage{epsfig}
\usepackage{dcolumn}
\usepackage{bm}

\newcommand{\remove}[1]{}
\newcommand{\dd}{\mathrm{d}}

\def\be{\begin{equation}}
\def\ee{\end{equation}}

\newcommand{\beq}{\begin{equation}}
\newcommand{\eeq}{\end{equation}}
\newcommand{\beqa}{\begin{eqnarray}}
\newcommand{\eeqa}{\end{eqnarray}}

\renewcommand{\pl}{\partial}
\newcommand{\lag}{\langle}
\newcommand{\rag}{\rangle}

\newcommand{\ii}{{\rm i}}
\newcommand{\inta}{\int_{-\ii\infty}^{+\ii\infty}}

\newcommand{\vx}{{\bf x}}
\newcommand{\vk}{{\bf k}}

\renewcommand{\vr}{{\bf r}}

\newcommand{\tdelta}{{\tilde{\delta}}}
\newcommand{\tdPhi}{{\tilde{\delta\Phi}}}
\newcommand{\tj}{{\tilde{j}}}

\newcommand{\tlambda}{{\tilde{\lambda}}}

\newcommand{\tpsi}{{\tilde{\psi}}}
\newcommand{\ttheta}{{\tilde{\theta}}}
\newcommand{\tu}{{\tilde{u}}}
\newcommand{\tW}{{\tilde{W}}}
\newcommand{\tzeta}{{\tilde{\zeta}}}

\newcommand{\cD}{{\cal D}}
\newcommand{\cF}{{\cal F}}
\newcommand{\cG}{{\cal G}}
\newcommand{\cH}{{\cal H}}

\newcommand{\cO}{{\cal O}}
\newcommand{\cP}{{\cal P}}

\newcommand{\rhob}{\overline{\rho}}
\newcommand{\Om}{\Omega_{\rm m}}
\newcommand{\Ode}{\Omega_{\rm de}}
\newcommand{\wde}{w}
\newcommand{\rhom}{\rho_{\rm m}}
\newcommand{\rhode}{\bar{\rho}_{\rm de}}
\newcommand{\bea}{\begin{array}}
\newcommand{\ea}{\end{array}}

\begin{document}

\title{Structure Formation in Modified Gravity Scenarios}

\author{Philippe Brax}
\affiliation{Institut de Physique Th\'eorique,\\
CEA, IPhT, F-91191 Gif-sur-Yvette, C\'edex, France\\
CNRS, URA 2306, F-91191 Gif-sur-Yvette, C\'edex, France}
\author{Patrick Valageas}
\affiliation{Institut de Physique Th\'eorique,\\
CEA, IPhT, F-91191 Gif-sur-Yvette, C\'edex, France\\
CNRS, URA 2306, F-91191 Gif-sur-Yvette, C\'edex, France}
\vspace{.2 cm}

\date{\today}
\vspace{.2 cm}

\begin{abstract}
We study the growth of structures in modified gravity models where the Poisson equation
and the relationship between the two Newtonian potentials are modified by explicit functions
of space and time. This parameterisation applies to the $f(R)$ models and more generally
to screened modified gravity models.
We investigate the linear and weakly nonlinear regimes using the ``standard'' perturbative
approach and a resummation technique, while we use the spherical dynamics to go beyond
low-order results.
This allows us to estimate the matter density power spectrum and bispectrum from linear
to highly nonlinear scales, the full probability distribution of the density contrast
on weakly nonlinear scales, and the halo mass function.
We analyse the impact of modifications of gravity on these quantities for a few realistic models.
In particular, we find that the standard one-loop perturbative approach is not sufficiently
accurate to probe these effects on the power spectrum and it is necessary to use
resummation methods even on  weakly nonlinear scales which provide the best observational window for modified gravity
as relative deviations from General Relativity do not grow significantly  on smaller
scales where theoretical predictions become increasingly difficult.

\end{abstract}

 \vskip2pc

\maketitle

\section{Introduction}
\label{Introduction}
The discovery of the acceleration of the expansion of the Universe cannot be explained using General Relativity and a matter content comprising only fluids with a positive equation of state. Seemingly, a new fluid with a negative equation of state, either a cosmological constant or dynamical dark energy, is required to generate the late time acceleration\cite{Copeland:2006wr}. Another plausible explanation could be that gravity itself is poorly understood on large scales and needs to be modified\cite{Clifton:2011jh}. As General Relativity (GR) is the unique Lorentz invariant low energy theory of spin two gravitons, any modification of gravity must include new degrees of freedom\cite{Weinberg:1965rz}. Hence, in both the dark energy and the modified gravity contexts, new fields need to be included, the simplest ones being of course  scalar fields. However, the presence of scalar fields is tightly constrained by fifth force and equivalence principle tests\cite{Hoyle:2004cw,Bertotti:2003rm}. This implies that the scalars leading to either dark energy or modified gravity must  be screened in local and dense environments such as on earth or in the solar system\cite{Khoury:2010xi}. Such models abound: chameleons\cite{Khoury:2003aq,Khoury:2003rn,Brax:2004qh}, dilatons\cite{Damour:1994zq,Brax:2011ja,Brax:2010gi}, Galileons\cite{Nicolis:2008in}, symmetrons\cite{Pietroni:2005pv,Olive:2007aj,Hinterbichler:2010es} and their generalisations\cite{Brax:2012gr}. In all these cases, the background cosmology coincides with a $\Lambda$Cold Dark Matter ($\Lambda$CDM) Universe. The only hope of observing non-trivial effects relies on the fact that perturbations in these models grow anomalously inside the Compton radius of the scalar field as first noticed in \cite{Brax:2004qh,Brax:2005ew}. This anomalous growth can only be effective on intermediate scales. Indeed, on very large scales outside the Compton radius, normal gravity is retrieved while screening effects imply that GR is also recovered on small scales in very dense regions of the Universe\cite{Khoury:2003aq}. This opens up the possibility that relevant effects may be present at  the mega parsec scale and that deviations from GR may be detectable by future galaxy surveys.

In the following, we will concentrate on a formulation of the perturbation equations involving two Newtonian potentials and a time and scale dependent relationship between them. In terms of scalar field models, this corresponds to the Jordan frame picture; the difference between the two Newtonian potentials being due to the scalar field perturbation. In this picture, we choose to capture the modified gravity effects using a single function $\epsilon(k,a)$ whose interpretation in the Einstein frame is obvious: it measures the deviations of the geodesics under the influence of the scalar field. This function is universally characterised in terms of the mass and the coupling function of the scalar field. Here, we will consider it as defining the modified gravity models which we will study.

Doing so, we neglect the nonlinear effects due to the presence of non-linear terms originating from the  scalar field modifying gravity. As such we only modify the Euler equation by including the effects of a new scalar force.  Hence, at this level of approximation, the models only differ from the GR treatment of $\Lambda$CDM perturbations by the inclusion of a time and scale dependent contribution to Newton's constant in the Euler equation. This simple modification of gravity is amenable to a quasi-linear and a fully non-linear treatment.

The precision that future galaxy surveys will reach implies that simple linear perturbation theory is not accurate enough. One must include higher order effects and at one-loop order
(i.e., next-to-leading order) we will find that the ``standard'' perturbative expansion is not
sufficiently accurate to probe the modified gravity effects we investigate here.
Therefore, we generalise a method derived using the saddle point of the generating functional of matter and velocity fluctuations. This resummation scheme was already tested in the GR
case and shown to be more accurate than the standard approach.

To go beyond these low-order results we also study the dynamics of spherical perturbations\cite{Borisov:2011fu,Brax:2010tj,Li2012a}.
This can be exactly solved until shell crossing and it provides the full probability distribution
of the matter density contrast on weakly nonlinear scales as well as the large-mass tail
of the halo mass function. The latter can then be used to build a phenomenological
halo model that also converges to the perturbative results on quasi-linear scales.
This provides a simple estimate of the matter density power spectrum and bispectrum
from linear to highly nonlinear scales,
and a global picture of structure formation in such modified-gravity scenarios. We discuss the relative deviations from GR of these various quantities as a function of scale.

However, let us  note that the analytical treatment of modified gravity developed here should only be taken as a first step, to indicate the type of effects one may expect,
because of our simplified parameterization of modified gravity.
First, more accurate modelizations would include some of the non-linearities due to the scalar potential at the one loop level\cite{Bernardeau2011a}, which modify the Euler equation in an effective way. Second, the screening effects of the scalar field force in dense regions would modify the spherical collapse of an initial over density\cite{Brax:2010tj,Li:2011qda}.
Here and as a first step, we will not consider these issues and treat the modification of gravity at the linear level in the scalar sector of the models. In \cite{Schmidt2009,Oyaizu2008,GilMarin:2011xq}, this corresponds to the ``no-chameleon'' regime which should be seen as a non-screening case here in as much as we are neglecting the screening effects of modified gravity in dense regions.  In the appendix, we compare our analytic treatment of the "no-chameleon" case with the simulations of \cite{Oyaizu2008} which shows a very convincing agreement.
Of course, in future work, we intend to include one-loop corrections in the scalar sector as well as screening effects in the spherical collapse.
Yet, it is useful to first develop the analytic formalism for the simpler parameterization
studied in this paper. This will serve as a basis for more complex models that involve
further ingredients (which are also more model-dependent, while the formalism
developed here can be applied to any function $\epsilon(k,a)$ in the Euler equation).

A similar approach was followed in \cite{Koyama2009} where $f(R)$ and DGP models where considered. These cases were treated in the Jordan frame where the effect of modified gravity appears, for instance, in the difference between the two Newtonian potentials due to the anisotropic stress resulting from the presence of an extra scalar degree of freedom. In this work, the non-linear terms up to third order in the scalar dynamics were included, allowing one to study the onset of the screening mechanism at the perturbative level. Moreover, only the standard one loop contribution was taken into account in the quasi-linear regime and a fitting PPF formula was used to analyse fully non-linear scales. In the present work, the non-linearities in the scalar sector are not taken into account. On the other hand,
we go beyond the standard one loop perturbative expansion and include a partial resummation of perturbation theory. Moreover, the highly non-linear regime is studied using the spherical collapse and a halo model taking into account shell coupling due to the scale dependence of modified gravity. One of the advantages of our approach resides also in its versatility. Indeed we work  in the Einstein frame where numerous models of modified gravity are defined\cite{Brax:2012gr}. Our treatment can be applied to chameleon and $f(R)$ models and easily extended to  other models like dilatons and symmetrons. These extensions are being currently investigated.

The paper is arranged as follows. In section~\ref{Modified-Gravity}, we describe the modified gravity models we will consider. We present the dynamical equations in the hydrodynamical approximation in section~\ref{Equations-of-motion}, and we study the perturbative regime
in section~\ref{Perturbative-regime}, for the density power spectrum and bispectrum.
Next, we analyse the spherical collapse in the no-screening case in
section~\ref{Spherical-collapse}.
This allows us to obtain the probability distribution of the density contrast on weakly nonlinear
scales in section~\ref{Probability-distribution} and the halo mass function in
section~\ref{mass-function}.
Finally, we use these ingredients to build a phenomenological halo model in
section~\ref{nonlinear}, which provides estimates of the power spectrum and bispectrum
from linear to highly nonlinear scales.
We conclude in section~\ref{Conclusion}.

\section{Modified Gravity}
\label{Modified-Gravity}

\subsection{The perturbed equations}
\label{perturbed-equations}

We consider models of modified gravity which can be defined by a change of the perturbation equations for Cold Dark Matter (CDM).
The modifications are usually parameterised by two time and scale dependent functions $\gamma (k,a)$ and $\mu(k,a)$\cite{Zhao:2008bn}. Other approaches have also been emphasized like in \cite{Baker:2011jy}. The $\gamma-\mu$ parameterisation
does not follow directly from a Lagrangian formulation where causality is automatically taken into account. In the following, we will use a restricted class of modified gravity models where the perturbed dynamics can be entirely specified by two time dependent functions only,
$m(a)$ and $\beta(a)$. These two functions enter as building blocks of a time and space dependent function $\epsilon (k,a)$. Finally, the knowledge of $\epsilon (k,a) $ defines
$\gamma(k,a)$ and $\mu (k,a)$ completely. The origin of this parameterisation springs from modified gravity models where a scalar field alters gravity on large scales and is screened in dense environments, leading to no modification of gravity in the solar system and in laboratory experiments. In turn, the dynamics of these models can be entirely reconstructed from the time evolution of the mass function $m(a)$ of the scalar field, and its coupling to matter particles $\beta (a)$. This way of describing modifying gravity applies to chameleons and $f(R)$ models, symmetrons and dilatons. Here, we will simply use the $\{m(a),\beta(a)\}$ parameterisation as a way of unambiguously defining modified gravity models at the level of the perturbations.

At the linear level, the perturbation equations of the CDM fluid follow from the conservation of
matter
\be
\theta= - \delta' ,
\label{continuity-0}
\ee
where the density contrast is $\delta= (\rho_{\rm m}-\rhob_{\rm m})/\rhob_{\rm m}$ and
$\theta=\partial^i v_i$ is the divergence of the velocity field. We denote by a prime the time
derivative in conformal time $\tau$, with $\dd\tau= \dd t/a$ and $a(t)$ is the scale factor.
The Euler equation involves the Newtonian potential $\Psi$ and reads in Fourier space as
\be
\ttheta' +{\cal H} \ttheta = k^2 \tilde\Psi ,
\label{Euler-0}
\ee
where we denote Fourier-space quantities with a tilde.
Here ${\cal H}= a' /a$ is the conformal expansion rate and
we are using the Newtonian gauge with two distinct potentials $\Psi$ and $\Phi$,
\begin{equation}
ds^2=-a^2(1+2\Psi)d\tau^2+a^2(1-2\Phi)d\vx^2,
\end{equation}
where $\vx$ are comoving coordinates.
The gravitational dynamics determine the evolution of $\Phi$ as
\begin{equation}
-k^2\tilde\Phi=4\pi\nu(k,a) {\cal G} \rhob_{\rm m} \tdelta /a ,
\end{equation}
which is a modification of the Poisson equation ($\rhob_{\rm m}$ is the mean comoving matter
density and $\cal G $ is Newton's constant). We also assume that there is a constitutional relation between the two potentials,
\begin{equation}
\tilde\Psi=\gamma (k,a) \tilde\Phi ,
\end{equation}
implying that
\begin{equation}
-k^2\tilde\Psi=4\pi\mu(k,a) {\cal G} \rhob_{\rm m} \tdelta/a ,
\label{Psi-deltam}
\end{equation}
where
\begin{equation}
\mu (k,a)= \gamma (k,a) \nu (k,a) .
\end{equation}
As a result, this implies that the density contrast obeys
\be
\tdelta'' + {\cal H} \tdelta' -\frac{3\Omega_{\rm m}}{2} {\cal H}^2 \mu(k,a) \tdelta=0 ,
\ee
where $\Omega_{\rm m}(a)$ is the matter density cosmological parameter.
The growth of structures depends on the choice of the function $\mu(k,a)$. We will define a large class of such models in the following section.

\subsection{Parameterised modified gravity}
\label{Parameterised-gravity}

The choice of function $\mu (k,a)$ seems to be unlimited. Here we focus on the simple choice
\be
\mu (k,a)= 1+ \epsilon (k,a)
\label{mu-eps}
\ee
and
\be
\gamma (k,a)= \frac{1+\epsilon (k,a)}{1-\epsilon (k,a)} ,
\ee
where $\epsilon$ measures the deviation from General Relativity and  is defined by two time dependent functions only, $m(a)$ and $\beta(a)$\cite{Brax:2012gr}.
 In modified gravity models with a screened scalar field in dense environments, $m(a)$ is the mass of the scalar field at the cosmological background level. Similarly $\beta (a)$ is the coupling function between the scalar field and CDM particles. The space and time dependent function $\epsilon (k,a)$ is expressed as
 \be
 \epsilon (k,a)= \frac{2 \beta^2(a)}{1+ \frac{m^2(a) a^2}{k^2}}
 \label{eps-def}
 \ee
 This parameterisation is valid for chameleons and $f(R)$ models, symmetrons and dilatons\cite{Brax:2012gr}. This implies in particular that
 \be
\mu(k,a)= \frac{(1+2\beta^2) k^2 + m^2 a^2}{k^2 +m^2a^2}
\ee
and
\be
\gamma(k,a)= \frac{(1+2\beta^2) k^2 + m^2 a^2}{(1-2\beta^2)k^2 +m^2a^2} .
\ee
This is an explicit parametrisation which shows that modified gravity effects only appear on scales such that $k\gtrsim am(a)$, i.e. when scales are within the Compton wavelength of the scalar field. Outside the Compton wavelength, General Relativity is retrieved.
 These expressions are valid in the Jordan frame where Newton's constant become time dependent too\cite{Brax:2012gr}. For the models we consider here with $m\gg H$, such a time variation can be safely neglected in the Jordan frame. In the Einstein frame, the particle masses vary accordingly in a negligible manner.

In the rest of this paper, we will only deal with one particular family of models defined by
the coupling constant
\be
\beta=\frac{1}{\sqrt 6}
\ee
and the mass of the scalar field which is given by
\be
m(a) = m_0 \, a^{-3(n+2)/2} ,
\label{ma-fR}
\ee
where $m_0$ is a free scale which will be chosen to be close to 1 Mpc$^{-1}$
and $n>0$.
In the matter dominated epoch, these models are equivalent to
$f(R)$ theories in the large curvature regime\cite{Brax:2012gr}
where the $f(R)$ correction to the Einstein-Hilbert action reads\cite{Hu:2007nk}
\be
f(R) \approx -16 \pi {\cal G} \rho_\Lambda - \frac{f_{R_0}}{n} \frac{R_0^{1+n}}{R^n}
\label{fR-def}
\ee
and  $\rho_\Lambda$ is the effective dark energy in the late time Universe. In the recent past of the Universe, the mass of the large curvature models differs slightly from (\ref{ma-fR}), see the appendix for more details. The mass $m_0$ is given by
the useful relationship
\be
m_0= \frac{H_0}{c}
\sqrt{\frac{\Omega_{m0} +4\Omega_{\Lambda0}}{(n+1) \vert f_{R_0}\vert }}
\label{m0-fR-def}
\ee
with $c/H_0 \approx 4 \ {\rm Gpc}$.
Modifications  of gravity must satisfy  $m_0 c/H_0\gtrsim 10^3$ to comply with  a loosely screened Milky Way \cite{Brax:2011aw}. This also corresponds to
$\vert f_{R_0}\vert $ less than $10^{-5}$, the case $\vert f_{R0}\vert =10^{-4}$ being marginal. When $m_0$ is too large, effects of modified gravity on large scale structure occur on very non-linear scales. In the following, we will use values of $m_0\sim 1 {\rm Mpc}^{-1}$ which satisfy the loose screening bound for the Milky Way and imply interesting effects on large scale structure.

We can also deduce now the two parametric functions
\be
\mu(k,a)= \frac{\frac{4}{3} \frac{k^2}{m_0^2}a^{s}  + 1}{\frac{k^2}{m_0^2}a^{s} +1}
\label{mu-def}
\ee
and
\be
\gamma(k,a)= \frac{\frac{4}{3} \frac{k^2}{m_0^2}a^{s}  + 1}{\frac{2}{3}\frac{k^2}{m_0^2}a^{s} +1} ,
\label{gamma-def}
\ee
where
\be
s=3n+4 .
\label{s-def}
\ee
We will use the parameterisation of $\epsilon (k,a)$ in the following when we give numerical examples.
More precisely, we will consider the four cases $(n,m_0)= (0,0.1), (0,1), (1,0.1)$, and
$(1,1)$, where $m_0$ is given in units of Mpc$^{-1}$.
This corresponds to the two scales $m_0=0.1$ and $1$ Mpc$^{-1}$ and to the two exponents
$n=0$ and $1$.
For these models we should have $n>0$, see Eq.(\ref{fR-def}), and the choice $n=0$
for our numerical computations is only meant to exemplify the case of small $n$, that is
$s\rightarrow 4$. The scales we consider are of the same order as  the ones used so far in N-body simulations where $\vert f_{R0}\vert = 10^{-4},10^{-5}, 10^{-6}$ and $n=1$. We will give a qualitative comparison with these numerical results, especially we will briefly analyse the difference between the full numerical simulations, the no-chameleon case where the chameleon effects in dense region is neglected and our resummation method in the appendix. There we analyse the $f(R)$ models where we take into account the late time effect of the cosmological constant on the mass function $m(a)$. A more quantitative comparison is left for future work.

\section{Perturbative Dynamics}
\label{Equations-of-motion}

\subsection{Hydrodynamical perturbations}
\label{Hydro}

As explained in the previous section and in the introduction,
we consider models where the continuity and the Euler equations are only modified by the non-trivial relationship between the two Newtonian potentials. Formally, these equations have the same structure as in GR. When interpreted in terms of scalar field models, new non-linearities
should appear in the Euler equation. However, the analysis of their role is left for future work.
Then, the continuity and Euler equations read in Fourier space as
\beqa
\frac{\pl\tdelta}{\pl\tau}(\vk,\tau) + \ttheta(\vk,\tau) & = &
- \int\dd\vk_1\dd\vk_2 \; \delta_D(\vk_1+\vk_2-\vk) \nonumber \\
&& \times \alpha(\vk_1,\vk_2) \ttheta(\vk_1,\tau) \tdelta(\vk_2,\tau) ,
\label{F-continuity-1}
\eeqa
\beqa
\frac{\pl\ttheta}{\pl\tau}(\vk,\tau) + \cH \ttheta(\vk,\tau) + \frac{3\Om}{2}
\cH^2 [1+\epsilon(k,\tau)]  \tdelta(\vk,\tau) = \nonumber \\
&& \hspace{-8cm} - \!\! \int\dd\vk_1\dd\vk_2 \; \delta_D(\vk_1+\vk_2-\vk)
\beta(\vk_1,\vk_2) \ttheta(\vk_1,\tau) \ttheta(\vk_2,\tau) ,
\nonumber\\
&&
\label{F-Euler-1}
\eeqa
which are the nonlinear generalizations of Eqs.(\ref{continuity-0}) and (\ref{Euler-0}),
with the parameterization (\ref{mu-eps}).
The kernels $\alpha$ and $\beta$ are given by
\beq
\alpha(\vk_1,\vk_2)= \frac{(\vk_1\!+\!\vk_2)\cdot\vk_1}{k_1^2} ,
\beta(\vk_1,\vk_2)= \frac{|\vk_1\!+\!\vk_2|^2(\vk_1\!\cdot\!\vk_2)}{2k_1^2k_2^2} .
\label{alpha-beta-def}
\eeq
In this paper we are mostly interested in the recent Universe on large scales, hence we do not
distinguish between the dark matter and the baryons that are treated as usual as a single
collisionless fluid.
These equations are only a first approximation of the dynamics of modified gravity on sub-horizon scales. Indeed, non-linearities in the potential and coupling function of the scalar field inducing the modification of gravity imply that the full dynamics should be described by the fluid equations for CDM particles and the Klein-Gordon equation for the scalar field. Here we consider only the linear part of the scalar field dynamics which is tantamount to treating the scalar field as massive with a linear coupling to matter. When the mass of the scalar field is large enough $m(a)\gg H$, this allows one to integrate out the scalar dynamics and reduce the equations of motion to the previous ones with a modified Newton constant. A priori, this procedure can be carried out to all orders taking into account the higher derivatives of the scalar field potential and coupling function at the minimum of the effective potential describing the background cosmology. Explicitly, this has been carried out to the one-loop level in the scalar field perturbation, resulting in an effective dynamics, once the scalar field effects have been integrated out, with a modified $\beta(\vk_1,\vk_2)$ \cite{Bernardeau2011a}. The effect of this new contribution will be taken into account in a forthcoming publication.

It is convenient to write the two fields $\delta$ and $\theta$ as a
two-component vector $\psi$ \citep{Crocce2006a}, which we define as
\beq
\psi \equiv \left(\bea{c} \psi_1 \\ \psi_2 \ea \right) \equiv
\left( \bea{c} \delta \\ -\theta/\dot{a} \ea \right) .
\label{psi-def}
\eeq
Because of the factor $\epsilon(k,\tau)$ in the Euler equation (\ref{F-Euler-1}) the
linear growing mode $D_+(k,t)$ depends on the wavenumber $k$. Therefore,
instead of using $D_+$ as the time coordinate we use the logarithm of the scale
factor,
\beq
\eta(t) = \ln a(t) .
\label{eta-def}
\eeq
This agrees with the standard choice used in most perturbative studies for the simpler
case of the Einstein-de-Sitter universe, where $D_+=a$
\cite{Crocce2006a,Valageas2007,Taruya2008,Pietroni2008}.
Then, the equations of motion (\ref{F-continuity-1})-(\ref{F-Euler-1}) read as
\beqa
\frac{\pl\tpsi_1}{\pl\eta} - \tpsi_2 & = & \int \dd\vk_1\dd\vk_2 \;
\delta_D(\vk_1\!+\!\vk_2\!-\!\vk) \alpha(\vk_1,\vk_2) \nonumber \\
&& \times \; \tpsi_2(\vk_1) \tpsi_1(\vk_2) ,
\label{F-continuity-2}
\eeqa
\beqa
\frac{\pl\tpsi_2}{\pl\eta} - \frac{3}{2} \Om (1+\epsilon) \tpsi_1
+ \left( \frac{1}{2} - \frac{3}{2} \wde \Ode \right) \tpsi_2 & =  & \nonumber \\
&& \hspace{-7cm} \int\!\! \dd\vk_1\dd\vk_2 \; \delta_D(\vk_1\!+\!\vk_2\!-\!\vk)
\beta(\vk_1,\vk_2) \tpsi_2(\vk_1) \tpsi_2(\vk_2) ,
\label{F-Euler-2}
\eeqa
where $\Ode(a)$ is the dark energy cosmological parameter and $\wde$ the dark
energy equation-of-state parameter.
As in \cite{Valageas2007,Valageas2007a,Valageas2011d}, this can be written
in a more concise form as
\beq
\cO(x,x') \cdot \tpsi(x') = K_s(x;x_1,x_2) \cdot \tpsi(x_1) \tpsi(x_2) ,
\label{eq-psi-1}
\eeq
where we have introduced the coordinate $x=(\vk,\eta,i)$, $i=1,2$ is the discrete index
of the two-component vector $\tpsi$, and repeated coordinates are integrated over.
The matrix $\cO$ reads as
\beqa
\cO(x,x') & = & \delta_D(\vk-\vk') \delta_D(\eta-\eta') \nonumber \\
&& \hspace{-1.8cm} \times \left( \bea{cc} \frac{\pl}{\pl\eta} & -1 \\  & \\
- \frac{3}{2} \Om(\eta) (1\!+\!\epsilon(k,\eta)) &  \frac{\pl}{\pl\eta} \!+\!
\frac{1}{2} \!-\! \frac{3}{2} \wde \Ode(\eta) \ea \right)
\label{O-def}
\eeqa
and the symmetric vertex $K_s$ is
\beqa
K_s(x;x_1,x_2) & = & \delta_D(\vk_1+\vk_2-\vk) \delta_D(\eta_1-\eta)
\delta_D(\eta_2-\eta) \nonumber \\
&& \times \; \gamma_{i;i_1,i_2}^s(\vk_1,\vk_2) ,
\label{Ks-def}
\eeqa
with
\beq
\gamma_{1;1,2}^s(\vk_1,\vk_2) = \frac{\alpha(\vk_2,\vk_1)}{2} , \;\;
\gamma_{1;2,1}^s(\vk_1,\vk_2) = \frac{\alpha(\vk_1,\vk_2)}{2} , \nonumber
\eeq
\beq
\gamma_{2;2,2}^s(\vk_1,\vk_2) = \beta(\vk_1,\vk_2) ,
\eeq
and zero otherwise.

The vertex $K_s$ does not depend on cosmology and it is not modified. Here modified gravity
only affects the linear operator $\cO$ through the term $\epsilon(k,\eta)$.
In the case of a $\Lambda$CDM universe, that is, for $\epsilon=0$, the matrix $\cO$
and the linear growing mode $D_+(t)$ only depend on time. Then, it is possible to
remove the explicit time-dependence of the equations of motion by using the
time-coordinate $\eta=\ln D_+$
and making the approximation $\Om/f^2 \simeq 1$, where $f=\dd\ln D_+/\dd\ln a$.
This is a good approximation that is used in most perturbative works
and it means that terms of order $n$ in perturbation theory scale with time as
$D_+(t)^n$ \cite{Bernardeau2002}.
Here we do not use this approximation because we consider the
case where the linear growing mode and the matrix $\cO$ also depend on wavenumber.
This also means that in the $\Lambda$CDM limit, $\epsilon\rightarrow 0$, our
approach is exact in the sense that it does not rely on the approximation
$\Om/f^2 \simeq 1$.

\subsection{Linear regime}
\label{Linear}

\subsubsection{Linear growing and decaying modes}
\label{Linear-growing-decaying}

The linear regime corresponds to the linearization of the equations of motion
(\ref{eq-psi-1}) or (\ref{F-continuity-2})-(\ref{F-Euler-2}). We have already discussed
the linear equations in section~\ref{perturbed-equations} to introduce modified-gravity
effects.
Here we present a more detailed analysis. The linear equations are
$\cO\cdot\psi_L = 0$ or
\beq
\frac{\pl\tpsi_{L1}}{\pl\eta} - \tpsi_{L2} = 0  ,
\label{F-continuity-L1}
\eeq
\beq
\frac{\pl\tpsi_{L2}}{\pl\eta} - \frac{3}{2} \Om (1+\epsilon) \tpsi_{L1}
+ \left( \frac{1}{2} - \frac{3}{2} \wde \Ode \right) \tpsi_{L2} =  0 ,
\label{F-Euler-L2}
\eeq
where the subscript ``L'' denotes the linear solutions.
Substituting Eq.(\ref{F-continuity-L1}) into Eq.(\ref{F-Euler-L2}) yields a second-order
equation for the linear modes $D(\eta)$,
\beq
\frac{\pl^2 D}{\pl\eta^2} + \left( \frac{1}{2} - \frac{3}{2} \wde \Ode \right)
\frac{\pl D}{\pl\eta} - \frac{3}{2} \Om (1+\epsilon) D =  0 .
\label{D-pm}
\eeq
As usual, we have a growing mode $D_+(\eta)$ and a decaying mode $D_-(\eta)$,
and we define the initial conditions by the growing mode $D_+$, so that in the
linear regime we have:
\beq
\tpsi_L(\vk,\eta) = \tdelta_{L0}(\vk) \left( \bea{c} D_+(k,\eta) \\
\frac{\pl D_+}{\pl\eta}(k,\eta) \ea \right) .
\label{psiL+}
\eeq
In other words, we assume the decaying mode has had time to decrease to a
negligible amplitude, which is the case in standard cosmologies.
Then, the initial conditions are fully determined by the linear density field
$\tdelta_{L0}(\vk)$.

\begin{figure}
\begin{center}
\epsfxsize=8.5 cm \epsfysize=6 cm {\epsfbox{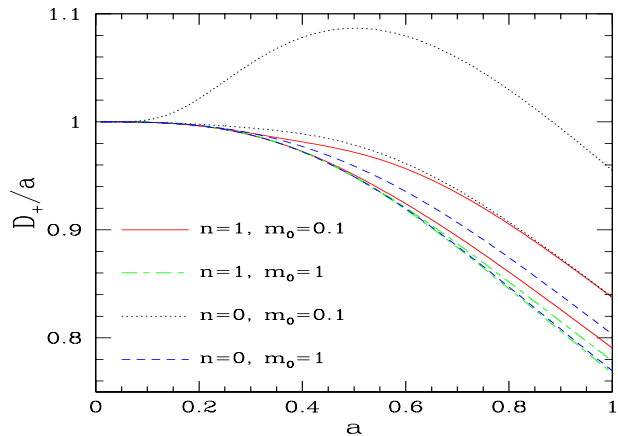}}
\end{center}
\caption{Linear growing mode $D_+(k,t)$ normalized to the scale factor
$a(t)$ for four $(n,m_0)$ models. In each case we show the results for
wavenumbers $k=1 h$Mpc$^{-1}$ (lower curve) and $k=5 h$Mpc$^{-1}$
(upper curve), as a function of $a(t)$. These two scales are in the non-linear regime and have only been chosen to exemplify the type of effects obtained in modified gravity.}
\label{fig-Dplin_z}
\end{figure}

\begin{figure}
\begin{center}
\epsfxsize=8.5 cm \epsfysize=6 cm {\epsfbox{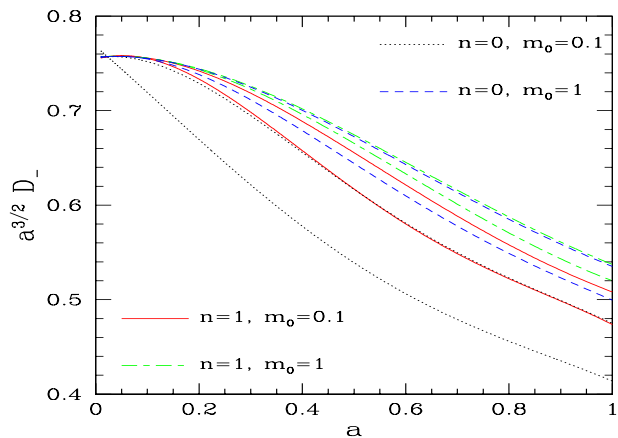}}
\end{center}
\caption{Linear decaying mode $D_-(k,t)$ normalized to
$a(t)^{-3/2}$ for four $(n,m_0)$ models. In each case we show the results for
wavenumbers $k=1 h$Mpc$^{-1}$ (upper curve) and $5 h$Mpc$^{-1}$
(lower curve), as a function of $a(t)$. These two scales are in the non-linear regime and have only been chosen to exemplify the type of effects obtained in modified gravity.}
\label{fig-Dmlin_z}
\end{figure}

\begin{figure}
\begin{center}
\epsfxsize=8.5 cm \epsfysize=6 cm {\epsfbox{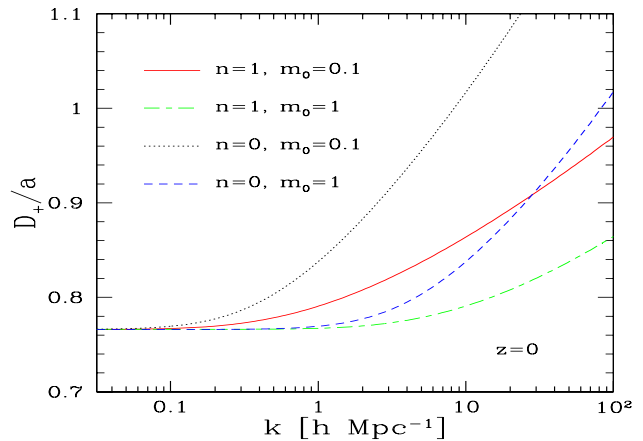}}
\end{center}
\caption{Linear growing mode $D_+(k,t)$ normalized to the scale factor
$a(t)$ for four $(n,m_0)$ models, at redshift $z=0$ up to non-linear scales. }
\label{fig-Dplin_k}
\end{figure}

\begin{figure}
\begin{center}
\epsfxsize=8.5 cm \epsfysize=6 cm {\epsfbox{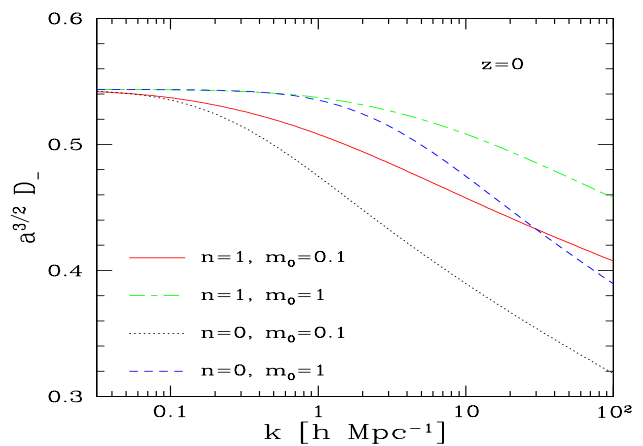}}
\end{center}
\caption{Linear decaying mode $D_-(k,t)$ normalized to $a(t)^{-3/2}$
for four $(n,m_0)$ models, at redshift $z=0$ up to non-linear scales.}
\label{fig-Dmlin_k}
\end{figure}

It is convenient to normalize the growing mode to the scale factor at early times.
Indeed, we consider modified-gravity models parameterized by a function
$\epsilon(k,a)$ such that $\epsilon\rightarrow 0$ for $a\rightarrow 0$.
Then, at early times we recover the Einstein-de Sitter universe (the dark energy
component also becomes negligible) and we have the usual behaviours:
\beq
t\rightarrow 0: \;\; D_+ \rightarrow a = e^{\eta} , \;\;
D_- \propto a^{-3/2} = e^{-3\eta/2} .
\label{D+-asymp}
\eeq
For numerical computations, it is convenient to introduce the reduced growing mode
$g_+(k,\eta) = D_+(k,\eta)/a$. From Eq.(\ref{D-pm}) it obeys
\beq
\frac{\pl^2 g_+}{\pl\eta^2} + \left( \frac{5}{2} \!-\! \frac{3}{2} \wde \Ode
\right) \frac{\pl g_+}{\pl\eta} + \frac{3}{2} \left[ (1\!-\!\wde) \Ode
\!-\! \Om\epsilon \right] g_+ =  0
\label{g+}
\eeq
with the initial conditions
\beq
\eta \rightarrow -\infty : \;\; g_+ \rightarrow 1 , \;\;
\frac{\pl g_+}{\pl\eta} \rightarrow 0 .
\label{g+init}
\eeq
The linear growing mode can be easily computed from Eqs.(\ref{g+})-(\ref{g+init}).
Although the linear decaying mode $D_-$ also obeys Eq.(\ref{D-pm}) it is not
convenient to use this for numerical computations (solving forward in time is
unstable because of the contamination by the growing mode).
It is better to use the Wronskian,
\beq
W = D_+ \frac{\pl D_-}{\pl\eta} - \frac{\pl D_+}{\pl\eta} D_- ,
\label{Wronskian-def}
\eeq
which in our case is still independent of $k$ and given by
\beq
W(\eta) =  - e^{- (1/2) \int_0^{\eta} \dd \eta' \, [1-3\wde\Ode(\eta')]} .
\label{Wronskian-1}
\eeq
This normalization of $W$ also defines the normalization of $D_-$, which reads
\beq
D_-(k,\eta) = - D_+(k,\eta) \int_{\eta}^{\infty} \dd \eta' \,
\frac{W(\eta')}{D_+(k,\eta')^2} .
\label{D_-}
\eeq
The integrals in Eqs.(\ref{Wronskian-1}) and (\ref{D_-}) allow a fast computation
of $D_-(k,\eta)$.

We show in Figs.~\ref{fig-Dplin_z} and \ref{fig-Dmlin_z} the linear growing
and decaying modes as a function of time (described by the scale factor $a(t)$).
The deviation from the General Relativity linear mode (which is almost identical
to the lower curve in Fig.~\ref{fig-Dplin_z} and to the upper curve in
Fig.~\ref{fig-Dmlin_z}) increases for higher wavenumber.
On these scales, the effects of modified gravity grow as we span the parameters
$(n,m_0)=(1,1), (0,1), (1,0.1), (0,0.1)$.
Indeed, as seen from Eqs.(\ref{mu-def})-(\ref{s-def}), deviations from
GR appear at lower $k$ for small mass $m_0$ and at earlier time
for smaller $n$.
We can see that a positive $\epsilon(k,a)$ in the Euler equation (\ref{F-Euler-1})
leads to a larger growing mode $D_+$ and a smaller decaying mode $D_-$.
This can be understood from the fact that a positive $\epsilon$ can also be
interpreted as a larger effective Newton constant in Eq.(\ref{Psi-deltam}).
This implies a faster development of gravitational clustering and both linear
modes evolve faster than in the $\Lambda$CDM cosmology.

These behaviours can also be seen in Figs.~\ref{fig-Dplin_k} and \ref{fig-Dmlin_k}
where we show the linear modes as a function of wavenumber at redshift $z=0$.
Although we plot our results up to $k=100 h$Mpc$^{-1}$ to allow a clear
separation between different curves, values beyond $1 h$Mpc$^{-1}$
do not describe the true quantitative difference between the models for
observables such as the power spectrum because they are in the nonlinear
regime, which is not described by these linear modes.
In addition, on small scales new ``screening'' mechanisms, which are not
described by the equations of motion (\ref{F-continuity-1})-(\ref{F-Euler-1}),
take place and lead to a convergence to General Relativity
and to the $\Lambda$CDM predictions.
In agreement with the parameterization (\ref{eps-def}), the linear modes deviate
from the GR result at a wavenumber $k \sim m_0$ (in the plots the values of
$m_0$ are given in units of 1 Mpc$^{-1}$). At high $k$ the deviation is larger for smaller
$n$ (whence smaller $s$) because modifications of gravity have had more time
to affect the dynamics, see Eq.(\ref{mu-def}).

\subsubsection{Linear growth rate}
\label{fz}

\begin{figure}
\begin{center}
\epsfxsize=8.5 cm \epsfysize=6 cm {\epsfbox{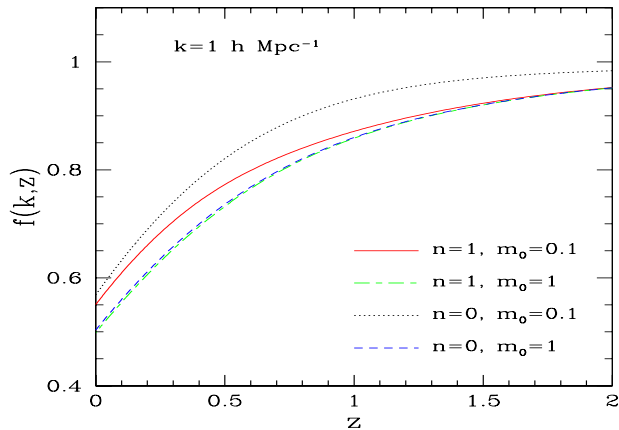}}
\end{center}
\caption{Linear growth rate $f(k,z)=\pl\ln D_+/\pl\ln a$ for wavenumber
$k=1 h$Mpc$^{-1}$, for four $(n,m_0)$ models.}
\label{fig-f_z}
\end{figure}

We plot in Fig.~\ref{fig-f_z} the linear growth rate $f(k,z)$ as a function of redshift,
defined as usual by
\beq
f(k,z) = \frac{\pl \ln D_+(k,a)}{\pl \ln a} .
\label{fz-def}
\eeq
Both the linear growing mode $D_+$ and the linear growth rate $f$ depend on
wavenumber and to avoid overcrowding the figure we only plot our results for
$k=1 h$Mpc$^{-1}$ (which is in the mildly nonlinear regime at $z=0$).
The $\Lambda$CDM prediction could not be distinguished from the results obtained
for $(n,m_0)=(1,1)$ and $(0,1)$ (lower curves). In agreement with Fig.~\ref{fig-Dplin_z},
the larger linear growing modes $D_+$ obtained for $(n,m_0)=(1,0.1)$ and $(0,0.1)$
lead to larger growth rates $f$. The deviation associated with the case
$(n,m_0)=(1,0.1)$ would be difficult to detect with future surveys such as Euclid
but the case $(n,m_0)=(0,0.1)$ should give a clear signal
(see Fig.2.5 in \cite{Laureijs2011}).

\subsubsection{Linear correlation and response functions}
\label{Linear-correlation}

From Eq.(\ref{psiL+}) the linear two-point correlation of the vector $\psi_L$, whence
of the linear density and velocity fields, reads as
\beqa
C_L(x_1,x_2) & = & \lag \tpsi_L(x_1) \tpsi_L(x_2)\rag \\
&& \hspace{-2cm} = \delta_D(\vk_1\!+\!\vk_2) P_{L0}(k_1)
\left( \bea{cc} D_{+1} D_{+2} & D_{+1} D_{+2}' \\ & \\ D_{+1}' D_{+2} &
D_{+1}' D_{+2}' \ea \right)
\label{CL-1}
\eeqa
where $D_{+i}=D_+(k_i,\eta_i)$ and $D_{+i}'=\frac{\pl D_+}{\pl\eta}(k_i,\eta_i)$.

In Sect.~\ref{steepest-descent} we will consider a perturbative resummation
scheme that goes beyond standard one-loop perturbation theory. It involves the
response function (or propagator) defined as the average of the functional derivative
\beq
R(x_1,x_2) = \left\lag \frac{\cD\tpsi(x_1)}{\cD\tzeta(x_2)}\right\rag_{\tzeta=0} ,
\label{R-def}
\eeq
where $\tzeta$ is a ``noise'' added to the right hand side of Eq.(\ref{eq-psi-1}).
Thus, $R(x_1,x_2)$ measures the response of the system at time $\eta_1$ to an
infinitesimal perturbation at an earlier time $\eta_2$. It also describes the
``propagation'' of infinitesimal fluctuations.
By causality, it satisfies
\beq
\eta_1 < \eta_2 : \;\;\; R(x_1,x_2) = 0 ,
\eeq
and it obeys the initial condition
\beq
\eta_1 \rightarrow \eta_2^+ : \;\;\; R(x_1,x_2) \rightarrow
\delta_D(\vk_1-\vk_2) \; \delta_{i_1,i_2} .
\label{R-init}
\eeq
In the linear regime, where the equation of motion (\ref{eq-psi-1}) reduces to
$\cO\cdot\psi_L=0$, the response function obeys
\beq
\eta_1 > \eta_2 : \;\;\; \cO \cdot R_L = 0 .
\eeq
Using the initial condition (\ref{R-init}), this gives
\beqa
R_L(x_1,x_2) & = & \frac{\Theta(\eta_1-\eta_2) \, \delta_D(\vk_1-\vk_2)}
{D_{+2}'D_{-2}-D_{+2}D_{-2}'} \nonumber \\
&& \hspace{-2.2cm} \times \left( \bea{cc} D_{+2}'D_{-1}\!-\!D_{-2}'D_{+1}
&  D_{-2}D_{+1}\!-\!D_{+2}D_{-1} \\ & \\ D_{+2}'D_{-1}'\!-\!D_{-2}'D_{+1}'
&  D_{-2}D_{+1}'\!-\!D_{+2}D_{-1}' \ea \right)
\label{RL-1}
\eeqa
which involves both the linear growing and decaying modes $D_+$ and $D_-$.
Here $\Theta(\eta_1-\eta_2)$ is the Heaviside function, which ensures causality.

\section{Perturbative regime}
\label{Perturbative-regime}

The equation of motion (\ref{eq-psi-1}) is nonlinear and it has no explicit general
solution. Therefore, it is usually solved by perturbative methods, which are sufficient
on large scales and at early times where the density and velocity fluctuations are
small. Within our parameterization, modified gravity only changes the linear operator
$\cO$ of Eq.(\ref{O-def}), through the factor $\epsilon(k,\eta)$. Thus, we keep
the same quadratic nonlinearity as in General Relativity, with the same vertex $K_s$ of
Eq.(\ref{Ks-def}). Therefore, we can use the same perturbative schemes as in
standard cosmologies.

We first describe the standard perturbative approach in
Sect.~\ref{Standard-expansion}
and next a more accurate resummation scheme in Sect.\ref{steepest-descent}.
Here we only go up to ``one-loop order'': our standard perturbative prediction only
includes the linear and one-loop (i.e., next-to-leading) terms, while our resummed
prediction only adds a partial resummation of higher-order terms.

We follow the approach described in detail in \cite{Valageas2008} (see also
\cite{Valageas2007,Valageas2007a}).

\subsection{Standard expansion}
\label{Standard-expansion}

Since the equation of motion (\ref{eq-psi-1}) is quadratic in $\tpsi$, it can be
solved through a perturbative expansion in powers of the linear solution $\tpsi_L$,
as
\beq
\tpsi(x) = \sum_{n=1}^{\infty} \tpsi^{(n)}(x) , \;\; \mbox{with} \;\; \tpsi^{(n)} \propto (\tpsi_L)^n .
\label{psi-n-def}
\eeq
Substituting this expansion into Eq.(\ref{eq-psi-1}) gives the recursion
\beq
\cO\cdot\tpsi^{(n)} =  K_s(x;x_1,x_2) \cdot \sum_{\ell=1}^{n-1}
\tpsi^{(\ell)}(x_1) \tpsi^{(n-\ell)}(x_2) ,
\label{recurs}
\eeq
which allows to compute terms of increasing order, starting with
$\tpsi^{(1)}=\tpsi_L$.
One usually writes the expansion (\ref{psi-n-def}) in terms of the density and velocity
fields, as \cite{Goroff1986,Bernardeau2002}
\beqa
\tdelta(\vk,\eta) & = & \sum_{n=1}^{\infty} \int \dd\vk_1 .. \vk_n
\delta_D(\vk_1+..+\vk_n-\vk) \nonumber \\
&& \times \; F_n^s(\vk_1,..,\vk_n;\eta) \; \tdelta_{L0}(\vk_1) .. \tdelta_{L0}(\vk_n) ,
\label{Fn}
\eeqa
and
\beqa
\ttheta(\vk,\eta) & = & \sum_{n=1}^{\infty} \int \dd\vk_1 .. \vk_n
\delta_D(\vk_1+..+\vk_n-\vk) \nonumber \\
&& \times \; E_n^s(\vk_1,..,\vk_n;\eta) \; \tdelta_{L0}(\vk_1) .. \tdelta_{L0}(\vk_n) ,
\label{En}
\eeqa
where $\tdelta_{L0}$ is the linear density field at some chosen time, as in
Eq.(\ref{psiL+}).
The symmetrized  kernels $F_n^s$ and $E_n^s$ are obtained from the recursion
(\ref{recurs}).
In General Relativity the time-dependence of these kernels factorizes
as $F_n^s \propto D_+^n F_n^s(\vk_1,..,\vk_n)$ and
$E_n^s \propto -a(\dd \ln D_+/\dd t) D_+^n E_n^s(\vk_1,..,\vk_n)$ upon using the
approximation $\Om/f^2 \simeq 1$ \cite{Bernardeau2002}.
In our case, where the linear growing mode $D_+(k,\eta)$ depends on wavenumber,
there is no such factorization and one must solve for the kernels
$F_n^s(\vk_1,..,\vk_n;\eta)$ and $E_n^s(\vk_1,..,\vk_n;\eta)$ for each time $\eta$
of interest.

Finally, from the expansion (\ref{psi-n-def}) one obtains the two-point correlation as
\beqa
C(x_1,x_2) & = & \lag\tpsi(x_1)\tpsi(x_2)\rag \\
& = & \lag \tpsi^{(1)} \tpsi^{(1)}\rag +
\lag \tpsi^{(3)} \tpsi^{(1)}\rag + \lag \tpsi^{(1)} \tpsi^{(3)}\rag \nonumber \\
&& + \lag \tpsi^{(2)} \tpsi^{(2)}\rag + ...
\label{Wick}
\eeqa
where we can use Wick's theorem to perform the average over the initial conditions
$\tpsi_{L0}$. In particular, up to one-loop order the density power spectrum reads as
\beq
P(k,\eta) = P^{\rm tree}(k,\eta) + P^{\rm 1loop}(k,\eta) ,
\label{P-tree-1loop}
\eeq
where $P^{\rm tree}$, associated with ``tree diagrams'', also
corresponds to the linear power spectrum,
\beq
P^{\rm tree}(k,\eta) = P_L(k,\eta) = D_+(k,\eta)^2 \, P_{L0}(k) ,
\label{Ptree}
\eeq
while $P^{\rm1loop}$, associated with ``one-loop'' diagrams, is also given by
\beq
P^{\rm 1loop}(k,\eta) = P^{(b)}(k,\eta) + P^{(c)}(k,\eta) ,
\label{P1loop}
\eeq
using the notations of \cite{Valageas2008}, with (see also
\cite{Goroff1986,Bernardeau2002,Valageas2007}),
\beq
P^{(b)}(k,\eta) = 6 P_{L0}(k) \int \dd\vk' \, P_{L0}(k') F_3^s(\vk',-\vk',\vk;\eta) ,
\label{F3}
\eeq
\beq
P^{(c)}(k,\eta) = 2 \int \dd\vk' P_{L0}(k') P_{L0}(|\vk-\vk'|)
F_2^s(\vk',\vk-\vk';\eta)^2 .
\label{F2}
\eeq

\subsection{Path-integral formulation}
\label{Path-integral formulation}

\subsubsection{General formulation}
\label{General-formulation}

The standard perturbative approach recalled in Sect.~\ref{Standard-expansion}
computes the density power spectrum, and more generally many-body correlation
functions, by first deriving an explicit expression for the nonlinear field $\tpsi$ in terms
of the initial field $\tpsi_L$, as in Eqs.(\ref{psi-n-def}) and (\ref{Fn})-(\ref{En}),
up to some order, and second taking the Gaussian average over the initial conditions,
as in Eq.(\ref{Wick}).

It is possible to work in the reverse order, by first taking the average over the initial
conditions and second writing an expansion in terms of the many-body correlations.
A well-known procedure in the context of plasma physics and the study of the
Vlasov equation is to use the BBGKY hierarchy, which gives a recursion between
successive correlation functions that may be truncated at some order
\cite{Peebles1980}.
A similar approach has also been used in \cite{Pietroni2008} to study the formation of
large-scale structures in the single-flow perturbative regime, as in
Eqs.(\ref{F-continuity-1})-(\ref{F-Euler-1}).
As described in \cite{Valageas2007,Valageas2007a,Valageas2008}, an alternative
approach, also used in field theory and statistical physics \cite{Martin1973,Phythian1977},
is based on a path-integral
formulation. There, it is shown that the statistical properties of the nonlinear field
$\tpsi$, which are fully defined by the equation of motion (\ref{eq-psi-1}) and the
Gaussian initial conditions (\ref{psiL+}), can be obtained from the generating functional
\beq
Z[\tj] = \lag e^{\tj\cdot\tpsi}\rag = \int \cD\tpsi \cD\tlambda \;
e^{\tj\cdot\tpsi - S[\tpsi,\tlambda]} ,
\label{Z-def}
\eeq
where $\tlambda(x)$ is a Lagrange multiplier and the action $S[\tpsi,\tlambda]$ reads as
\beq
S[\tpsi,\tlambda] = \tlambda\cdot(\cO\cdot\tpsi-K_s\cdot\tpsi\tpsi)
- \frac{1}{2} \tlambda\cdot \Delta_I \cdot\tlambda
\label{S-def}
\eeq
Here $\Delta_I$ is the two-point correlation of the initial conditions, taken at a time
$\eta_I$. This matrix disappears in the final equations when we take the limit
$\eta_I\rightarrow-\infty$. Whereas moments of the field $\tpsi$ generate the
many-body correlations of the density and velocity fields, such as the density
power spectrum $P(k)$, moments that involve the auxiliary field $\tlambda$
generate the response functions \cite{Valageas2007a,Phythian1977}.
In particular, we have
\beq
\lag\tlambda\rag=0, \;\; \lag\tlambda\tlambda\rag=0 ,
\;\; \lag\tpsi(x_1)\tlambda(x_2)\rag=R(x_1,x_2) .
\eeq

As explained in \cite{Valageas2007,Valageas2008}, the standard perturbative results
of Sect.~\ref{Standard-expansion} can be recovered from the generating functional
(\ref{Z-def}).
Indeed, one can see at once from Eq.(\ref{recurs}) that the expansion
(\ref{psi-n-def}) is also an expansion over powers of the vertex $K_s$, with
$\tpsi^{(n)} \propto K_s^{n-1}$ and $F_n^s \propto K_s^{n-1}$.
Therefore, the standard expansion in powers of $\tdelta_{L0}$ for $\tpsi$, which
leads to the usual expansion in powers of $P_{L0}$ for averaged quantities, such
as the density power spectrum (\ref{P-tree-1loop}), is identical to an expansion in
$K_s$.
Then, this expansion can be directly obtained from Eq.(\ref{Z-def}) by expanding in
the cubic part $\tlambda\cdot K_s\cdot\tpsi\tpsi$ of the action (\ref{S-def}).
This gives an alternative expression of the expansion (\ref{Wick}) in terms of
Feynman's diagrams\footnote{Although the result at each order $P_{L0}^n$, or
$K_s^{2(n-1)}$, is identical whether one uses either of these two methods, this term
of order $n$ is split in different manners in the two methods as they involve different
types of diagrams, see \cite{Valageas2008} for details.}.

\subsubsection{Direct steepest-descent expansion}
\label{steepest-descent}

One interest of the expression (\ref{Z-def}) is that it can also serve as the basis of
other approximation schemes. Here we focus on the ``direct steepest-descent''
method described in \cite{Valageas2007,Valageas2008}, which is compared with
numerical simulations for the density power spectrum and bispectrum in
\cite{Valageas2011d,Valageas2011e}.
In this approach, instead of expanding the cubic part of the action to write
Eq.(\ref{Z-def}) as a series of Gaussian integrals, one expands around a saddle-point
(which depends on $\tj$) as in a semi-classical or ``large-N'' expansion
\cite{Valageas2004,Zinn-Justin1989}.
This yields the Schwinger-Dyson equations
\beqa
\cO \cdot C & = & \Sigma \cdot C + \Pi \cdot R^T ,
\label{C-SD}\\
\cO \cdot R & = & \delta_D + \Sigma \cdot R ,
\label{R-SD}
\eeqa
for the nonlinear two-point correlation $C$ and response $R$, where $\Sigma$ and
$\Pi$ are ``self-energy'' terms (there are two ``correlations'', $C$ and $R$, and
two ``self-energies'', $\Sigma$ and $\Pi$, because there are two fields, the physical
field $\psi$ and the auxiliary field $\lambda$).

These equations are exact and define $\Sigma$ and $\Pi$.
The ``direct steepest-descent'' or ``large-N'' expansion scheme
corresponds to writing the self-energy terms $\Sigma$ and $\Pi$ as  series in
powers of the linear correlation $C_L$ and response $R_L$.
Then, the order of the approximation is set by the order of the truncation chosen for
these expansions of $\Sigma$ and $\Pi$. Because the truncation is made
on $\Sigma$ and $\Pi$, rather than on $C$ and $R$, this automatically yields a partial
resummation of higher-order terms (e.g., formally $R$ would be given by the highly
nonlinear expression $(\cO -\Sigma)^{-1}$ whose expansion in  $P_{L0}$ contains
terms of all orders as soon as $\Sigma$ contains at least one power of $P_{L0}$).
As described in \cite{Valageas2004,Valageas2007,Valageas2008}, the result obtained
for the correlation $C$ at a given order (e.g., at one-loop order as in this paper)
agrees with the result obtained by the standard perturbative expansion at the same
order, and only differs by additional higher-order terms (which are only partially
resummed).

Then, this ``direct steepest-descent'' scheme gives at the one-loop order
\beqa
\Sigma^{\rm 1loop}(x,y) & = & 4 K_s(x;x_1,x_2) K_s(z;y,z_2) R_L(x_1,z) \nonumber \\
&& \times \, C_L(x_2,z_2) ,
\label{Sigma-1loop}
\eeqa
\beqa
\Pi^{\rm 1loop}(x,y) & = & 2 K_s(x;x_1,x_2) K_s(y;y_1,y_2) C_L(x_1,y_1) \nonumber\\
&& \times \, C_L(x_2,y_2) .
\label{Pi-1loop}
\eeqa
This corresponds to a one-loop diagram
\cite{Valageas2007,Valageas2008,Valageas2011d} and at this order
$\Sigma \propto P_{L0}$ while $\Pi\propto P_{L0}^2$.
Substituting into
Eqs.(\ref{C-SD})-(\ref{R-SD})  gives the nonlinear correlation complete up to
order $P_{L0}^2$, as in (\ref{P-tree-1loop}), with the addition of a partial
resummation of higher-order terms.
Equation (\ref{C-SD}) can be solved as
\beq
C(x_1,x_2) = R\times C_L(\eta_I)\times R^T + R \cdot \Pi \cdot R^T ,
\label{C-Pi}
\eeq
where the first product does not contain any integration over time,
and we take $\eta_I\rightarrow-\infty$.
Thus, to compute the density power spectrum up to one-loop order within the
 direct steepest-descent resummation, we first compute the linear correlation
$C_L$ and $R_L$, given by Eqs.(\ref{CL-1}) and (\ref{RL-1}).
This provides the self-energies $\Sigma$ and $\Pi$ from Eqs.(\ref{Sigma-1loop})
and (\ref{Pi-1loop}). Next, we compute $R$ by solving the integro-differential
equation (\ref{R-SD}) and $C$ from the explicit expression (\ref{C-Pi}).

The formalism used for the $\Lambda$CDM cosmology still applies to our
modelization of modified gravity. However, the numerical computation is somewhat
heavier. Indeed, as described in \cite{Valageas2007,Valageas2011d}, in the
$\Lambda$CDM case, the approximation $\Om/f^2\simeq 1$ allows us to explicitly
factor the time-dependence of the linear correlation and response functions, and of
the self-energies.
Here this is no longer possible, because of the arbitrary function $\epsilon(k,\eta)$
in the linear operator (\ref{O-def}).
This makes the numerical implementation slightly more complex, as we can
no longer use these factorizations to simplify the algorithms and we must keep track
of the complex dependence on time and wavenumber of all linear modes and
two-point functions. However, the method remains exactly the same, as described
above, and it is still possible to devise efficient and reasonably fast numerical codes.

\subsubsection{Recovering the standard one-loop results}
\label{Recovering}

Since we compute the self-energies $\Sigma$ and $\Pi$ for the one-loop
steepest-descent scheme, we can also use them to recover the standard perturbative
expansion instead of using the standard procedure recalled in
Sect.~\ref{Standard-expansion}.
Indeed, the solution of Eq.(\ref{R-SD}) can be written as the expansion over powers
of $\Sigma$,
\beqa
R & = & R_L + R_L \cdot \Sigma \cdot R \\
& = & R_L + R_L \!\cdot\! \Sigma \!\cdot\! R_L + R_L \!\cdot\! \Sigma \!\cdot\!
R_L \!\cdot\! \Sigma \!\cdot\! R_L + ...
\eeqa
Therefore, up to order $P_{L0}$ we can write
\beq
R= R^{(0)} + R^{(1)} ,
\eeq
with
\beq
R^{(0)} = R_L , \;\; R^{(1)} = R_L \cdot \Sigma^{\rm 1loop} \cdot R_L .
\eeq
Then, from (\ref{C-Pi}) the two-point correlation reads up to order $P_{L0}^2$ as
\beq
C= C^{(1)} + C^{(2)} ,
\label{C-1loop-standard}
\eeq
with
\beq
C^{(1)} = R_L \!\times\! C_L(\eta_I) \!\times\! R_L^T = C_L ,
\label{C1-RLCL}
\eeq
and
\beqa
C^{(2)} & = & R^{(1)} \times C_L(\eta_I) \times R_L^T + R_L \times C_L(\eta_I) \times R^{(1)T}
\nonumber \\
&& + R_L \cdot \Pi^{\rm1loop} \cdot R_L^T .
\label{C2-1loop}
\eeqa
This expression is equivalent to Eqs.(\ref{P-tree-1loop})-(\ref{F2}) for the density
power spectrum \citep{Valageas2008}.
Therefore, since we have already computed $\Sigma$ and $\Pi$ we can compute
the standard one-loop power spectrum through Eqs.(\ref{C1-RLCL})-(\ref{C2-1loop}), instead
of using Eqs.(\ref{F3})-(\ref{F2}).
This avoids explicitly computing the $n-$point kernels $F_n^s$ of the
standard expansion (\ref{Fn}).

A similar procedure, based on the closure approximation \citep{Taruya2008},
which is equivalent (at one-loop order) to the ``2PI'' effective action method of
\cite{Valageas2007}, was used in \cite{Koyama2009} to obtain the standard perturbative
predictions for several modified gravity models.
However, while \cite{Koyama2009} included quadratic and cubic nonlinearities in
the scalar field, associated with the onset of the chameleon mechanism,
in this paper we only consider modifications to the Poisson equation at the linear level.
On the other hand, within our simpler formulation of modified gravity we go beyond the standard
perturbative approach by computing the ``steepest-descent'' resummation
presented in the previous section.

\subsubsection{Alternative resummations}
\label{Alternative}

Finally, the path-integral (\ref{Z-def}) can also lead to alternative
resummation schemes, such as the ``1PI'' and ``2PI'' effective action methods
described in \cite{Valageas2004}. The 2PI effective action still leads to the
Schwinger-Dyson equations (\ref{C-SD})-(\ref{R-SD}) but the self-energy terms
are given in terms of the nonlinear two-point functions $C$ and $R$, instead of the
expansion over $C_L$ and $R_L$ used in the direct steepest-descent scheme.
At one-loop order, this amounts to replacing $C_L$ and $R_L$ by $C$ and $R$ in
Eqs.(\ref{Sigma-1loop})-(\ref{Pi-1loop}).
However, already for the $\Lambda$CDM case this makes the computation more
complex since Eqs.(\ref{C-SD})-(\ref{R-SD}) become coupled nonlinear equations
over $C$ and $R$ \citep{Valageas2007,Taruya2008}. Then, one needs to solve for
the four quantities
$C$, $R$, $\Sigma$, and $\Pi$ by simultaneously moving forward with time.
This numerical computation was performed in \cite{Valageas2007} and it appeared
that it did not provide a significant improvement over the simpler direct
steepest-descent scheme (although a more precise comparison with numerical
simulations may remain of interest). Therefore, we do not investigate this scheme further.

The direct steepest-descent method of Sect.~\ref{steepest-descent} is not necessarily
the most accurate resummation scheme.
In particular, it yields a response function that does not
decay at high $k$ or late times, but shows increasingly fast oscillations with an
amplitude that follows the linear response function. This is not realistic,
since one expects a Gaussian-like decay for Eulerian response functions,
as can be seen from theoretical arguments and numerical simulations
\citep{Crocce2006a,Crocce2006b,Valageas2007a,Bernardeau2010b,Bernardeau2012}.
However, the fast oscillations still provide an effective damping in a weak
sense (that is when the response function is integrated over).
Alternative resummation schemes have also been studied in the literature, such as
the ``renormalized perturbation theory'' \citep{Crocce2006a,Crocce2006b}
and several related approaches \citep{Bernardeau2008,Bernardeau2012a,Anselmi2012},
which rely on a response function that interpolates between its
low-$k$ standard perturbative expression and a resummed high-$k$ limit,
methods based on path-integral formulations \citep{Matarrese2008}, on closures
of the hierarchies satisfied by the correlation functions \citep{Taruya2008,Pietroni2008},
or on Lagrangian-space formulations \citep{Matsubara2008}.

The reason why we consider the direct steepest-descent method here is that
it provides a simple and efficient method, which has already been shown to be
reasonably accurate for $\Lambda$CDM cosmology \citep{Valageas2011d,Valageas2011e}.
An advantage with respect to some alternative approaches, which can show similar
levels of accuracy, is that it is fully systematic and contains no free parameter or
interpolation procedure. Therefore, the generalization from the $\Lambda$CDM cosmology
to modified-gravity scenarios is straightforward, as described in Sect.~\ref{steepest-descent},
and we can expect a similar accuracy.

\subsection{Bispectrum}
\label{Bispectrum}

Because the gravitational dynamics is nonlinear, the density field becomes
increasingly non-Gaussian in the course of time. The most popular measure of
these non-Gaussianities, which can be used to break degeneracies between
cosmological parameters or to constrain primordial non-Gaussianities, is
the three-point correlation function \citep{Sefusatti2006}.
In Fourier space this is the so-called bispectrum,
\beq
\lag \tdelta(\vk_1) \tdelta(\vk_2) \tdelta(\vk_3) \rag =
\delta_D(\vk_1+\vk_2+\vk_3) \; B(k_1,k_2,k_3) .
\label{bispec-def}
\eeq
This can be computed by the standard perturbative approach
\cite{Bernardeau2002}. Substituting the expansion (\ref{Fn}) yields the standard
tree-order result
\beq
B^{\rm tree}(k_1,k_2,k_3) =  2 F_2^s(\vk_2,\vk_3;\eta) P_{L0}(k_2) P_{L0}(k_3)
+ 2 \; {\rm cyc}.
\label{B-tree}
\eeq
where ``2 cyc.'' stands for two terms obtained by circular permutations over
$\{k_1,k_2,k_3\}$.

Within the path-integral formalism of Sect.~\ref{General-formulation}, expanding
Eq.(\ref{Z-def}) in powers of $K_s$, that is, in the cubic part of the action
$S$, yields for the three-point correlation at tree-order \cite{Valageas2008}
\beq
C_3^{\rm tree} = R_L \cdot K_s \cdot C_L C_L + 5 \; {\rm perm}.
\label{C3-tree}
\eeq
This gives for the equal-time density bispectrum:
\beqa
B^{\rm tree}(k_1,k_2,k_3;\eta) & = & 2 \int_{-\infty}^{\eta} \dd\eta'
\sum_{i_1',i_2',i_3'} R_{L;1,i_1'}(k_1;\eta,\eta') \nonumber \\
&& \hspace{-1cm} \times \; C_{L;1,i_2'}(k_2;\eta,\eta')
C_{L;1,i_3'}(k_3;\eta,\eta') \nonumber \\
&& \hspace{-1cm} \times \; \gamma^s_{i_1';i_2,i_3}(\vk_2,\vk_3) + 2 \; {\rm cyc}.
\label{B-tree-Z}
\eeqa
which is again equivalent to Eq.(\ref{B-tree}).
In practice, instead of Eq.(\ref{B-tree}) we use Eq.(\ref{B-tree-Z}) to compute
the standard tree-order bispectrum.
The effects of the modified-gravity function $\epsilon(k,a)$ are included through
the linear correlation and response $C_L$ and $R_L$, which depend on the
modified linear modes $D_+(k,a)$ and $D_-(k,a)$ as described in
Sect.~\ref{Equations-of-motion}.
As in Sect.~\ref{Recovering}, this allows us to obtain the ``standard'' perturbative
predictions without computing the kernels $F_n^s$ of Eq.(\ref{Fn}).

At one-loop order the expressions involve more terms. They can be found in
\cite{Valageas2008} (for the $\Lambda$CDM cosmology) for the standard
approach as in (\ref{B-tree}), the equivalent path-integral formulation
as in (\ref{B-tree-Z}), and the direct steepest-descent method used in
Sect.~\ref{steepest-descent} for the power spectrum.
Contrary to the power spectrum, a detailed comparison with numerical simulations
\cite{Valageas2011e} shows that at one-loop order the steepest-descent
resummation for the bispectrum is not more accurate than the standard result.
Therefore, we do not investigate this resummation for the bispectrum here.

Because the linear modes depend on wavenumber, computing the one-loop order
terms is significantly more difficult than in the $\Lambda$CDM case, even
within standard perturbation theory.
Using the scalings $B^{\rm tree} \propto D_+^4 P_{L0}^2$ and
$B^{\rm 1loop} \propto D_+^6 P_{L0}^3$, we consider the following
approximation:
\beq
B^{\rm 1loop} \simeq \left( \frac{B^{\rm tree}}
{B^{\rm tree}_{\Lambda \rm CDM}} \right)^{3/2}
B^{\rm 1loop}_{\Lambda \rm CDM} .
\label{B1loop}
\eeq
Thus, we simply rescale the one-loop correction obtained in the $\Lambda$CDM
scenario by the prefactor
$(B^{\rm tree}/B^{\rm tree}_{\Lambda \rm CDM})^{3/2}$.
This would be exact if the ratio of the linear modes were constant.
We choose this prefactor, rather than $(D_+(k)/D_{+,\Lambda \rm CDM}(k))^6$,
because it includes an integration over the past history and over the appropriate
range of wavenumbers of the linear modes.
This should be sufficient for our purpose, which is simply to estimate the
magnitude of these one-loop corrections.

\subsection{Numerical results}
\label{pert-num}

\begin{figure*}
\begin{center}
\epsfxsize=5.9 cm \epsfysize=5.4 cm {\epsfbox{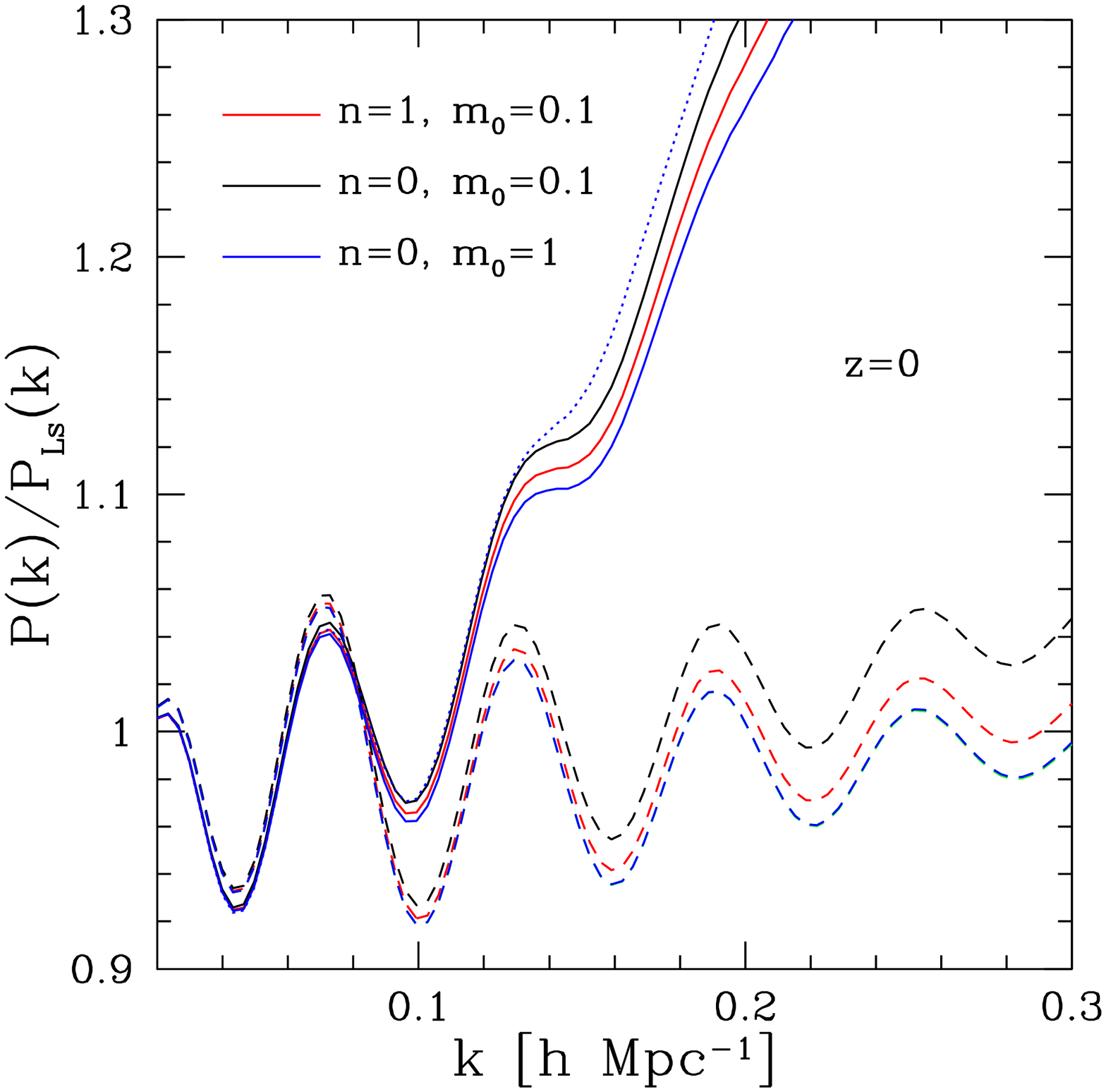}}
\epsfxsize=5.9 cm \epsfysize=5.4 cm {\epsfbox{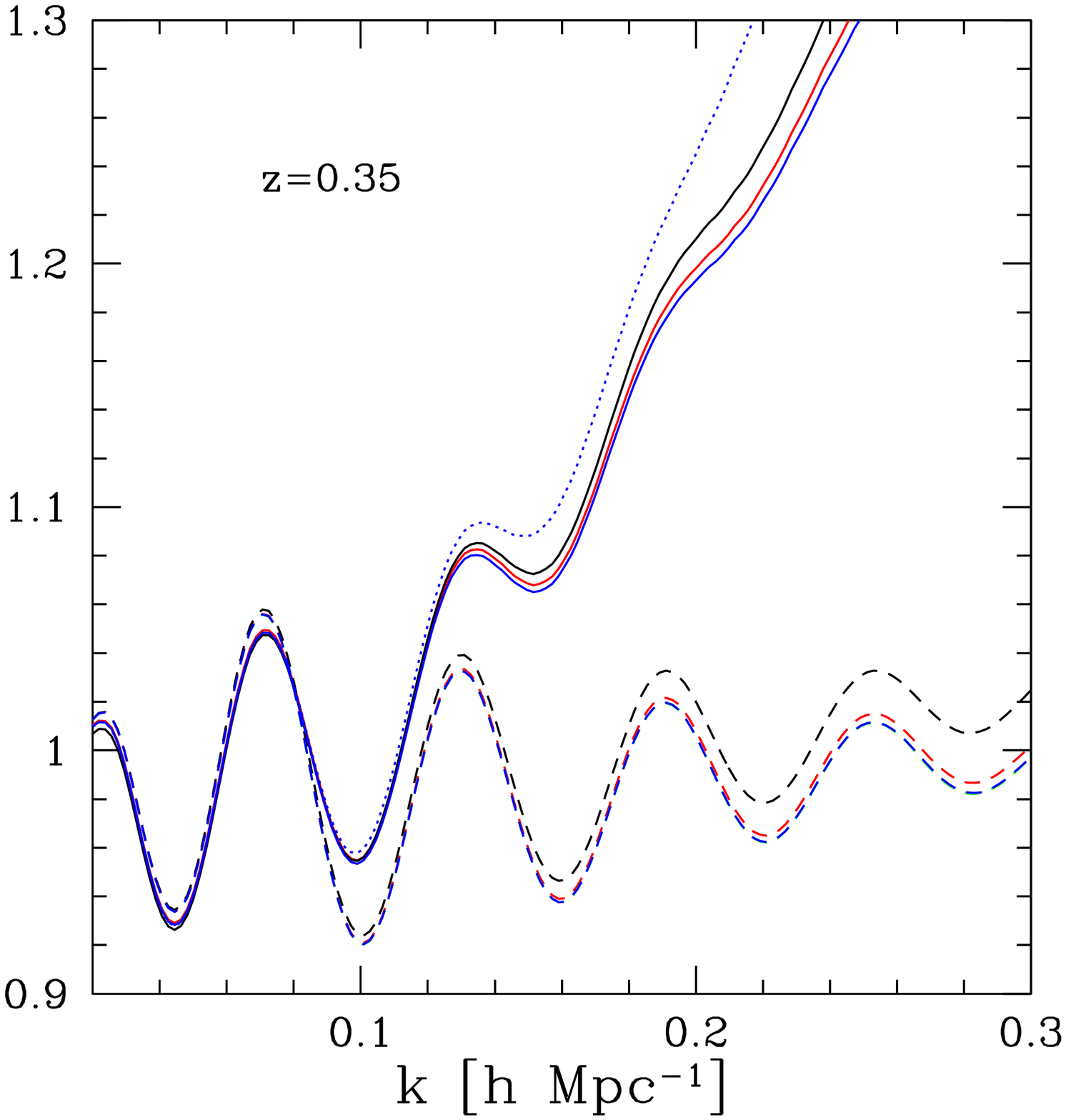}}
\epsfxsize=5.9 cm \epsfysize=5.4 cm {\epsfbox{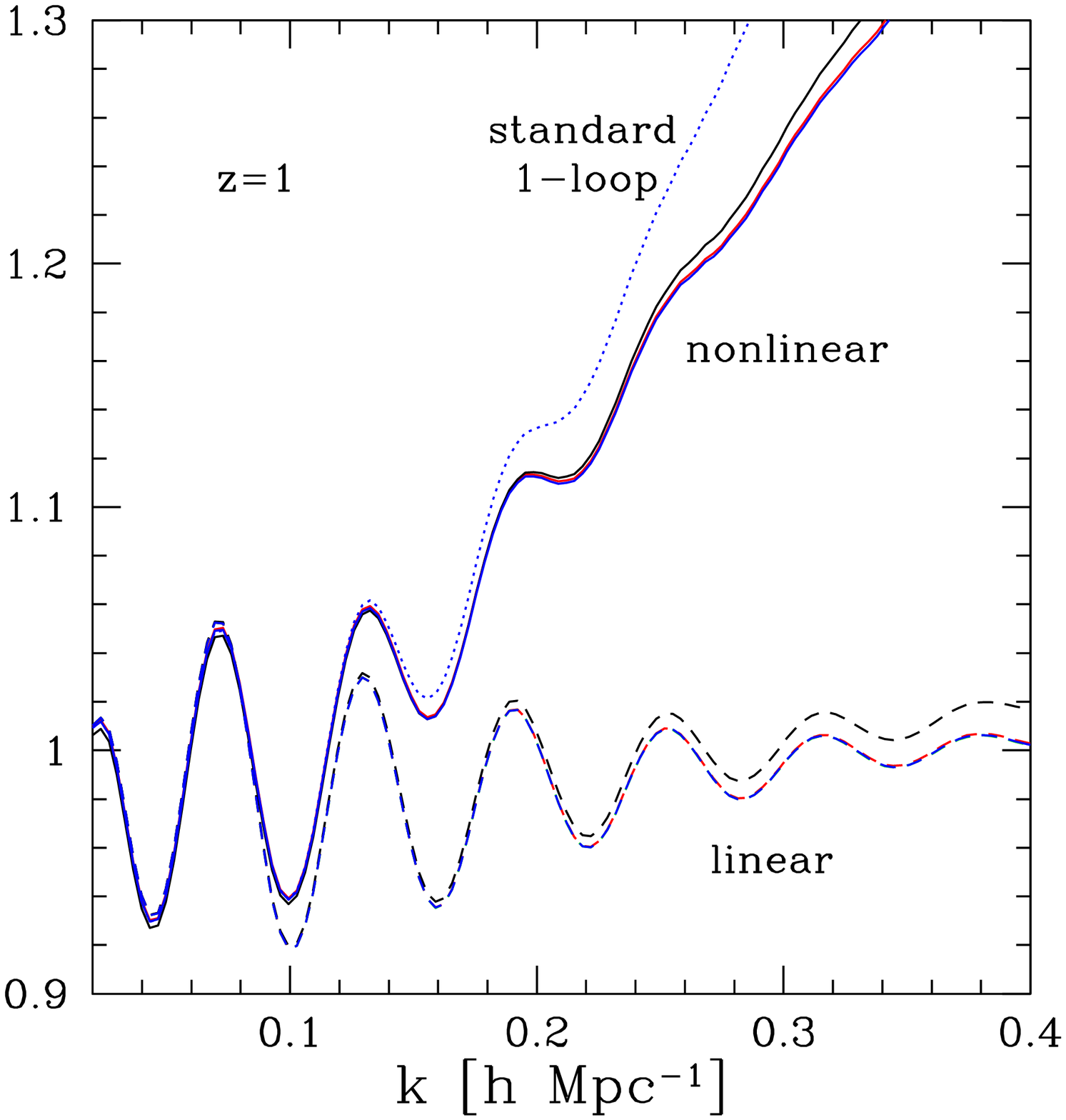}}
\end{center}
\caption{Ratio of the power spectrum $P(k)$ to a smooth
$\Lambda$CDM linear power spectrum $P_{Ls}(k)$ without baryonic oscillations,
from \cite{Eisenstein1999}. We show our results for three models with
$(n,m_0)=(1,0.1)$ (middle red lines), $(0,0.1)$ (upper black lines),
and $(0,1)$ (lower blue lines).
In each case, we plot both the linear power (dashed line) and our nonlinear result
(solid line) from Eq.(\ref{P-2H-1H}), which is based on Eq.(\ref{C-Pi}).
For comparison, we also plot the standard 1-loop result from
Eq.(\ref{C-1loop-standard}) for the case $(0,1)$ (upper blue dotted line).}
\label{fig-BAO}
\end{figure*}

\begin{figure*}
\begin{center}
\epsfxsize=5.9 cm \epsfysize=5.4 cm {\epsfbox{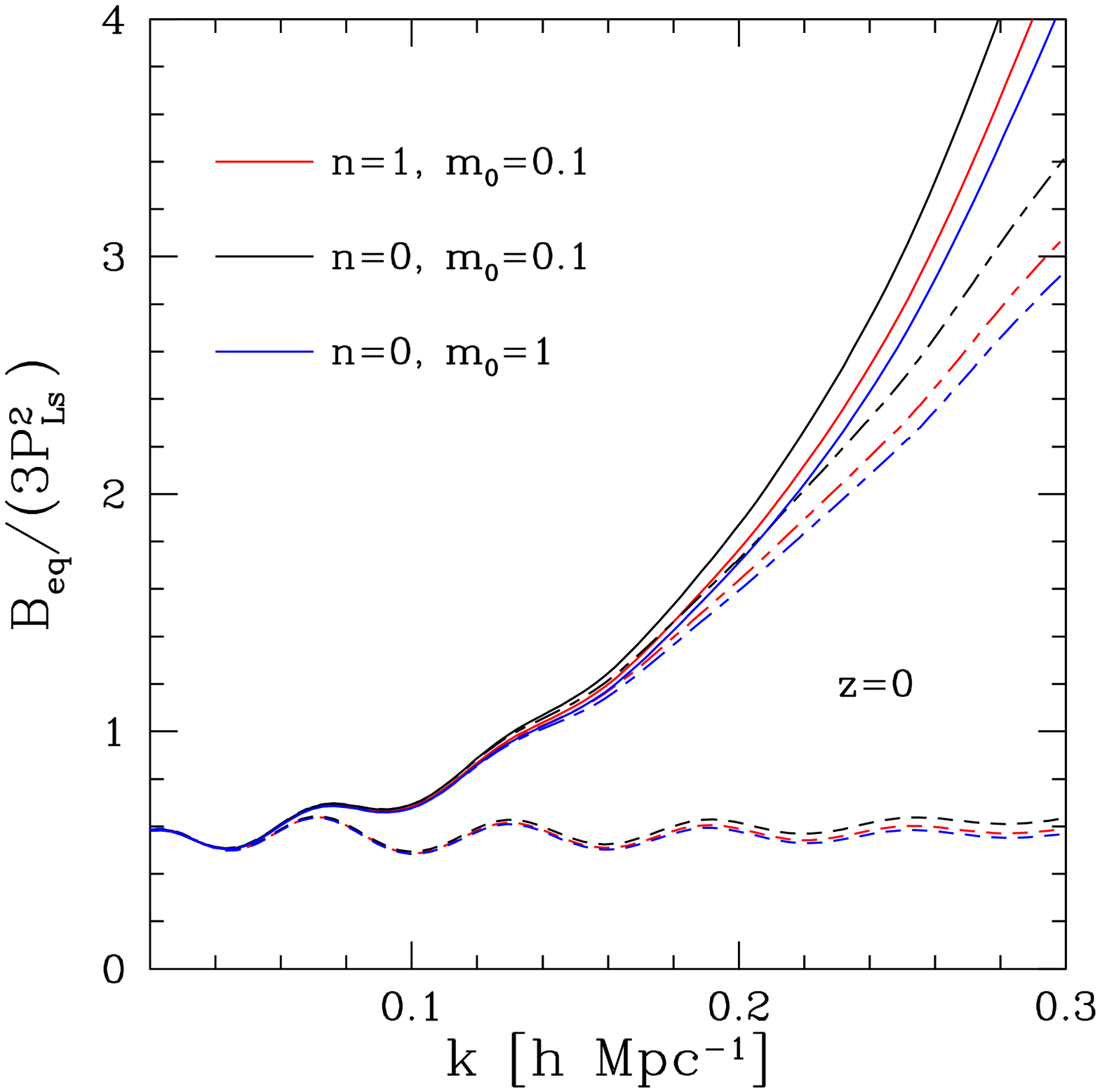}}
\epsfxsize=5.9 cm \epsfysize=5.4 cm {\epsfbox{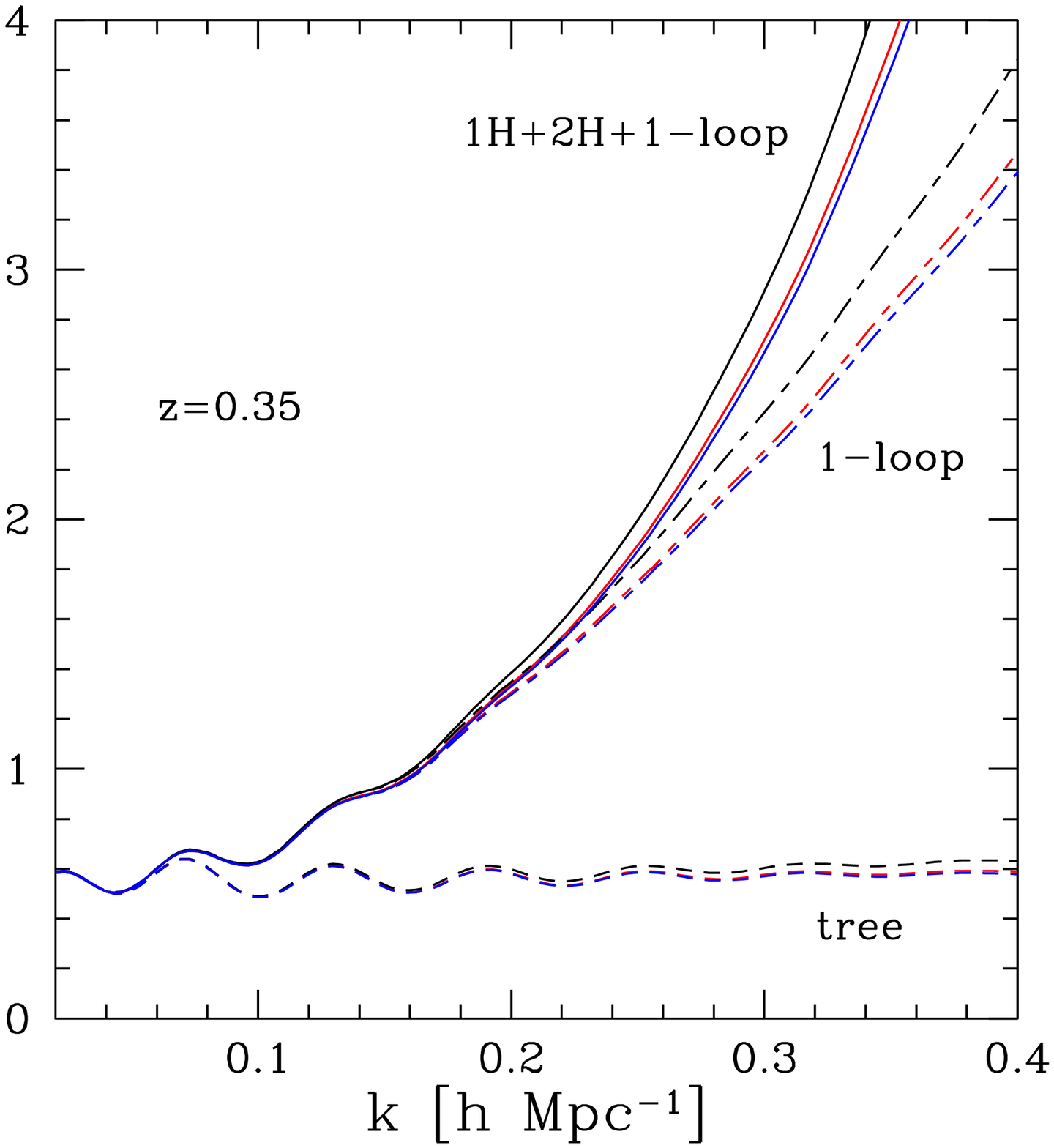}}
\epsfxsize=5.9 cm \epsfysize=5.4 cm {\epsfbox{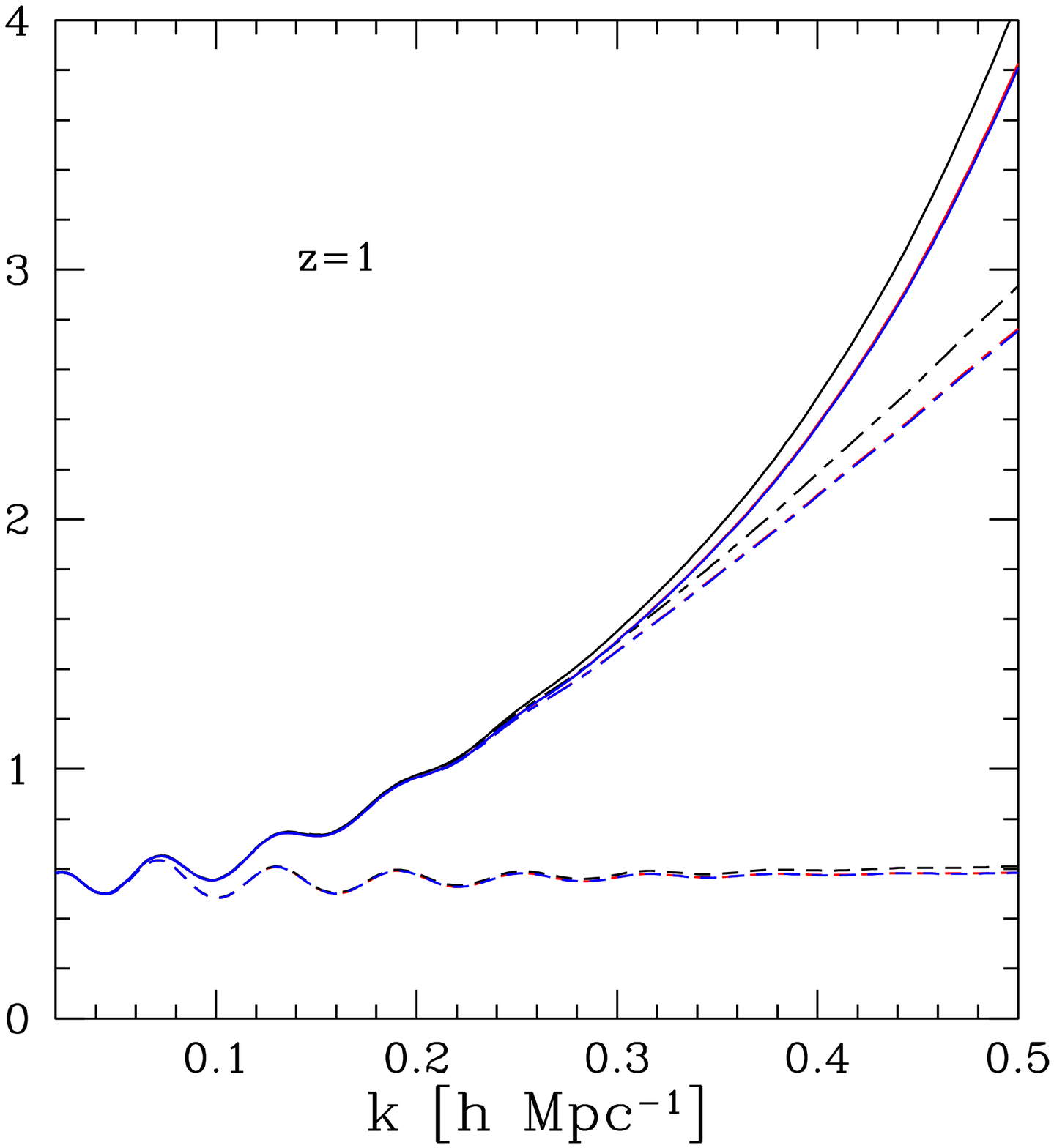}}
\end{center}
\caption{Ratio of the equilateral bispectrum, $B_{\rm eq}(k)=B(k,k,k)$, to the
product $3 P_{Ls}(k)^2$, where $P_{Ls}(k)$ is a smooth $\Lambda$CDM
linear power spectrum without baryonic oscillations, from \cite{Eisenstein1999}.
As in Fig.~\ref{fig-BAO}, we show our results for three models with
$(n,m_0)=(1,0.1)$ (middle red lines), $(0,0.1)$ (upper black lines),
and $(0,1)$ (lower blue lines).
In each case, we plot the tree-level bispectrum (dashed line) from
Eq.(\ref{B-tree-Z}), the 1-loop bispectrum (dash-dotted line) from
Eq.(\ref{B1loop}), and our nonlinear result (solid line) from Eq.(\ref{B-halo}).}
\label{fig-Bk-BAO}
\end{figure*}

\subsubsection{Set up}
\label{setup}

For our numerical computations, we adopt in this paper a flat $\Lambda$CDM reference
model with cosmological parameters
$(\Omega_{\rm m},\Omega_{\rm b},h,\sigma_8,n_{\rm s})
= (0.279, 0.046035, 0.701, 0.817, 0.96)$,
which is consistent with WMAP 5-year observations \citep{Komatsu2009}.
We use a publicly available code, {\tt CAMB} \citep{Lewis2000}, to compute the
linear power spectrum including baryon acoustic oscillations.
This is the same cosmology as used in \cite{Valageas2011d,Valageas2011e}, which
allows a clear comparison with their $\Lambda$CDM results.
Then, the four  models that we consider in this paper, defined by the parameters
$(n,m_0)=(1,0.1), (1,1), (0,0.1)$, and $(0,1)$, as described in
Sect.~\ref{Parameterised-gravity}, are defined by the same initial conditions as this
reference $\Lambda$CDM model. This means that they all coincide at early times and
on large scales, because $\epsilon(k,a)\rightarrow 0$ for $a\rightarrow 0$ or
$k\rightarrow 0$, but their linear variance $\sigma_8$ today on scale $8 h^{-1}$Mpc
slightly differs.

For later use, let us note $\delta_{L(\Lambda)}(\vx,\eta)$ the linear density field within the
reference $\Lambda$CDM cosmology,
\beq
\tdelta_{L(\Lambda)}(\vk,\eta) = D_{+(\Lambda)}(\eta) \, \tdelta_{L0}(\vk) ,
\label{deltaL-LCDM}
\eeq
where $D_{+(\Lambda)}$ is the $\Lambda$CDM linear growing mode,
which does not depend on wavenumber. Then, the actual linear density field can be
written in terms of this reference $\Lambda$CDM linear field as
\beq
\tdelta_L(\vk,\eta) = \frac{D_+(k,\eta)}{D_{+(\Lambda)}(\eta)} \, \tdelta_{L(\Lambda)}(\vk,\eta) .
\eeq
This is merely a re-writing of the initial conditions, which we choose to express at any time
$\eta$ through the reference $\Lambda$CDM growing mode.

\subsubsection{Power spectrum}
\label{BAO-Pk}

We show our results for the matter density power spectrum $P(k)$ on
BAO (baryon acoustic oscillations \cite{Eisenstein2005}) scales in Fig.~\ref{fig-BAO}.
To clearly distinguish the different curves and
the baryon acoustic oscillations we normalize $P(k)$ by a smooth
$\Lambda$CDM linear power spectrum $P_{Ls}(k)$ without baryon
oscillations, from \cite{Eisenstein1999}.
Our nonlinear prediction includes both the perturbative ``two-halo'' part
$P_{2\rm H}(k)$, based on the steepest-descent resummation (\ref{C-Pi}),
and the nonperturbative ``one-halo'' part $P_{\rm 1H}(k)$,
as described in Sect.~\ref{nonlinear} and Eq.(\ref{P-2H-1H}) below.
However, on these scales the power spectrum is dominated by the perturbative
contributions and the full nonlinear result is very close to the resummed
perturbative part (\ref{C-Pi}).

As explained above, all our results converge at low $k$ to the
same reference $\Lambda$CDM power, $P_{(\Lambda)}(k)$, because of our
common choice of initial conditions.
Moreover, on the scales shown in Fig.~\ref{fig-BAO}, this $\Lambda$CDM power
spectrum cannot be distinguished from the
$(n=0,m_0=1)$ result, where the effects of modified gravity are the weakest amongst
the models that we consider here.
As in the $\Lambda$CDM cosmology, the nonlinear evolution amplifies the
power spectrum but erases most of the oscillations.
The difference between the various modified gravity models and General Relativity is rather
small and it is not amplified by the nonlinear evolution.
We clearly see that to probe these deviations it is necessary to go beyond
linear theory and to include at least one-loop corrections.
Moreover, the comparison with the upper dotted curve, which shows the
standard one-loop result for the case $(n=0,m_0=1)$ (which cannot be distinguished
from GR), shows that these modified-gravity effects are at  the order
of or smaller than the accuracy of the standard one-loop prediction.
This means that to probe modified gravity on these scales it is necessary to use
more accurate analytical formalisms, such as the resummation scheme
described in Sect.~\ref{steepest-descent} and used in this paper, or to include higher-order
corrections within the standard perturbative approach
(but this latter option may not be very efficient because the standard perturbative expansion
does not converge very well).
This provides another motivation for the development of efficient perturbative
schemes, which re-sum high-order contributions.

\subsubsection{Bispectrum}
\label{BAO-Bk}

We show our results for the matter density bispectrum on BAO scales in
Fig.~\ref{fig-Bk-BAO}.
Here we only consider equilateral configurations, $B_{\rm eq}(k)=B(k,k,k)$,
and we normalize the bispectrum by $3 P_{Ls}(k)^2$.
Because $P_{Ls}(k)$ is not the actual power spectrum but a smooth
$\Lambda$CDM linear power spectrum without baryon acoustic oscillations, this
ratio is not identical to the usual ``reduced bispectrum''
$Q_{\rm eq}= B_{\rm eq}/(3 P^2)$. However, this allows us to clearly distinguish
the baryon acoustic oscillations of the tree-level bispectrum
(\ref{B-tree})-(\ref{B-tree-Z}).
Again, on these scales the $\Lambda$CDM bispectrum cannot be distinguished
from the $(n=0,m_0=1)$ result.

As for the power spectrum shown in Fig.~\ref{fig-BAO}, the nonlinear evolution
amplifies the bispectrum but erases most of the oscillations.
The difference between the various  models and GR
is again rather small and it is necessary to go beyond the tree-level
prediction. Unfortunately, the comparison between our approximate one-loop
prediction and our full nonlinear model, which includes the nonperturbative
``two-halo'' and ``one-halo'' contributions as described in Sect.~\ref{nonlinear}
below, suggests that one-loop terms are not sufficient
to obtain reliable measures of such modified-gravity effects and that nonperturbative
contributions cannot be neglected.
Since the theoretical accuracy of such nonperturbative terms is lower than the one
of perturbative terms (which can be computed in a systematic and rigorous
fashion), this means that the bispectrum is not a very efficient probe of these
modified-gravity models (unless one can run dedicated N-body simulations
for each modified-gravity scenario).
Thus, the power spectrum studied in Sect.~\ref{BAO-Pk}
should provide a better tool, as the accuracy of its theoretical predictions is better
controlled.

\section{Spherical collapse}
\label{Spherical-collapse}

\subsection{General case}
\label{General}

To go beyond low-order perturbation theory, the main analytical tool that can provide
exact nonlinear results is the study of the spherical collapse. This allows an explicit
computation of the nonlinear dynamics (restricted to spherical symmetry) that can
also serve as a basis to evaluate several quantities of cosmological interest,
such as the halo mass functions and the probability distributions of the density contrast.
We describe in this section the equations that govern the spherical
dynamics and give a simple approximation for typical fluctuations.

Following the usual approach for $\Lambda$CDM or quintessence cosmologies
\cite{Valageas2010,Wang1998}, the physical radius $r(t)$, which contains a constant
mass $M$ until shell-crossing, evolves as
\beq
\ddot{r} = - \frac{\pl \Psi}{\pl r} = - \frac{1}{a} \frac{\pl \Psi}{\pl x}, \;\;\; \mbox{with} \;\;\;
\Psi= \Phi_{\rm N} + \Psi_{\epsilon} ,
\eeq
where $\Psi$ is the total potential seen by massive particles.
Here we note with a dot derivatives with respect to time $t$, physical coordinates
by $\vr$ and comoving coordinates by $\vx$.
Within our framework, defined by Eqs.(\ref{F-continuity-1})-(\ref{F-Euler-1}), the
potential $\Psi$ contains two parts, the usual Newtonian potential
$\Psi_{\rm N}=\Phi_{\rm N}$, associated with General Relativity, and the effective component
$\Psi_{\epsilon}$, associated with the modification of gravity.

In physical coordinates, we have
\beq
\nabla_{\vr}^2 \Phi_{\rm N} = 4\pi\cG \, \left( \rhom^{(\rm phys.)} + (1+3\wde)
\rhode^{(\rm phys.)} \right) ,
\label{Phi-N}
\eeq
where we note with a superscript ``$(\rm phys.)$'' densities in physical coordinates
and we again assumed an uniform dark energy component.
Using Gauss' theorem, this yields the usual part $(\ddot r)_{\rm N}$ of the acceleration of
the shell at radius $r$ \cite{Valageas2010,Wang1998},
\beq
(\ddot r)_{\rm N} = - \frac{4\pi\cG}{3} \, r \left[ \rhom^{(\rm phys.)}(<r)
+ (1+3\wde) \rhode^{(\rm phys.)} \right] ,
\label{r-N-1}
\eeq
where $\rhom^{(\rm phys.)}(<r)$ is the mean physical density within radius $r$,
\beq
\rhom^{(\rm phys.)}(<r) = \frac{3M}{4\pi r^3} .
\eeq

In comoving coordinates (with the background Hubble flow), the effective component $
\Psi_{\epsilon}$ only depends on the matter density fluctuations,
$\delta\rhom=\rhom-\rhob_{\rm m}$, through
\beq
\tilde \Psi_{\epsilon} =  \epsilon(k,t) \, \tdPhi_{\rm N} \;\;\; \mbox{with} \;\;\;
\nabla^2 (\delta\Phi_{\rm N}) = 4\pi\cG \delta\rhom/a ,
\label{Phi-eps-1}
\eeq
whence
\beq
\tilde\Psi_{\epsilon}(\vk,t) =  - \frac{4\pi\cG\rhob_{\rm m}}{a k^2} \, \epsilon(k,t)
\tdelta(\vk,t) .
\eeq
This is a linear approximation in the spherical collapse dynamics which is only valid as long
as the screening effects of modified gravity are not taken into account. When the screening effects appear, the scalar force leading to the extra contribution in Newton's equation is highly suppressed and the spherical over density collapses like in GR. These effects can be modeled out in the top-hat approximation like in  \cite{Brax:2010tj} or using the exclusion set theory \cite{Li:2011qda}. Taking into account these effects is left for future work.

Going back to configuration space, this yields the additional part
$(\ddot r)_{\epsilon}$ due to this ``fifth force'',
\beq
(\ddot r)_{\epsilon} = - \frac{4\pi\cG}{3} \, r \, \rhob_{\rm m}^{(\rm phys.)}
\int_0^{\infty} \dd k \, 4\pi k^2 \epsilon(k) \, \tdelta(k) \, \tW(kx) ,
\label{r-eps-1}
\eeq
where the integral is written in terms of comoving quantities and $x=r/a$.
Here we introduced the Fourier transform of the 3D top-hat of radius $x$ and volume $V$,
\beq
\tW(kx) = \int_V\ \frac{\dd\vx'}{V} \, e^{\ii \vk\cdot\vx'}
= 3 \frac{\sin(kx)-kx\cos(kx)}{(kx)^3} .
\label{W-3D-def}
\eeq
If $\epsilon$ does not depend on wavenumber we can check that
Eq.(\ref{r-eps-1}) gives
\beqa
(\ddot r)_{\epsilon} & = & - \epsilon \frac{4\pi\cG}{3} \, r \, \rhob_{\rm m}^{(\rm phys.)}
\, \delta(<x) \\
& = & - \epsilon \frac{4\pi\cG}{3} \, r \left[ \rhom^{(\rm phys.)}(<r)
- \rhob_{\rm m}^{(\rm phys.)} \right] .
\label{eps-const}
\eeqa
In agreement with Eq.(\ref{r-N-1}), an uniform $\epsilon(t)$ gives rise to a fifth force
that is proportional to the Newtonian gravitational force where we subtract the
background part (associated with the mean density of the universe).

Collecting Eqs.(\ref{r-N-1}) and (\ref{r-eps-1}) we obtain the equation of motion
\beqa
\ddot r & = & - \frac{4\pi\cG}{3} \, r \biggl [ \rhom^{(\rm phys.)}(<r)
+ (1+3\wde) \rhode^{(\rm phys.)}  \nonumber \\
&& + \rhob_{\rm m}^{(\rm phys.)} \int_0^{\infty} \dd k \, 4\pi k^2 \epsilon(k) \,
\tdelta(k) \, \tW(kx) \biggl ] .
\label{r-tot-1}
\eeqa
As in \cite{Valageas2010,Wang1998}, it is convenient to introduce the normalized
radius $y(t)$ defined as
\beq
y(t) = \frac{r(t)}{a(t) q} \;\; \mbox{with} \;\;  q= \left(\frac{3M}{4\pi\rhob_{\rm m}}\right)^{1/3} ,
\;\; y(t=0) =1 .
\label{y-def}
\eeq
Thus, $q$ is the Lagrangian comoving coordinate of the shell $r(t)$, that is, the comoving
radius that would enclose the same mass $M$ in a uniform universe with the same cosmology.
This also implies
\beq
\frac{\rhom^{(\rm phys)}(<r)}{\rhob_{\rm m}^{(\rm phys)}} =  y^{-3} , \;\;\;\;
\delta_r \equiv \delta(<r) = y^{-3} -1 .
\eeq
Choosing again $\eta=\ln a(t)$ as the time coordinate, as in the previous sections,
Eq.(\ref{r-tot-1}) reads as
\beqa
\frac{\pl^2 y}{\pl \eta^2} + \left( \frac{1}{2}-\frac{3}{2} \wde \Ode \right) \frac{\pl y}{\pl \eta}
+ \frac{\Om}{2} \left( y^{-3} - 1\right) y & = & \nonumber \\
&& \hspace{-5.9cm} - \frac{\Om}{2} y \int_0^{\infty} \dd k \, 4\pi k^2 \epsilon(k) \,
\tdelta(k) \, \tW(kx) .
\label{y-1}
\eeqa
The left hand side is the usual result in $\Lambda$CDM cosmology
\cite{Valageas2010,Wang1998} and the right hand side is the new term associated
with the ``fifth force''.
If $\epsilon$ does not depend on wavenumber, the integral reduces to
$\epsilon(a) \delta(<x)= \epsilon (y^{-3}-1)$, as in the usual third term of the left hand
side. Then, the motion of each mass shell, described by $y(M,\eta)$ or $r(M,\eta)$,
is independent of the other shells before shell crossing.
If $\epsilon(k,a)$ depends on wavenumber, the integral does not reduce
to a simple function of $y$ at the same mass scale and it explicitly depends on
the whole density profile, $\delta(x)$ or $\tdelta(k)$ in Fourier space, of the matter
perturbation.
Then, the dynamics of all mass shells are coupled at all times, even before shell crossing,
and we must solve for the evolution of the full density profile with time, $y(M,\eta)$,
as a function of $M$ and $\eta$.

In previous works \citep{Schmidt2009,Li2012a}, the spherical collapse dynamics was often
approximated
through an effective rescaling of Newton's constant (this corresponds to a function
$\epsilon(a)$ that does not depend on $k$). This allows one to recover the
usual form of the equations of motion where all shells are decoupled before shell crossing.
By varying this effective Newton constant \cite{Schmidt2009}, or making it a dynamical
variable that
depends on the environment \citep{Li2012a}, one may capture screening effects.
Here we do not include such screening effects but Eq.(\ref{y-1}) takes into account the
dependence of the dynamics on the density profile.
This allows us to include the effects associated with the dependence on wavenumber
of $\epsilon(k,a)$.
As we will check in Fig.~\ref{fig-deltac_M_z0} below, this already yields a dependence
on mass of the linear density threshold $\delta_c(M)$ associated with halo formation.

Thus, the modified-gravity term makes the equation of motion significantly more complex,
because it is no longer local and it turns the usual ordinary differential equation into a partial
integro-differential equation.

\subsection{Approximation for typical profiles}
\label{typical}

Let us assume we are interested in the dynamics of a single mass shell $M$. Then,
we wish to obtain from Eq.(\ref{y-1}) a closed approximate equation for
$y_M(\eta) \equiv y(M,\eta)$, which does not involve the other shells $M'$.
The simplest method is to use an ansatz for the density profile $\delta(x,\eta)$,
or $\tdelta(k,\eta)$, that is parameterized by $y_M(\eta)$.
This will simplify numerical computations because it will transform Eq.(\ref{y-1}) into
a single ordinary differential equation.
Then, let us recall that the mean conditional profile of the linear density contrast
$\delta_L(\vx)$, under the constraint that the mean density contrast within a comoving
radius $R$ is equal to $\delta_{LR}$, reads as \cite{Bernardeau1994a}
\beq
\delta_L(\vx) = \frac{\delta_{LR}}{\sigma_R^2} \int_V \frac{\dd\vx'}{V}
\, C_{\delta_L\delta_L}(\vx,\vx') ,
\label{prof-lin-1}
\eeq
where $C_{\delta_L\delta_L}$ is the matter density linear correlation,
\beqa
C_{\delta_L\delta_L}(\vx_1,\vx_2) & = & \lag \delta_L(\vx_1) \delta_L(\vx_2) \rag
\nonumber \\
&& \hspace{-2cm} = \int_0^{\infty} \dd k \; 4\pi k^2 P_L(k) \;
\frac{\sin(k|\vx_2-\vx_1|)}{k|\vx_2-\vx_1|} ,
\label{CL-delta-def}
\eeqa
and $\sigma_R^2$ is the variance of the linear density contrast at scale $R$,
\beq
\sigma_R^2 = \lag \delta_{LR}^2\rag = \int_0^{\infty} \dd k \; 4\pi k^2 P_L(k)
\tW(kR)^2 .
\label{sigL-def}
\eeq
This only relies on the assumption that the linear density field is Gaussian.
Then, we consider the approximation where the density profile used in Eq.(\ref{y-1})
is set to
\beq
\delta(\vx) = \frac{\delta_{x_M}}{\sigma^2_{x_M}} \int_{V_M} \frac{\dd\vx'}{V_M} \, C_{\delta_L\delta_L}(\vx,\vx') ,
\label{prof-approx-1}
\eeq
which reads in Fourier space as
\beq
\tdelta(k) = \frac{y_M^{-3}-1}{\sigma^2_{x_M}} \, P_L(k) \, \tW(k x_M) ,
\label{prof-approx-k}
\eeq
where we used $\delta_{x_M}=y_M^{-3}-1$.
Here $x_M(\eta)= r_M(\eta)/a(\eta)=y_M(\eta) q_M$ is the comoving radius of the shell $M$
and it follows its spherical dynamics.
Substituting the ansatz (\ref{prof-approx-k}) into Eq.(\ref{y-1}) gives the equation of motion
\beqa
\frac{\dd^2 y_M}{\dd\eta^2} + \left( \frac{1}{2}-\frac{3}{2} \wde \Ode \right)
\frac{\dd y_M}{\dd\eta} + \frac{\Om}{2} \left( y_M^{-3} \! - \! 1\right) y_M && \nonumber \\
&& \hspace{-7.3cm} \times \left( \! 1 + \frac{1}{\sigma_{x_M}^2}
\int_0^{\infty} \!\! \dd k \, 4\pi k^2 \epsilon(k) P_L(k) \tW(kx_M)^2 \! \right) = 0 . \nonumber \\
&& \label{y-approx-1}
\eeqa
The equation (\ref{y-approx-1}) is exact if $\epsilon$ does not depend on
wavenumber, in which case the parenthesis is equal to $(1+\epsilon(\eta))$ and we
recover the behaviour of Eq.(\ref{eps-const}).
It is also valid at order one over $\delta_L$ and $\epsilon$ when the initial perturbation has
the linear profile (\ref{prof-lin-1}) at early time.
Thus, it agrees with the typical profile (\ref{prof-lin-1}), under the constraint $\delta_{Lx_M}$
at mass shell $M$, in the linear regime, at zeroth order over $\epsilon$.
It is no longer exact at higher orders over $\delta_L$ because the nonlinear dynamics
changes the shape of the density profile in a complex fashion.
It is not valid at order $\epsilon $, even in the linear regime, because the mean profile
(\ref{prof-lin-1}) is not a solution of the linear dynamics, as the linear growing
mode $D_+(k,a)$ depends on wavenumber.
In our case, where $\epsilon \ll 1$, this is a negligible effect and we would actually
obtain similar results by using in Eqs.(\ref{prof-approx-1}) and (\ref{prof-approx-k})
the reference $\Lambda$CDM linear correlation $C_{\delta_L\delta_L(\Lambda)}$
and power $P_{L(\Lambda)}$.

\subsection{Spherical-collapse mapping}
\label{mapping}

\begin{figure}
\begin{center}
\epsfxsize=8.5 cm \epsfysize=6 cm {\epsfbox{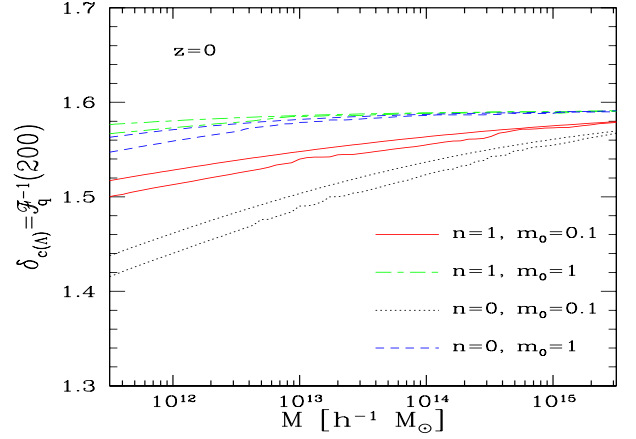}}
\end{center}
\caption{Reference linear density contrast $\delta_{c(\Lambda)}=\cF_q^{-1}(200)$ associated
with a nonlinear density threshold of $200$ at redshift $z=0$. We show our results
as a function of the halo mass $M$ for four $(n,m_0)$ models, for typical initial profiles
of the form (\ref{saddle-prof}). In each case, the upper curve is the approximate result
from Eq.(\ref{y-approx-1}) and the lower curve the exact result from Eq.(\ref{y-1}).}
\label{fig-deltac_M_z0}
\end{figure}

In the linear regime we can check that Eq.(\ref{y-1}) agrees with Eq.(\ref{D-pm}) for the linear
growing mode. Indeed, using $y_L=1-\delta_{Lq}/3$,
$\delta_{Lq} = \int_V \dd\vx  \delta_L(\vx)/V$, and $x=q$ at lowest order,
Eq.(\ref{y-1}) becomes at linear order:
\beqa
\int_V \frac{\dd\vx}{V} \int \dd\vk \, e^{\ii\vk\cdot\vx} \biggl \lbrace
\frac{\pl^2\tdelta_L}{\pl\eta^2}(\vk) + \left( \frac{1}{2}-\frac{3}{2} \wde \Ode \right)
\frac{\pl\tdelta_L}{\pl\eta}(\vk) && \nonumber \\
&& \hspace{-5.5cm} - \frac{3\Om}{2} (1+\epsilon(k))\tdelta_L(\vk)
\biggl \rbrace = 0 .
\label{y-L2}
\eeqa
This agrees with Eq.(\ref{D-pm}) and we recover the linear solution
$\tdelta_L(\vk,\eta) = D_+(k,\eta) \tdelta_{L0}(\vk)$.

At linear order, the ansatz (\ref{prof-approx-k}) reads in Fourier-space as
$\tdelta_L(\vk)= (\delta_{Lq_M}/\sigma_{q_M}^2) P_L(k) \tW(k q_M)$.
Substituting into Eq.(\ref{y-L2}) remains exact if the profile of the perturbation
is given by Eq.(\ref{prof-approx-1}) (or for the shell $M$, whatever the initial profile, if
$\epsilon$ does not depend on wavenumber).

We now consider the spherical dynamics of typical initial perturbations, of the form
(\ref{prof-lin-1}) at early times, which we write as
\beq
\delta_{Lq'(\Lambda)} = \delta_{Lq(\Lambda)}
\frac{\sigma^2_{q,q'(\Lambda)}}{\sigma_{q(\Lambda)}^2} ,
\label{saddle-prof}
\eeq
for the mean initial density contrast within arbitrary radius $q'$.
Here, as explained in Sect.~\ref{setup}, we choose to write the initial conditions
in terms of the reference $\Lambda$CDM linear field, which is simply an
``update'' at arbitrary time $\eta$ of the initial field $\delta_{L0}$ given at a fixed time.
This is more convenient than using the actual linear field $\delta_L$, which depends
on the modified-gravity growing mode $D_+(k,\eta)$ and mixes dependences on the
initial conditions and on the modified gravity parameters.
In this fashion, Eq.(\ref{saddle-prof}) describes the same initial condition
for all our  models.
Here $\sigma^2_{q_1,q_2(\Lambda)}$ is the cross-correlation of the smoothed
reference linear density contrast at scales $q_1$ and $q_2$,
\beqa
\sigma^2_{q_1,q_2(\Lambda)} & = &
\lag \delta_{Lq_1(\Lambda)} \delta_{Lq_2(\Lambda)} \rag \nonumber \\
&& \hspace{-1cm} = \int_0^{\infty} \dd k \; 4\pi k^2 P_{L(\Lambda)}(k) \tW(kq_1) \tW(kq_2) ,
\eeqa
and $\sigma_{q(\Lambda)}^2=\sigma_{q,q(\Lambda)}^2$.
For each mass scale $q$, with $M=(4\pi/3) \rhob_{\rm m}q^3$, and initial amplitude
$\delta_{Lq(\Lambda)}$, which define the initial condition (\ref{saddle-prof}), we can solve the
spherical dynamics (\ref{y-1}) or the approximate dynamics (\ref{y-approx-1}).
For the ``exact'' dynamics (\ref{y-1}) we consider for simplicity that inner shells
that have already collapsed to the center of the halo remain at the center.
(After shell crossing we should modify Eq.(\ref{y-1}) to take into account the
change with time of the mass enclosed within a given shell. However, we do not
consider this effect because radial orbits suffer from a strong instability, which diverges
at the time of collapse to the center \citep{Valageas2002b}, and after that time one
should include transverse motions that lead to virialization.)
As long as shell crossing is restricted to inner shells, within the mass scale $M$ of
interest, this is not a very serious problem because the dynamics is mostly sensitive
to the total mass enclosed within a given radius (as in the usual Newtonian case
or for $\epsilon$ that does not depend on wavenumber) or to the local slope of the
density profile (for the low-$k$ behaviour $\epsilon(k) \propto k^2$).

At a given mass scale $q$ and time $\eta$, this defines a mapping,
$\delta_{Lq(\Lambda)}\mapsto\delta_x=\cF_q(\delta_{Lq(\Lambda)})$, from the reference
linear density contrast $\delta_{Lq(\Lambda)}$ to the nonlinear density contrast $\delta_x$.
Here $x$ is again the Eulerian comoving radius of the shell $M$, with $x=r/a=y q$
as in (\ref{y-def}).

If $\epsilon$ does not depend on wavenumber, this mapping does not depend
on the scale $q$ nor on the shape of the initial profile.
If $\epsilon$ depends on wavenumber, this mapping depends both on the mass scale $q$
(whence the subscript $q$ in $\cF_q$) and on the initial shape of the profile (which is why
we had to choose a specific case, such as the typical shape (\ref{saddle-prof})).
This implies that if we choose for instance a given nonlinear density threshold,
such as 200, to define halos, the associated linear density contrast
$\delta_{c(\Lambda)}=\cF_q^{-1}(200)$ depends on the mass of the halo (through the
scale $q$).

We show our results for this linear density threshold $\cF_q^{-1}(200)$
at redshift $z=0$ in Fig.~\ref{fig-deltac_M_z0}. For each  model we plot
both the exact result from Eq.(\ref{y-1}) and the approximate result from
Eq.(\ref{y-approx-1}).
We clearly see the mass dependence associated with the modification of
gravity. For positive $\epsilon$ gravitational clustering is more efficient and
a lower value of $\delta_{L(\Lambda)}$ is required to reach the nonlinear density contrast
$\delta=200$. Because we recover General Relativity on large scales
($\epsilon \rightarrow 0$ for $k\rightarrow 0$) all curves converge to the
$\Lambda$CDM threshold at large mass and show increasingly large deviations
from GR at smaller mass.
 The asymptotic value is $\delta_c \simeq 1.59$ rather than $1.67$ as we define
$\delta_c$ as $\cF_q^{-1}(200)$ instead of $\cF_q^{-1}(\infty)$, that is, by a
nonlinear density contrast of $200$ rather than by the full collapse to the center,
as in \cite{Valageas2009}.

Similar trends were obtained in \cite{Li2012a}, using a simplified dynamics described by
an effective Newton constant that depends on the ``environment'' density, which allowed
them to include screening effects. Thus, because the latter are more important for large
mass they obtained a mass-dependent threshold $\delta_c$ that decreases at small
mass and converges to the GR value at large mass.
We can see in Fig.~\ref{fig-deltac_M_z0} that even without such screening
effects, a dependence on mass is already present because of the dependence on
wavenumber of $\epsilon(k,a)$. Since both effects show similar trends, including them
both would give a steeper dependence on mass than in Fig.~\ref{fig-deltac_M_z0}.
Nevertheless, it is interesting to also investigate both mechanisms separately,
as their relative amplitude depends on the details of the modified-gravity model.

We can see that the approximation (\ref{y-approx-1}) somewhat underestimates the
departure from the GR result. This can be understood from the fact that the
dynamics steepens the density profile, which amplifies the right hand side in
Eq.(\ref{y-1}).
Nevertheless, the approximation (\ref{y-approx-1}), which is much easier to compute,
gives a reasonable estimate of the modified-gravity effect.
Because inner shells have already collapsed when the shell at mass $M$ reaches the
nonlinear threshold $\delta_x=200$, we should include virialization effects which
smooth out the inner density profile. Therefore, the difference seen in
Fig.~\ref{fig-deltac_M_z0}
should actually be somewhat overestimated. Moreover, for smaller nonlinear
density contrast $\delta_x$ the relative deviation decreases, because the ansatz
(\ref{prof-approx-1}) is exact at linear order (for our initial conditions).
Thus, for practical estimates the approximation (\ref{y-approx-1}) should be sufficient,
at least in a first step.

\section{Density Contrast Probability in the Quasilinear Regime}
\label{Probability-distribution}

Following \cite{Valageas2002,Valageas2009}, we can use the spherical collapse
dynamics described in Sect.~\ref{Spherical-collapse} to derive the probability
distribution of the matter density contrast in the quasi-linear regime.

To compute the probability distribution, $\cP(\delta_x)$, of the nonlinear density
contrast within a sphere of comoving radius $x$, it is convenient to introduce the cumulant
generating function
\beqa
e^{-\varphi(y)/\sigma_{x(\Lambda)}^2}
& \equiv & \left\lag e^{-y\delta_x/\sigma_{x(\Lambda)}^2}\right \rag
\label{phi-average-1} \\
& = & \int_{-1}^{\infty} \dd\delta_x \; e^{-y\delta_x/\sigma_{x(\Lambda)}^2} \; \cP(\delta_x) .
\label{phi-P-def}
\eeqa
This determines the distribution $\cP(\delta_x)$ through the inverse Laplace transform
\beq
\cP(\delta_x) = \inta \frac{\dd y}{2\pi \ii \sigma_{x(\Lambda)}^2} \;
e^{[y\delta_x-\varphi(y)]/\sigma_{x(\Lambda)}^2} .
\label{P-phi-def}
\eeq
In Eqs.(\ref{phi-P-def})-(\ref{P-phi-def}) we rescaled the cumulant generating function
by a factor $\sigma_{x(\Lambda)}^2$ so that it has a finite limit in the quasilinear regime,
$\sigma_{x(\Lambda)}\rightarrow 0$, for the case of Gaussian initial fluctuations
\cite{Bernardeau2002}. In particular, its expansion at $y=0$ reads
\beq
\varphi(y) = - \sum_{n=2}^{\infty} \frac{(-y)^n}{n!} \;
\frac{\lag\delta_x^n\rag_c}{\sigma_{x(\Lambda)}^{2(n-1)}} .
\label{phi-cum}
\eeq
The average (\ref{phi-average-1}) can be written as the path-integral
\beq
e^{-\varphi(y)/\sigma_{x(\Lambda)}^2} = (\det C_{\delta_L\delta_L(\Lambda)}^{-1})^{1/2} \!\!
\int \!\! \cD \delta_{L(\Lambda)} \; e^{-S[\delta_{L(\Lambda)}]/\sigma_{x(\Lambda)}^2} ,
\label{phi-path}
\eeq
where $C_{\delta_L\delta_L(\Lambda)}^{-1}$ is the inverse matrix of the two-point correlation
of the reference linear density field and the action $S$ reads as
\beq
S[\delta_{L(\Lambda)}] = y \, \delta_x[\delta_{L(\Lambda)}] + \frac{\sigma_{x(\Lambda)}^2}{2} \, \delta_{L(\Lambda)}
\cdot C_{\delta_L\delta_L(\Lambda)}^{-1} \cdot \delta_{L(\Lambda)}
\label{S-dL-def}
\eeq
Here $\delta_x[\delta_{L(\Lambda)}]$ is the nonlinear functional which assigns to the initial
condition, defined by the reference linear density field $\delta_{L(\Lambda)}(\vx')$, the
nonlinear density contrast $\delta_x$ within the sphere of radius $x$.

As in Sect.~\ref{mapping}, we choose to define the initial conditions through the
reference $\Lambda$CDM linear field $\delta_{L(\Lambda)}$. We could also write all
expressions above in terms of the actual linear field $\delta_L$, its correlation
$C_{\delta_L\delta_L}$, and the variance $\sigma_x^2$.
Here we prefer the formulation (\ref{phi-path}) because it clearly separates the initial
conditions from the modified-gravity effects. Thus, in the action (\ref{S-dL-def}) all
modified-gravity effects are enclosed in the functional $\delta_x[\delta_{L(\Lambda)}]$,
which describes the gravitational dynamics, whereas if we express the initial conditions
in terms of the $\epsilon$-dependent linear field $\delta_L$ these modified gravity effects
would appear in all terms of the action.
Of course, we adopt this formulation because we wish to compare with this $\Lambda$CDM
reference several models that only show small deviations.

The action $S$ does not depend on the normalization of the linear power spectrum
since both $\sigma_{x(\Lambda)}^2$ and $C_{\delta_L\delta_L(\Lambda)}$ are proportional
to $P_{L(\Lambda)}$.
Then, in the quasilinear limit, $\sigma_{x(\Lambda)}\rightarrow 0$, the path integral
(\ref{phi-path}) is dominated by the minimum of the action \cite{Valageas2002},
\beq
\sigma_{x(\Lambda)} \rightarrow 0 : \;\; \varphi(y) \rightarrow
\min_{\delta_{L(\Lambda)}(\vx')} S[\delta_{L(\Lambda)}] .
\label{min}
\eeq
Using the spherical symmetry of the top-hat window $W$ that defines the spherical
average $\delta_x$, one obtains a spherical saddle-point \cite{Valageas2002}.
In General Relativity its linear radial profile is given by Eq.(\ref{saddle-prof}),
where $q$ is the Lagrangian radius that corresponds to the Eulerian
radius $x$,
\beq
q^3 = (1+\delta_x) x^3 .
\label{q-r}
\eeq
Then, the amplitude $\delta_{Lq(\Lambda)}$ of the saddle-point (\ref{saddle-prof}), which also
sets the scale $q$ through Eq.(\ref{q-r}), is given by the spherical-collapse mapping,
\beq
\delta_x = \cF(\delta_{Lq(\Lambda)}) .
\label{F-coll}
\eeq
This derivation agrees with the results that can be obtained from a perturbative
computation of the cumulants $\lag\delta_x^n\rag_c$ at leading order and a
resummation of the series (\ref{phi-cum}) \cite{Bernardeau1994a}.
It also extends these results to the case where the series (\ref{phi-cum}) has a zero
radius of convergence, which occurs when $\cP(\delta_x)$ decreases more slowly
than a simple exponential at large densities \citep{Valageas2002}
\footnote{In the context of $\Lambda$CDM cosmologies, this feature appears for
power-law initial power spectra $P_L(k) \propto k^n$ with $n<0$, see
\cite{Valageas2002} for details.}.

A nice feature of this derivation is that it bypasses the computation of the cumulants
$\lag\delta_x^n\rag_c$ through the kernels $F_n^s$ of Eq.(\ref{Fn}),
as all spherically-averaged quantities are given by the spherical-dynamics
mapping $\cF(\delta_{L(\Lambda)})$ (which includes terms at all orders by expanding
over $\delta_{L(\Lambda)}$).
However, the problem is more complex in our case because of the dependence
of $\epsilon(k,a)$ on wavenumber. Indeed, this means that the nonlinear density
contrast $\delta_x$ at radius $x$ does not depend on the linear density
contrast $\delta_{Lq(\Lambda)}$ at the Lagrangian radius $q$, associated with the same
mass $M$ only. Indeed, as discussed in Sect.~\ref{Spherical-collapse}, the spherical
dynamics (\ref{y-1}) depends on the full shape of the initial perturbation.
Taking into account this modification changes the profile $\delta_{L(\Lambda)}(\vx')$ of the
minimum of the action $S[\delta_{L(\Lambda)}]$ in Eq.(\ref{min}), because the functional
$\delta_x[\delta_{L(\Lambda)}(\vx')]$ is no longer of the form
$\delta_x=\cF(\delta_{Lq(\Lambda)})$.

To simplify the analysis we neglect this change of the profile of the saddle-point.
This is actually valid to first order over $\epsilon$.
Indeed, let us write the action
$S$ as $S=S_0+\hat{\epsilon} S_1$, where $S_0$ is the usual $\Lambda$CDM
action (where $\epsilon=0$), and $S_1$ is the modification due to a nonzero
$\epsilon(k,a)$ kernel, where we factored out a normalization parameter
$\hat{\epsilon}$ that scales as $\epsilon$.
Because of this new term $\hat{\epsilon} S_1$, the saddle-point $\delta_{L(\Lambda)}$
is changed to
$\delta_{L(\Lambda)}=\delta_{L0(\Lambda)}+\hat{\epsilon} \delta_{L1(\Lambda)}$, where
$\delta_{L0(\Lambda)}$ is the GR saddle-point (\ref{saddle-prof}).
Then, the generating function is changed to
$\varphi(y) \rightarrow S_0[\delta_{L0(\Lambda)}+\hat{\epsilon} \delta_{L1(\Lambda)}]
+\hat{\epsilon} S_1[\delta_{L0(\Lambda)}+\hat{\epsilon} \delta_{L1(\Lambda)}]$.
Because $\delta_{L0(\Lambda)}$ is a saddle-point of the action $S_0$, we have
$S_0[\delta_{L0(\Lambda)}+\hat{\epsilon} \delta_{L1(\Lambda)}]=
S_0[\delta_{L0(\Lambda)}]+\cO(\hat{\epsilon}^2)$, that is, $S_0[\delta_{L(\Lambda)}]$ is only
modified by terms of order $\epsilon^2$.
Because of the prefactor $\hat{\epsilon}$ we also have
$\hat{\epsilon} S_1[\delta_{L0(\Lambda)}+\hat{\epsilon} \delta_{L1(\Lambda)}]=
\hat{\epsilon} S_1[\delta_{L0(\Lambda)}]+\cO(\hat{\epsilon}^2)$.
Therefore, $S[\delta_{L(\Lambda)}]=S[\delta_{L0(\Lambda)}]+\cO(\hat{\epsilon}^2)$ and we can
neglect the change of the saddle-point up to first order over $\epsilon$.
In fact, we do better than this because we only neglect the change of the radial profile
but we keep track of the dependence on $\epsilon$ of the amplitude $\delta_{Lq(\Lambda)}$
of the saddle-point.

On the other hand, if we use the approximation (\ref{y-approx-1}) instead of Eq.(\ref{y-1}),
the functional $\delta_x[\delta_{L(\Lambda)}(\vx')]$ is again of the form
$\delta_x=\cF_q(\delta_{Lq(\Lambda)})$
and the saddle-point profile (\ref{saddle-prof}) becomes exact within this
approximation.

\begin{figure}
\begin{center}
\epsfxsize=8.5 cm \epsfysize=6 cm {\epsfbox{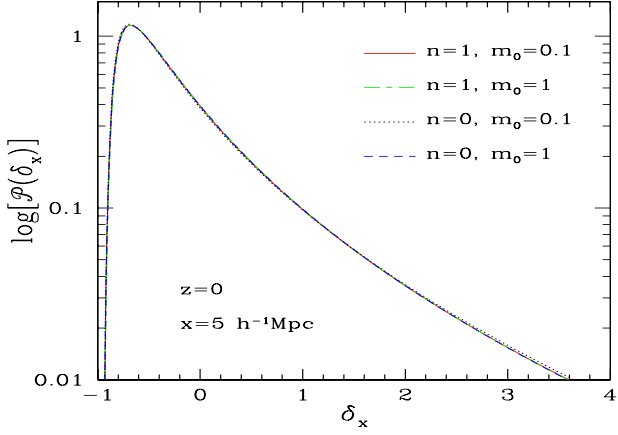}}
\end{center}
\caption{Probability distribution of the matter density contrast within spherical
cells of radius $5h^{-1}$Mpc at $z=0$ (all curves almost fall on each other).}
\label{fig-lPrho_r5_z0}
\end{figure}

\begin{figure}
\begin{center}
\epsfxsize=8.5 cm \epsfysize=6 cm {\epsfbox{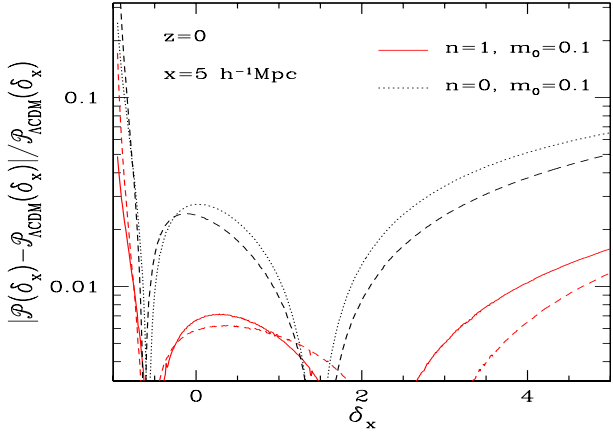}}
\end{center}
\caption{Relative deviation from General Relativity of the probability distribution
$\cP(\delta_x)$, at redshift $z=0$ for a radius $x=5h^{-1}$Mpc.
For each $(n,m_0)$ model the deviation from GR is positive at low
and high densities and negative around $\delta\sim 0$. The solid and dotted lines
are the exact results from Eq.(\ref{y-1}) for $(n,m_0)=(1,0.1)$ and $(0,0.1)$.
The closest dashed line of the same color is the result from the approximation
(\ref{y-approx-1}), for the same value of $(n,m_0)$.}
\label{fig-dlPrho_r5_z0}
\end{figure}

In both cases, whether we use the approximation (\ref{y-approx-1}) or the
exact equation (\ref{y-1}), the function $\cF_q$ now also depends on the scale $q$,
in contrast to the usual Newtonian case.

Then, from this spherical-collapse mapping $\cF_q(\delta_{Lq(\Lambda)})$,
described in Sect.~\ref{mapping}, we obtain the generating function $\varphi(y)$
as follows \cite{Valageas2002,Valageas2009}.
Substituting the profile (\ref{saddle-prof}) into Eq.(\ref{S-dL-def}) and using
Eq.(\ref{F-coll}) the minimum (\ref{min}) reads as
\beq
\varphi(y) = \min_{\delta_{Lq(\Lambda)}} \left[ y \cF_q(\delta_{Lq(\Lambda)}) + \frac{1}{2} \,
\frac{\sigma_{x(\Lambda)}^2}{\sigma_{q(\Lambda)}^2} \, \delta_{Lq(\Lambda)}^2 \right] .
\label{phi-min-F}
\eeq
Defining the function $\tau(\zeta)$ through the parametric system
\cite{Valageas2002,Bernardeau1994},
\beq
\zeta=\delta_x=\cF_q(\delta_{Lq(\Lambda)}) \;\;\;  \mbox{and} \;\;\;
\tau = - \delta_{Lq(\Lambda)} \frac{\sigma_{x(\Lambda)}}{\sigma_{q(\Lambda)}} ,
\label{tau-G-def}
\eeq
the minimum (\ref{phi-min-F}) also writes as
\beq
\varphi(y) = \min_{\zeta} \left[ y \zeta + \frac{\tau(\zeta)^2}{2} \right] .
\label{phi-min-G}
\eeq
This corresponds to the implicit equations (Legendre transform)
\beq
y= -\tau\frac{\dd\tau}{\dd\zeta} \;\;\;  \mbox{and} \;\;\;
\varphi = y \zeta + \frac{\tau^2}{2} .
\label{phi-y}
\eeq
Finally, this gives the probability distribution $\cP(\delta_x)$ through
Eq.(\ref{P-phi-def}).
The probability distribution $\cP(\delta_x)$ depends on the spherical-collapse
dynamics and on the shape of the initial power spectrum $P_{L(\Lambda)}(k)$, through the
ratio $\sigma_{x(\Lambda)}/\sigma_{q(\Lambda)}$ in the second Eq.(\ref{tau-G-def}).
This second effect, sometimes called a ``smoothing effect'' \cite{Bernardeau1994},
is due to the collapse
(or expansion) of the mass shell $M$ from the Lagrangian scale $q$ to the Eulerian
scale $x$. This mixes scales and implies that the distribution $\cP(\delta_x)$ at scale
$x$ is sensitive to the initial power over all scales.
In our modified-gravity case, a second dependence on the shape of the
linear power spectrum appears through the mapping $\cF_q$ itself, because of the
$\epsilon$-dependent terms in Eqs.(\ref{y-1}) and (\ref{y-approx-1}).

We show in Fig.~\ref{fig-lPrho_r5_z0} the probability distribution $\cP(\delta_x)$
at redshift $z=0$ and radius $x=5 h^{-1}$Mpc.
Here we use the exact dynamics (\ref{y-1}) but using the approximation (\ref{y-approx-1})
gives very close results that would not be distinguished in this figure.
We recover the usual asymmetric shape due to nonlinear gravitational clustering,
which builds an extended high-density tail and shifts the peak of the distribution
towards low densities before a sharp low density cutoff at $\delta_x \rightarrow -1^+$
(on small scales, most of the matter lies in overdensities
but most of the volume lies in underdense regions).

Since it is difficult to distinguish different curves on this figure we plot the
relative deviation from GR in Fig.~\ref{fig-dlPrho_r5_z0}, for the
two  models where it is the largest.
(The two other cases would fall below the range plotted in the figure for the most
part.)
We plot our results using either the exact equation (\ref{y-1}) or the approximation
(\ref{y-approx-1}). We can see that both curves are very close. Indeed, as explained
in Sect.~\ref{mapping}, for smaller density fluctuations the ansatz (\ref{prof-approx-1})
becomes more accurate as it is exact to linear order and the profile has not
had time to be strongly modified by the dynamics
(moreover, the collapse is not very sensitive to the exact shape of the profile).

As we consider models with a positive value of $\epsilon$, which leads to
an effective amplification of gravity, it is easier to build large nonlinear density
fluctuations. This was also apparent in Fig.~\ref{fig-deltac_M_z0} for the
specific case of $\delta_x=200$.
For Gaussian initial conditions the tails of the probability distribution $\cP(\delta_x)$
are of the form $\cP(\delta_x) \sim e^{-\delta_{Lq(\Lambda)}^2/(2\sigma_{q(\Lambda)}^2)}$,
where $\delta_{Lq(\Lambda)}=\cF_q^{-1}(\delta_x)$, and the lower value of
$|\delta_{Lq(\Lambda)}|$ that is needed to reach a given $|\delta_x|$ yields a slower decay
of the rare-event tails. This is why we recover a positive deviation from
GR (i.e., a higher probability $\cP$) at both very low and very high
densities in Fig.~\ref{fig-dlPrho_r5_z0}.
Of course, since probability distributions are always normalized to unity this implies
that the relative deviation shows a change of sign and that the probability distribution
$\cP(\delta_x)$ obtained in these  models is smaller than the
$\Lambda$CDM one for moderate densities.
This explains the behaviours seen in Fig.~\ref{fig-dlPrho_r5_z0}.

These features are in qualitative agreement with the results obtained in numerical
simulations of various modified gravity models \citep{Hellwing2009,Li2012},
which also find that an effective amplification of gravity generically leads to
more numerous very low density and high density regions, while shifting the peak of
the probability distribution towards lower densities.

The relative deviation from GR does not necessarily grow to unity at high
densities (and may even decline).
This is due to the fact that high densities at a given Eulerian radius $x$ correspond
to large masses, hence to large Lagrangian (i.e. initial) radius $q$.
Then, because we recover General Relativity on large scales the linear threshold
$\delta_{Lq(\Lambda)}=\cF_q^{-1}(\delta_x)$ converges to the one obtained in the
$\Lambda$CDM cosmology, as in Fig.~\ref{fig-deltac_M_z0}.
Therefore, depending on the rate of convergence towards General Relativity on
large scales (as compared with the increasingly high sensitivity of the rare tail)
the large-density tail may or may not converge back to the GR prediction.
In modified gravity scenarios with a screening mechanism that implies convergence
to GR in high-density environments, such as the chameleon mechanism, the high-density
tail is expected to show a faster convergence back to the GR prediction.

These effects do not appear at very low densities, which correspond to
increasingly small mass $M$ and Lagrangian radius $q$, where the modifications
from General Relativity do not vanish within our framework. In this limit, the relative
deviation of $\cP(\delta_x)$ from the $\Lambda$CDM reference can grow up to unity.
However, this appears far in the low-density tail, which is characterized by a
very sharp cutoff, and this may not be a very efficient tool to probe
modified-gravity effects.

\section{Halo mass function}
\label{mass-function}

\begin{figure}
\begin{center}
\epsfxsize=8.5 cm \epsfysize=6 cm {\epsfbox{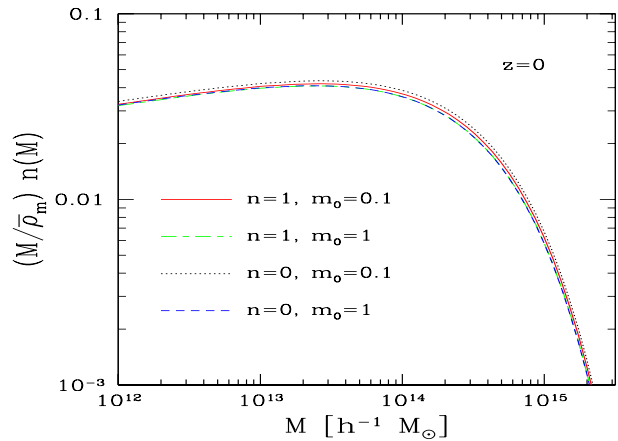}}
\end{center}
\caption{Halo mass function at redshift $z=0$.}
\label{fig-lnM_z0}
\end{figure}

\begin{figure}
\begin{center}
\epsfxsize=8.5 cm \epsfysize=6 cm {\epsfbox{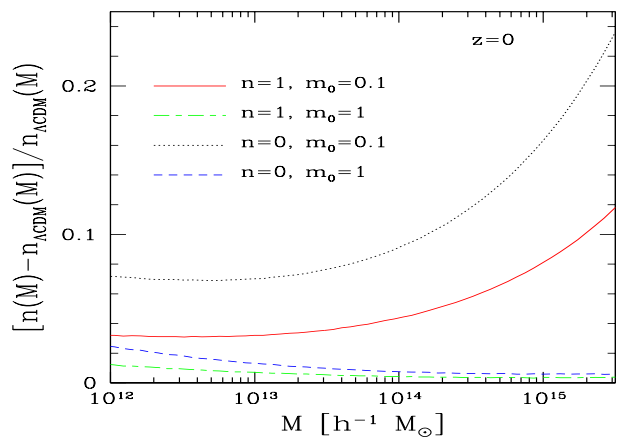}}
\end{center}
\caption{Relative deviation from $\Lambda$CDM of the halo mass function
at redshift $z=0$.}
\label{fig-rnM_z0}
\end{figure}

The computation of the probability distribution $\cP(\delta_x)$ was described
in the previous section for the quasilinear regime,
$\sigma_{x(\Lambda)}\rightarrow 0$. However, this result is more general and actually applies
to rare events, where the path integral (\ref{phi-path}) is peaked around the minimum
of the action $S$. In the quasilinear limit any finite nonzero density contrast $\delta_x$
becomes a rare event, which is why Eq.(\ref{phi-min-G}) determines the full
probability distribution in this regime.
For arbitrary values of $\sigma_x$, Eq.(\ref{phi-min-G}) applies to rare events,
that is, to the tails of the probability distribution $\cP(\delta_x)$
\cite{Valageas2002b} (this again allows one to recover the results obtained from
a perturbative analysis \cite{Bernardeau1994a}).
However, for large overdensities shell crossing appears at some stage
(typically for $\delta_x > 200$), after which Eq.(\ref{phi-min-G}) no longer holds
\cite{Valageas2002b,Valageas2009}.
Nevertheless, for lower densities one obtains the asymptotic behaviour
$\cP(\delta_x) \sim e^{-\delta_{Lq(\Lambda)}^2/(2\sigma_{q(\Lambda)}^2)}$.
This also determines the large-mass tail of the halo mass function $n(M)\dd M/M$,
where we define halos as spherical objects with a fixed density contrast threshold
$\delta=200$,
\beq
M\rightarrow \infty : \;\; \ln[n(M)] \sim -
\frac{\delta_{L(\Lambda)}(M)^2}{2\sigma_{(\Lambda)}(M)^2} ,
\label{n-M-tail}
\eeq
with
\beq
\delta_{L(\Lambda)}(M) = \cF_q^{-1}(\delta) ,
\eeq
where $\sigma_{(\Lambda)}(M)=\sigma_{q(\Lambda)}$ with $M=\rhob_{\rm m} 4\pi q^3/3$.

As in \cite{Valageas2009,Valageas2010}, a simple approximation for the mass
function that satisfies the large-mass asymptote (\ref{n-M-tail}) can be obtained
using the Press \& Schechter scaling variable $\nu$ \cite{Press1974},
\beq
n(M) \frac{\dd M}{M} = \frac{\rhob_{\rm m}}{M} \; f(\nu) \; \frac{\dd\nu}{\nu}
\label{n-M}
\eeq
with
\beq
\nu = \frac{\cF_q^{-1}(200)}{\sigma_{(\Lambda)}(M)} ,
\label{nu-def}
\eeq
where we choose to define halos by the nonlinear density threshold $\delta=200$.
The scaling function $f(\nu)$ is obtained from a fit to $\Lambda$CDM
numerical simulations that satisfies the exponential tail $f(\nu) \sim e^{-\nu^2/2}$
\cite{Valageas2009}
\beq
f(\nu)= 0.502 \left[ (0.6 \nu)^{2.5} + (0.62 \nu)^{0.5} \right] \, e^{-\nu^2/2} .
\label{f-nu}
\eeq
This ensures that the halo mass function is always normalized to unity and
obeys the large-mass tail (\ref{n-M-tail}), for any spherical-collapse mapping
$\cF_q$.
The only change from the $\Lambda$CDM cosmology is that the linear threshold
$\cF_q^{-1}(200)$ in Eq.(\ref{nu-def}) now depends on the mass $M$ through
the scale $q(M)$. The approximation (\ref{f-nu}) only ensures that the large-mass
tail is correct, but it may happen that the low-mass power-law tail should depend on
$\epsilon$. An analysis of such effects would require numerical simulations because
analytical methods cannot predict the low-mass tail of the halo mass function
(which is sensitive to mergers and non-local effects). Nevertheless, we can expect
modifications for moderate masses to be less important and partly taken into account
through the normalization constraint of the mass function.

As compared with the excursion set approach presented in
\cite{Li2012a,Li2012b,Lam2012}, we do not include screening effects but we take into
account the dependence on wavenumber of the modified-gravity
kernel $\epsilon(k,a)$. As explained in Sect.~\ref{Spherical-collapse}, this leads to
a mass-dependent linear threshold $\delta_L(M)$ whence to deviations from the
$\Lambda$CDM mass function that will depend on mass.

We show the halo mass function in Fig.~\ref{fig-lnM_z0}, and its relative deviation
from the $\Lambda$CDM mass function in Fig.\ref{fig-rnM_z0}.
Here we use the approximation (\ref{y-approx-1}) for the mapping
$\cF_q(\delta_{Lq(\Lambda)})$
but we checked that using Eq.(\ref{y-1}) yields close results.
For the models that we consider here the mass functions are very close to each
other and relative deviations are on the order of $10\%$ or less.
In agreement with the behaviour of the probability distribution $\cP(\delta_x)$
discussed in the previous section, a positive $\epsilon(k,a)$
leads to more numerous high density fluctuations and to a larger number of
massive collapsed halos. This explains why the ratio to the $\Lambda$CDM mass
function is greater than unity for $\nu > 1$, which corresponds to rare halos.
Again, this relative deviation grows for lower $n$ and smaller $m_0$.

The same trends appear in numerical simulations of similar modified gravity scenarios
\cite{Hellwing2009,Schmidt2009,Li2011,Ferraro2011}, with an increase of the large-mass
tail for models with an effective amplification of gravity. We show our results for $f(R)$ models
with $\vert f_{R0}\vert=10^{-4},10^{-5},10^{-6}$, as in \cite{Schmidt2009,Li2011,Ferraro2011},
in Appendix~\ref{fR-models}.

On the mass scales shown in Fig.\ref{fig-rnM_z0}, the ratio keeps growing at high
masses for $m_0=0.1$ while it decreases for $m_0=1$. As in the high-density
tail shown in Fig.~\ref{fig-dlPrho_r5_z0}, this is due to two competiting effects:
i) the exponential tail (\ref{n-M-tail}) of the halo mass function amplifies the
sensitivity to modified-gravity effects at large masses, but ii) these deviations
from General Relativity decrease at large scale whence at large mass
($\epsilon(k,a)\rightarrow 0$ for $k\rightarrow 0$), as seen in Fig.~\ref{fig-deltac_M_z0}.
Then, depending on the relative importance of both effects, the ratio of the mass
function to its $\Lambda$CDM reference may or may not grow with mass on the
scales that are considered.
As expected, a lower parameter $m_0$ (which implies a modification of gravity up
to larger scales, $k\sim m_0$ and $q\sim 1/m_0$, see Eq.(\ref{eps-def}))
yields a slower convergence to General Relativity at high mass, whence a larger
weight to the first effect i) above.
This explains why on the mass scales shown in Fig.\ref{fig-rnM_z0} the ratio
keeps growing at high masses for $m_0=0.1$ while it decreases for $m_0=1$.

\section{From linear to highly nonlinear scales}
\label{nonlinear}

Following \cite{Valageas2011d,Valageas2011e}, we can combine the perturbative
results of Sect.~\ref{Perturbative-regime} with the halo mass function of
Sect.~\ref{mass-function} to obtain the matter density power spectrum and
bispectrum from linear to highly nonlinear scales.
As in the usual halo model \cite{Cooray2002}, we write the nonlinear
power spectrum as the sum of ``two-halo'' and ``one-halo'' terms,
\beq
P(k) = P_{\rm 2H}(k) + P_{\rm 1H}(k) ,
\label{P-2H-1H}
\eeq
where $P_{\rm 2H}$ is the contribution from pairs of particles that are located in
two different halos and $P_{\rm 1H}$ is the contribution from pairs located in the
same halo.
As explained in \cite{Valageas2011d}, $P_{\rm 2H}$ contains the perturbative
contribution to the power spectrum and we write
\beq
P_{\rm 2H}(k) = F_{\rm 2H}(2\pi/k) \, P_{\rm pert}(k) ,
\label{P2H}
\eeq
where $F_{\rm 2H}(q)$ is the fraction of pairs, with initial (i.e. Lagrangian) separation
$q$, that belong to two distinct halos, and $P_{\rm pert}(k)$ is the power spectrum
obtained by perturbation theory.
It is not possible to use the standard one-loop prediction, unless one adds
a high-$k$ cutoff, because it grows too fast at high $k$ and leads to unphysical
results at high $k$ for the sum (\ref{P-2H-1H}).
Here we consider the one-loop prediction $P_{\rm pert}(k)$ given by the
resummation (\ref{C-Pi}) with Eqs.(\ref{Sigma-1loop})-(\ref{Pi-1loop}).
Indeed, at this order it yields $P_{\rm pert}(k) \sim P_L(k)$ at high $k$
\cite{Valageas2007}, so that the two-halo term is subdominant with respect to the
one-halo term and one obtains a good match to numerical simulations
\cite{Valageas2011d,Valageas2011e}.
Next, the one-halo contribution, which is fully nonperturbative, reads
\cite{Valageas2011d}
\beq
P_{\rm 1H}(k) = \int_0^{\infty} \frac{\dd\nu}{\nu} \, f(\nu)
\frac{M}{\rhob_{\rm m}(2\pi)^3} \left( \tu_M(k)^2 - \tW(k q)^2 \right) ,
\label{P-1H}
\eeq
where $\tW$ is the Fourier transform of the 3D top-hat, defined in
Eq.(\ref{W-3D-def}), and $\tu_M$ is the normalized Fourier transform of the density
profile $\rho_M(x)$ of halos of mass $M$,
\beq
\tu_M(k) = \frac{1}{M} \int \dd\vx \; e^{-\ii\vk\cdot\vx} \; \rho_M(x) .
\label{uM-def}
\eeq
We use the usual ``NFW'' halo profile \cite{Navarro1997}, with the
mass-concentration relation from \cite{Valageas2011d}.
Therefore, we do not take into account the effects of the modified gravity
on the shape of the profiles of the dark matter halos.
Our one-halo term $P_{\rm 1H}$ only depends on $\epsilon(k,a)$ through the
change of the halo mass function described in Sect.~\ref{mass-function}.
The counterterm $\tW^2$ in Eq.(\ref{P-1H})
ensures that the one-halo contribution decays as $P_{\rm 1H}(k) \propto k^2$
at low $k$, so that the total power (\ref{P-2H-1H}) converges to the linear
power on large scales. This follows from the conservation of matter and the
fact that halo formation corresponds to a small-scale redistribution of matter
\cite{Peebles1974,Valageas2011d} \footnote{Taking into account momentum conservation
would give an even steeper $k^4$ tail \cite{Peebles1974} but the form
(\ref{P-1H}) is sufficient for practical purposes.}.

In a similar fashion, the matter density bispectrum can be written as the sum
of three-halo, two-halo, and one-halo terms,
\beq
B= B_{\rm 3H} + B_{\rm 2H} + B_{\rm 1H} ,
\label{B-halo}
\eeq
with \cite{Valageas2011e},
\beq
B_{\rm 3H}(k_1,k_2,k_3) = B_{\rm pert}(k_1,k_2,k_3) ,
\label{B3H}
\eeq

\beqa
B_{\rm 2H}(k_1,k_2,k_3) & = & P_L(k_1) \int\frac{\dd\nu}{\nu}
\frac{M}{\rhob_{\rm m}(2\pi)^3} f(\nu) \nonumber \\
&& \hspace{-1.5cm} \times \prod_{j=2}^3 \left( \tu_M(k_j) - \tW(k_j q) \right)
+ 2 \; {\rm cyc.} ,
\label{B2H}
\eeqa
\beqa
B_{\rm 1H}(k_1,k_2,k_3) & = & \int\frac{\dd\nu}{\nu} f(\nu)
\left(\frac{M}{\rhob_{\rm m}(2\pi)^3}\right)^3 \nonumber \\
&& \hspace{-1.5cm} \times \prod_{j=1}^3 \left( \tu_M(k_j) - \tW(k_j q) \right)
+ 2 \; {\rm cyc.} ,
\label{B1H}
\eeqa
Again, the counterterms $\tW$ in Eqs.(\ref{B2H}) and (\ref{B1H}) ensure that the
two-halo and one-halo contributions decay on large scales so that the
bispectrum converges to the perturbative prediction $B_{\rm pert}$.
As found in \cite{Valageas2011e} and contrary to the situation encountered for the
power spectrum, the standard one-loop perturbation theory prediction for
$B_{\rm pert}$ is well-behaved at high $k$ (i.e., it is significantly smaller than the
one-halo contribution) and it is more accurate than the resummation schemes that
have already been studied.
Therefore, we only consider the standard perturbative approach for the three-halo
contribution (\ref{B3H}). More precisely, we use the exact tree-level result
(\ref{B-tree-Z}) and the approximate one-loop correction (\ref{B1loop})
by setting $B_{\rm pert}=B^{\rm tree}+B^{\rm 1loop}$.

\begin{figure}
\begin{center}
\epsfxsize=8.5 cm \epsfysize=6 cm {\epsfbox{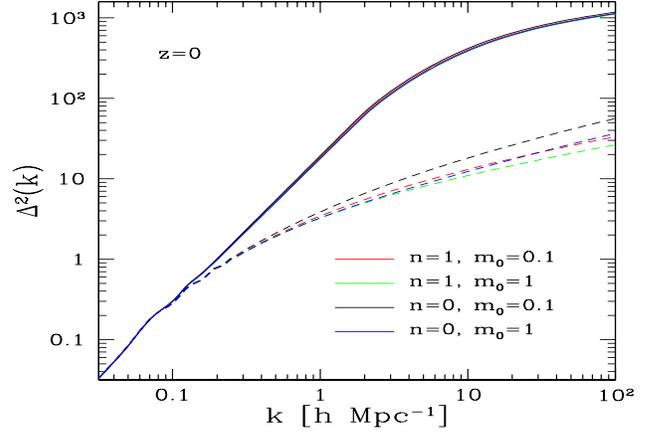}}
\end{center}
\caption{Logarithmic power, $\Delta^2(k)=4\pi k^3 P(k)$, at $z=0$ for four $(n,m_0)$
models. In each case we plot the linear power (dashed line) and the nonlinear
power (solid line).}
\label{fig-lDk_z0}
\end{figure}

\begin{figure}
\begin{center}
\epsfxsize=8.5 cm \epsfysize=6 cm {\epsfbox{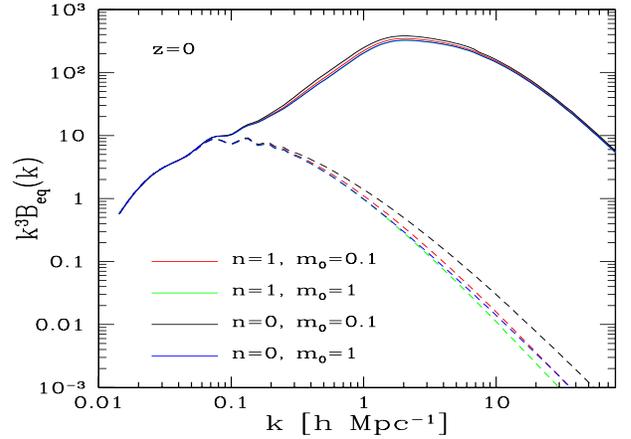}}
\end{center}
\caption{Equilateral bispectrum $B_{\rm eq}(k)=B(k,k,k)$, at $z=0$ for four $(n,m_0)$
models. The bispectrum is multiplied by a factor $k^3$ in this plot to decrease
the range spanned by the vertical axis and to make the figure easier to read.
In each case we plot the tree-level bispectrum (dashed line) and the full nonlinear
bispectrum (solid line).}
\label{fig-lBk_z0}
\end{figure}

While Eq.(\ref{B-halo}) yields a reasonably good match to numerical simulations
($\sim 10\%$) over all scales for the bispectrum \cite{Valageas2011e},
Eq.(\ref{P-2H-1H}) significantly underestimates the power spectrum on the transition
scales (by $\sim 20-30\%$), even though it gives a good accuracy on larger scales
($\sim 1\%$ below $k\simeq 0.3 h$ Mpc$^{-1}$ at $z=1$) and smaller scales
($\sim 10\%$ above $k \simeq 5 h$ Mpc$^{-1}$ at $z=1$). Following
\cite{Valageas2011e}, we consider a simple power-law interpolation
$P_{\rm tang}$between large and small scales,
\beq
P_{\rm tang}(k) = P_{\rm 2H+1H}(k) \;\; \mbox{for} \;\; k\leq k_- \;\;
\mbox{and} \;\; k\geq k_+'
\label{Ptang-1}
\eeq
and
\beq
P_{\rm tang}(k) \;\; \mbox{is a power law within} \;\; k_-\leq k \leq k_+' .
\label{Ptang-2}
\eeq
The transition range $[k_-,k_+']$ is automatically determined from the shape of
$P_{\rm 2H+1H}(k)$ and $B(k,k,k)$ and it depends on the shape of the
linear power spectrum and on redshift. This improves the agreement with
numerical simulations in the $\Lambda$CDM cosmology \cite{Valageas2011e}
while keeping the perturbative and 1-halo behaviours on large and small scales.

\begin{figure}
\begin{center}
\epsfxsize=8.5 cm \epsfysize=6 cm {\epsfbox{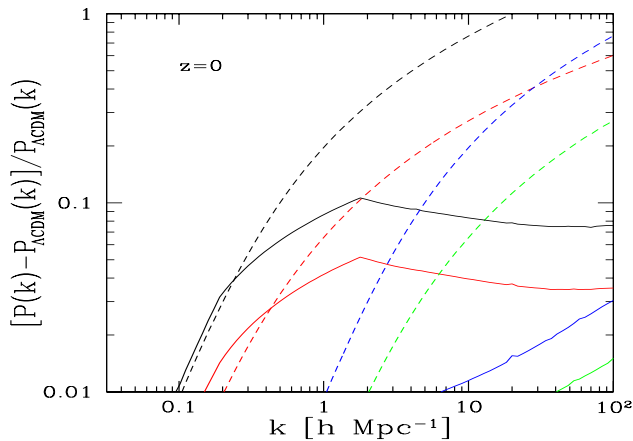}}
\end{center}
\caption{Relative deviation from $\Lambda$CDM of the power spectrum obtained
in four models at redshift $z=0$.
In each case, we plot both the relative deviation of the linear power (dashed line)
and of the nonlinear power (solid line).
From left to right we consider the models $(n,m_0)=(0,0.1), (1,0.1), (0,1)$, and
$(1,1)$.}
\label{fig-dlDk}
\end{figure}

\begin{figure}
\begin{center}
\epsfxsize=8.5 cm \epsfysize=6 cm {\epsfbox{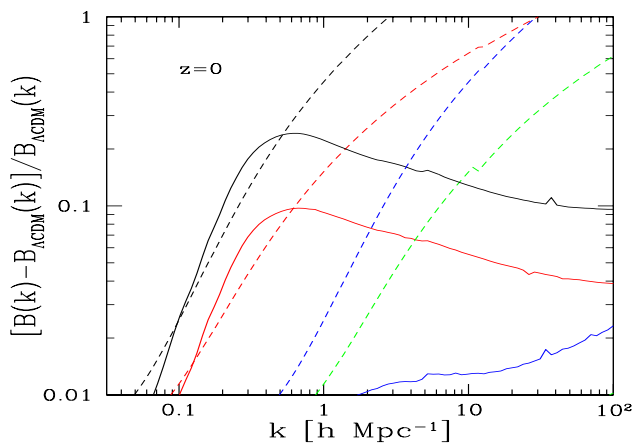}}
\end{center}
\caption{Relative deviation from $\Lambda$CDM of the bispectrum obtained
in four models at redshift $z=0$.
In each case, we plot both the relative deviation of the tree-level bispectrum
(dashed line) and of the nonlinear bispectrum (solid line).
From left to right we consider the models $(n,m_0)=(0,0.1), (1,0.1), (0,1)$, and
$(1,1)$.}
\label{fig-dlBk}
\end{figure}

We show in Figs.~\ref{fig-lDk_z0} and \ref{fig-lBk_z0} the matter density
power spectrum and bispectrum that we obtain at redshift $z=0$, from linear
to highly nonlinear scales. The various curves are very close and we can see
that at high $k$ the deviations are actually damped by nonlinear effects.
Within our framework, this is because we neglected any impact of modified
gravity on the halo profile (\ref{uM-def}) and the only influence of modified
gravity appears through the halo mass function $n(M)$.
This may not be such a bad approximation because in more realistic models
modifications to gravity vanish on small scales (e.g., through chameleon or
Vainshtein mechanisms) so that the density profiles of small halos are expected
to converge to the GR behaviour.
Then, we expect that our modelization provides a similar accuracy to the one
found in $\Lambda$CDM cosmology by comparison with numerical
simulations \cite{Valageas2011d,Valageas2011e}.
As in Sect.~\ref{pert-num}, we clearly see that nonlinear gravitational
clustering amplifies both the power spectrum and bispectrum at high $k$
but damps the baryon acoustic oscillations.
As in \cite{Valageas2011d,Valageas2011e}, our approach allows us to
describe the power spectrum and bispectrum from large linear scales
down to small highly nonlinear scales.

We show in Figs.~\ref{fig-dlDk} and \ref{fig-dlBk} the relative deviations from
the $\Lambda$CDM reference of the power spectrum and of the equilateral
bispectrum.
In the weakly nonlinear regime the relative deviations grow with $k$,
following the behaviour of $\epsilon(k,a)$. In agreement with the discussions above,
they reach a maximum on transition scales, starting to deviate from the
$\Lambda$CDM growth for $k\sim m_0$, and then slowly declining on highly nonlinear
scales.
On these nonlinear scales, the relative deviations at the level of the linear or
tree-order contributions are no longer a good estimate of the actual signal
and greatly overestimate the effects of modified gravity.
Since the theoretical accuracy is greater on weakly nonlinear scales
(which can be analyzed by systematic perturbative approaches) than on highly
nonlinear scales (which require phenomenological ingredients such as halo
profiles), these behaviours suggest that it is more efficient to focus on weakly
nonlinear scales to probe such modifications of gravity.

It is is also worth emphasizing that the deviations from $\Lambda$CDM which we have calculated with the steepest descent resummation method together with the halo model show the same trends as the N-body results \cite{Oyaizu2008,Schmidt2009}
obtained for models with $n=1$ and $|f_{R0}|=10^{-4},10^{-5},10^{-6}$.
Indeed, numerical results show that the deviation from $\Lambda$-CDM reaches a
peak at weakly non-linear scales before decreasing on highly non-linear scales.
Simple fitting procedures designed for $\Lambda$CDM cosmology \citep{Smith2003}
have been shown not to provide good results and to miss this high-$k$ behavior
\cite{Oyaizu2008}.
This shows the advantage of approaches like ours that are closer to physical modeling.
Even though they may be less accurate than
a specific fitting formula , their
behaviour as cosmological parameters and scenarios are modified is more reliable.

\section{Conclusion}
\label{Conclusion}

We have considered the dynamics of structure formation in modified gravity models analytically. To do so, we have used a steepest descent technique
for the generating functional of density and velocity perturbations as well as the spherical collapse dynamics. The models we have considered correspond to screened modifications of gravity due to a scalar field. In numerical examples we have focused on models defined by a power law mass function and a constant coupling to matter, which coincide with $f(R)$ models in the large curvature limit and in the matter era, although the techniques developed here are general. The results we have presented comprise the power spectrum, the bispectrum, the probability distribution of the density contrast, and the large-mass tail of the halo mass function. Modified gravity has interesting features astrophysically when the ratio of the mass of the scalar field over the Hubble rate now $m_0/H_0$ is of order $10^3$. In this case, deviations can be substantial and larger than a few percent. In this paper, we do not attempt to give precise predictions, we are more interested in indications that can be obtained relatively fast using our analytical tools without the need for large N-body simulations.

After a description of the linear growing and decaying modes, which become $k$-dependent
in these modified-gravity scenarios, we have obtained the associated linear growth rate $f(k,z)$.
For the realistic parameters $(n,m_0)$ studied here measuring its deviation from the
General Relativity prediction remains challenging, but future surveys such as Euclid
should give a clear signal for the most favorable cases (e.g., $(n,m_0)=(0,0.1)$).

Next, we have described how higher-order perturbative contributions can be computed
in the weakly nonlinear regime. The dependence on wavenumber of the linear
modes makes numerical implementations of these perturbative schemes significantly more
complex than in the usual General Relativity case, because time- and scale-dependences
no longer factor out. We have presented the generalization of the ``standard'' perturbative
approach as well as a ``steepest descent'' approach that performs partial resummations
of higher-order diagrams. The path-integral formalism that underlies this second method
also provides an efficient route to recover the standard perturbative approach and avoids
the need to compute the $n-$point kernels $F_n^s$.
We find that for realistic modified-gravity scenarios, such as the ones investigated here,
the deviations of the power spectrum from General Relativity on BAO scales are quite
modest (typically less than $6\%$) and below the accuracy of the standard perturbative
approach at one-loop order.
This means that one must use more accurate schemes, such as the one-loop
steepest-descent approach presented here, or possibly include higher-order terms
within the standard approach (but its convergence is not very well behaved).

For the bispectrum we find that nonperturbative contributions (associated with one-halo
and two-halo terms) cannot be neglected on the weakly nonlinear scales where the
deviations from General Relativity can be detected. This suggests that for practical
purposes the power spectrum is a more reliable probe of such modified-gravity effects,
because its deviations from the GR predictions are larger than for the
bispectrum in the perturbative regime, where rigorous and systematic approaches can
be developed.

To go beyond these low-order perturbative approaches, we have described the dynamics
of spherical density fluctuations, which can be exactly solved before shell crossing.
Again, modifications to gravity make the analysis significantly more complex, because the
motions of different shells no longer decouple, even before any shell crossing.
This means that one must solve the evolution with time of the full density profile.
Nevertheless, we have introduced a simple approximation for typical profiles that allows
to decouple the motion of the mass shell of interest. We find this provides a reasonable
approximation to the exact dynamics (but slightly underestimates the effects of
modified gravity).
This analysis provides the characteristic dependence on mass of the critical linear density
threshold $\delta_c(M)$ associated with a given nonlinear threshold (such as
$\delta=200$). In the cases studies here, where the function $\epsilon(k,a)$ is positive
and corresponds to a time- and scale-dependent effective amplification of gravity,
this threshold $\delta_c(M)$ decreases at low mass (because this amplification
is larger on smaller scales) and converges to the constant GR prediction
at large mass (because we recover General Relativity on large scales).

In contrast to some previous works, this dependence on mass does not arise from
screening effects (that depend on mass through the depth of the gravitational potential,
which triggers the screening mechanism) but from the $k$-dependence of the
modified-gravity kernel $\epsilon(k,a)$.

This also allows us to obtain the probability distribution, $\cP(\delta_x)$, of the nonlinear
density contrast within spherical cells, in the weakly nonlinear regime.
Because of this effective amplification of gravity, the tails of $\cP(\delta_x)$ grow with respect
to the General Relativity prediction (and by conservation of the probability normalization
to unity $\cP(\delta_x)$ decreases for moderate density fluctuations).
This growth is smaller and the relative ratio to GR does not necessarily goes to infinity in the
large-density tail, as opposed to the low-density tail, because on large scale the dynamics
converges to General Relativity.

The same effect amplifies the large-mass tail of the halo mass function. Again, the ratio to the
GR prediction may increase or decrease with mass in the range of interest
depending on how fast modifications to gravity vanish on large scales.

Finally, combining perturbative approaches with halo models, we have computed a
simple estimate of the power spectrum and bispectrum from linear to highly nonlinear scales.
Within this modelisation, we find that the relative deviation from General Relativity
is the largest on the transition scales between the linear and the highly nonlinear regimes,
for both the power spectrum and bispectrum. Since nonlinear scales are difficult to predict
with a high accuracy (because of the complex nonperturbative dynamics associated with
shell crossings and because one should include baryon and galaxy formation effects),
this suggests that weakly nonlinear scales, in particular in the perturbative regime,
are the best probes of these modified-gravity models.

 As a summary, our new results can be listed as follows:

- a comparison of the accuracy of one-loop perturbative expansions (by using two such
schemes and by estimating non-perturbative one-halo contributions) with realistic
deviations from GR, for the matter power spectrum and the bispectrum.

- an analysis of the spherical collapse that includes shell-coupling and the
scale-dependence of the modified-gravity kernel $\epsilon(k,a)$.

- the dependence on mass, due to the scale-dependence of $\epsilon$ (and not to
screening effects), of the deviation from GR on the halo mass function.

- an analytical model for the probability distribution $\cP(\delta_x)$ in the rare-event
regime.

- a combination of one-loop perturbative expansions with halo models for the
matter power spectrum and the bispectrum up to highly non-linear scales.

Our methods call for improvements to reach the needs of precision cosmology. Indeed we have neglected, for ease of treatment and as a first step, two major effects. The first one consists in including non-linearities in the scalar field sector of the models. Here the scalar field dynamics are only linear and non-linear effects in both the potential and the coupling to matter ought to be considered. Technically, this can be done at the one loop level by self-consistently modifying the Euler equation with non-linear terms coming from the scalar field interaction with matter particles. A second
ingredient we have not considered so far is the screening of the scalar force in dense environments. This will modify the spherical collapse of over densities and therefore the halo statistics. Eventually this will have an impact on the growth of non-linear structures. As a result, the effects described in this paper can only be taken as indications on quasi-linear scales. Work on all these aspects is in progress. We also intend to carry out a comparison of our analytical results with the N-body simulations which use the same mass and coupling parameterisation of modified gravity.  Doing so, and for a greater variety of models including dilatons and symmetrons, we hope to validate our analytical approach which could then be used for models that will appear in the future and be analysed without the need for large N-body simulations.

\appendix

\section{The case of $f(R)$ models}
\label{fR-models}

\begin{figure}
\begin{center}
\epsfxsize=8.5 cm \epsfysize=6 cm {\epsfbox{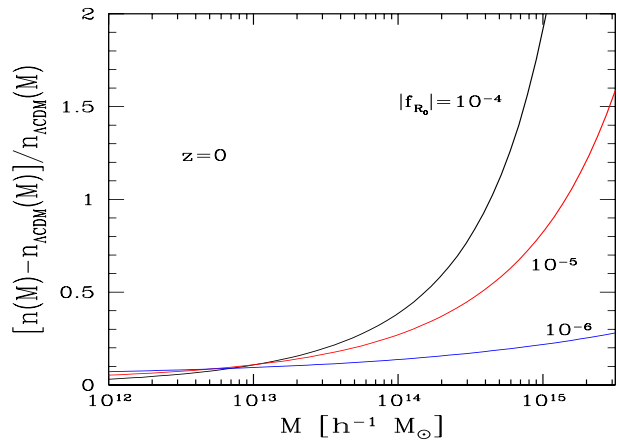}}
\end{center}
\caption{Relative deviation from $\Lambda$CDM of the halo mass function
at redshift $z=0$, for $n=1$ and $|f_{R_0}|=10^{-4}, 10^{-5},$ and $10^{-6}$,
from top to bottom.}
\label{fig-rnM_z0_fR}
\end{figure}

\begin{figure}
\begin{center}
\epsfxsize=8.5 cm \epsfysize=6 cm {\epsfbox{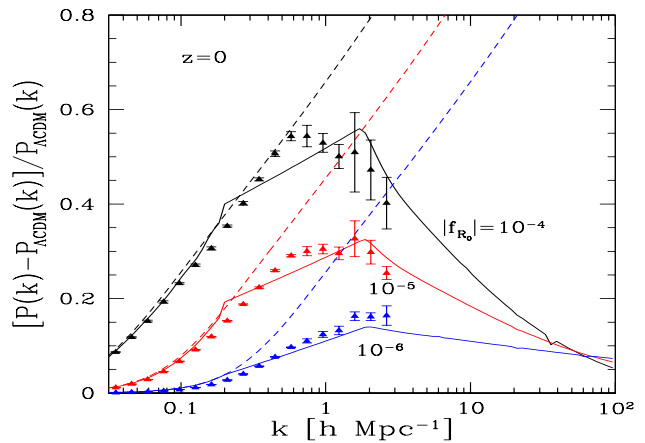}}
\end{center}
\caption{Relative deviation from $\Lambda$CDM of the power spectrum
at redshift $z=0$, for $n=1$ and $|f_{R_0}|=10^{-4}, 10^{-5}$, and $10^{-6}$.
In each case, we plot both the relative deviation of the linear power (dashed line)
and of the nonlinear power (solid line). The points are the results of the
``no-chameleon simulations'' from \cite{Oyaizu2008}.}
\label{fig-dlDk_fR}
\end{figure}

\begin{figure}
\begin{center}
\epsfxsize=8.5 cm \epsfysize=6 cm {\epsfbox{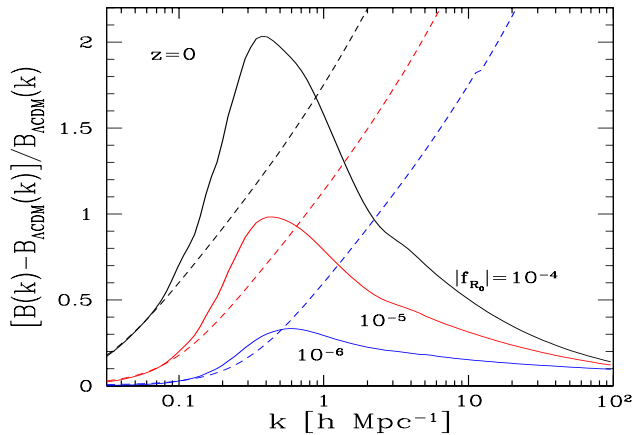}}
\end{center}
\caption{Relative deviation from $\Lambda$CDM of the bispectrum at redshift
$z=0$, for $n=1$ and $|f_{R_0}|=10^{-4}, 10^{-5}$, and $10^{-6}$.
In each case, we plot both the relative deviation of the tree-level bispectrum
(dashed line) and of the nonlinear bispectrum (solid line).}
\label{fig-dlBk_fR}
\end{figure}

We consider in this appendix the case of $f(R)$ models which have
also been studied through numerical simulations, with a power-law form
as in Eq.(\ref{fR-def}).
The mass of the scalar field evolves with time as \cite{Oyaizu2008}
\beq
m(a) = m_0 \; \left( \frac{\Omega_{\rm m0} (1+z)^3 + 4 \Omega_{\Lambda 0}}
{\Omega_{\rm m0}+ 4 \Omega_{\Lambda 0}} \right)^{(n+2)/2} ,
\label{ma-fR-n}
\eeq
where $m_0$ is given by Eq.(\ref{m0-fR-def}). This gives the approximate scaling
(\ref{ma-fR}) at high redshift but for accurate computations it is necessary to
use the more precise expression (\ref{ma-fR-n}).

To compare with the numerical results of
\cite{Oyaizu2008,Schmidt2009,Li2011,Ferraro2011,Zhao:2010qy} we adopt the same
WMAP3 $\Lambda$CDM reference model \cite{Spergel2007}, with
cosmological parameters
$(\Omega_{\rm m},\Omega_{\rm b},h,\sigma_8,n_{\rm s})
= (0.24, 0.04181, 0.73, 0.76, 0.958)$. We focus on the case $n=1$, with the amplitudes
$|f_{R_0}|=10^{-4}, 10^{-5}$, and $10^{-6}$.
We show in Figs.~\ref{fig-rnM_z0_fR}, \ref{fig-dlDk_fR}, and \ref{fig-dlBk_fR},
the relative deviations from the $\Lambda$CDM reference of the halo mass function,
the matter power spectrum, and the bispectrum.
Our results are similar to the ones obtained in Figs.~\ref{fig-rnM_z0},
\ref{fig-dlDk}, and \ref{fig-dlBk}, in the main text, for our power-law models
parameterized by $(n,m_0)$.
We can check that our results also show a reasonable agreement with the
``no-chameleon'' numerical simulations of
\cite{Oyaizu2008,Schmidt2009,Li2011,Ferraro2011}
for the halo mass function and the power spectrum, although we may overestimate the
large-mass tail for $M>10^{15} h^{-1} M_{\odot}$.
The almost straight lines on transition scales in Fig.~\ref{fig-dlDk_fR} correspond to
the interpolation (\ref{Ptang-2}) and should not be considered as an accurate
prediction. However, they correctly reproduce the saturation of the relative deviation
and the transition towards the highly nonlinear regime (dominated by the one-halo
contribution) where the relative deviation declines (within our framework, where we
neglect modifications of halo profiles).
The same behaviour is found in numerical simulations,  as can be seen  in figure 18 where we compare our results to the no-chameleon simulation
of \cite{Oyaizu2008}, with
a reasonably good quantitative match.
It is interesting to note that using simple fitting procedures designed
for $\Lambda$CDM cosmology, such as the halo-fit from \cite{Smith2003},
has been shown not to provide good results and to miss this high-$k$ behavior
\cite{Oyaizu2008}.
This is not fully surprising, since such fitting formulae were not designed for these
scenarios.
This shows the advantage of using approaches such as the one presented in this paper
that are closer to physical modeling (using both systematic perturbative expansions
and phenomenological halo models). Even though they may be less accurate than
a specific fitting formula for the class of models the latter was built from, their
behaviour as cosmological parameters and scenarios are modified is more reliable.

\begin{acknowledgments}

We thank W. Hu for sending us the numerical data shown for comparison in
Fig.~\ref{fig-dlDk_fR}.

\end{acknowledgments}

\bibliography{ref1}   

\begin{thebibliography}{80}
\expandafter\ifx\csname natexlab\endcsname\relax\def\natexlab#1{#1}\fi
\expandafter\ifx\csname bibnamefont\endcsname\relax
  \def\bibnamefont#1{#1}\fi
\expandafter\ifx\csname bibfnamefont\endcsname\relax
  \def\bibfnamefont#1{#1}\fi
\expandafter\ifx\csname citenamefont\endcsname\relax
  \def\citenamefont#1{#1}\fi
\expandafter\ifx\csname url\endcsname\relax
  \def\url#1{\texttt{#1}}\fi
\expandafter\ifx\csname urlprefix\endcsname\relax\def\urlprefix{URL }\fi
\providecommand{\bibinfo}[2]{#2}
\providecommand{\eprint}[2][]{\url{#2}}

\bibitem[{\citenamefont{Copeland et~al.}(2006)\citenamefont{Copeland, Sami, and
  Tsujikawa}}]{Copeland:2006wr}
\bibinfo{author}{\bibfnamefont{E.~J.} \bibnamefont{Copeland}},
  \bibinfo{author}{\bibfnamefont{M.}~\bibnamefont{Sami}}, \bibnamefont{and}
  \bibinfo{author}{\bibfnamefont{S.}~\bibnamefont{Tsujikawa}},
  \bibinfo{journal}{Int.J.Mod.Phys.} \textbf{\bibinfo{volume}{D15}},
  \bibinfo{pages}{1753} (\bibinfo{year}{2006}), \eprint{hep-th/0603057}.

\bibitem[{\citenamefont{Clifton et~al.}(2012)\citenamefont{Clifton, Ferreira,
  Padilla, and Skordis}}]{Clifton:2011jh}
\bibinfo{author}{\bibfnamefont{T.}~\bibnamefont{Clifton}},
  \bibinfo{author}{\bibfnamefont{P.~G.} \bibnamefont{Ferreira}},
  \bibinfo{author}{\bibfnamefont{A.}~\bibnamefont{Padilla}}, \bibnamefont{and}
  \bibinfo{author}{\bibfnamefont{C.}~\bibnamefont{Skordis}},
  \bibinfo{journal}{Phys.Rept.} \textbf{\bibinfo{volume}{513}},
  \bibinfo{pages}{1} (\bibinfo{year}{2012}), \eprint{1106.2476}.

\bibitem[{\citenamefont{Weinberg}(1965)}]{Weinberg:1965rz}
\bibinfo{author}{\bibfnamefont{S.}~\bibnamefont{Weinberg}},
  \bibinfo{journal}{Phys.Rev.} \textbf{\bibinfo{volume}{138}},
  \bibinfo{pages}{B988} (\bibinfo{year}{1965}).

\bibitem[{\citenamefont{Hoyle et~al.}(2004)\citenamefont{Hoyle, Kapner, Heckel,
  Adelberger, Gundlach et~al.}}]{Hoyle:2004cw}
\bibinfo{author}{\bibfnamefont{C.}~\bibnamefont{Hoyle}},
  \bibinfo{author}{\bibfnamefont{D.}~\bibnamefont{Kapner}},
  \bibinfo{author}{\bibfnamefont{B.~R.} \bibnamefont{Heckel}},
  \bibinfo{author}{\bibfnamefont{E.}~\bibnamefont{Adelberger}},
  \bibinfo{author}{\bibfnamefont{J.}~\bibnamefont{Gundlach}},
  \bibnamefont{et~al.}, \bibinfo{journal}{Phys.Rev.}
  \textbf{\bibinfo{volume}{D70}}, \bibinfo{pages}{042004}
  (\bibinfo{year}{2004}), \eprint{hep-ph/0405262}.

\bibitem[{\citenamefont{Bertotti et~al.}(2003)\citenamefont{Bertotti, Iess, and
  Tortora}}]{Bertotti:2003rm}
\bibinfo{author}{\bibfnamefont{B.}~\bibnamefont{Bertotti}},
  \bibinfo{author}{\bibfnamefont{L.}~\bibnamefont{Iess}}, \bibnamefont{and}
  \bibinfo{author}{\bibfnamefont{P.}~\bibnamefont{Tortora}},
  \bibinfo{journal}{Nature} \textbf{\bibinfo{volume}{425}},
  \bibinfo{pages}{374} (\bibinfo{year}{2003}).

\bibitem[{\citenamefont{Khoury}(2010)}]{Khoury:2010xi}
\bibinfo{author}{\bibfnamefont{J.}~\bibnamefont{Khoury}}
  (\bibinfo{year}{2010}), \eprint{1011.5909}.

\bibitem[{\citenamefont{Khoury and
  Weltman}(2004{\natexlab{a}})}]{Khoury:2003aq}
\bibinfo{author}{\bibfnamefont{J.}~\bibnamefont{Khoury}} \bibnamefont{and}
  \bibinfo{author}{\bibfnamefont{A.}~\bibnamefont{Weltman}},
  \bibinfo{journal}{Phys.Rev.Lett.} \textbf{\bibinfo{volume}{93}},
  \bibinfo{pages}{171104} (\bibinfo{year}{2004}{\natexlab{a}}),
  \eprint{astro-ph/0309300}.

\bibitem[{\citenamefont{Khoury and
  Weltman}(2004{\natexlab{b}})}]{Khoury:2003rn}
\bibinfo{author}{\bibfnamefont{J.}~\bibnamefont{Khoury}} \bibnamefont{and}
  \bibinfo{author}{\bibfnamefont{A.}~\bibnamefont{Weltman}},
  \bibinfo{journal}{Phys.Rev.} \textbf{\bibinfo{volume}{D69}},
  \bibinfo{pages}{044026} (\bibinfo{year}{2004}{\natexlab{b}}),
  \eprint{astro-ph/0309411}.

\bibitem[{\citenamefont{Brax et~al.}(2004)\citenamefont{Brax, van~de Bruck,
  Davis, Khoury, and Weltman}}]{Brax:2004qh}
\bibinfo{author}{\bibfnamefont{P.}~\bibnamefont{Brax}},
  \bibinfo{author}{\bibfnamefont{C.}~\bibnamefont{van~de Bruck}},
  \bibinfo{author}{\bibfnamefont{A.-C.} \bibnamefont{Davis}},
  \bibinfo{author}{\bibfnamefont{J.}~\bibnamefont{Khoury}}, \bibnamefont{and}
  \bibinfo{author}{\bibfnamefont{A.}~\bibnamefont{Weltman}},
  \bibinfo{journal}{Phys.Rev.} \textbf{\bibinfo{volume}{D70}},
  \bibinfo{pages}{123518} (\bibinfo{year}{2004}), \eprint{astro-ph/0408415}.

\bibitem[{\citenamefont{Damour and Polyakov}(1994)}]{Damour:1994zq}
\bibinfo{author}{\bibfnamefont{T.}~\bibnamefont{Damour}} \bibnamefont{and}
  \bibinfo{author}{\bibfnamefont{A.~M.} \bibnamefont{Polyakov}},
  \bibinfo{journal}{Nucl.Phys.} \textbf{\bibinfo{volume}{B423}},
  \bibinfo{pages}{532} (\bibinfo{year}{1994}), \eprint{hep-th/9401069}.

\bibitem[{\citenamefont{Brax et~al.}(2011{\natexlab{a}})\citenamefont{Brax,
  van~de Bruck, Davis, Li, and Shaw}}]{Brax:2011ja}
\bibinfo{author}{\bibfnamefont{P.}~\bibnamefont{Brax}},
  \bibinfo{author}{\bibfnamefont{C.}~\bibnamefont{van~de Bruck}},
  \bibinfo{author}{\bibfnamefont{A.-C.} \bibnamefont{Davis}},
  \bibinfo{author}{\bibfnamefont{B.}~\bibnamefont{Li}}, \bibnamefont{and}
  \bibinfo{author}{\bibfnamefont{D.~J.} \bibnamefont{Shaw}},
  \bibinfo{journal}{Phys.Rev.} \textbf{\bibinfo{volume}{D83}},
  \bibinfo{pages}{104026} (\bibinfo{year}{2011}{\natexlab{a}}),
  \eprint{1102.3692}.

\bibitem[{\citenamefont{Brax et~al.}(2010{\natexlab{a}})\citenamefont{Brax,
  van~de Bruck, Davis, and Shaw}}]{Brax:2010gi}
\bibinfo{author}{\bibfnamefont{P.}~\bibnamefont{Brax}},
  \bibinfo{author}{\bibfnamefont{C.}~\bibnamefont{van~de Bruck}},
  \bibinfo{author}{\bibfnamefont{A.-C.} \bibnamefont{Davis}}, \bibnamefont{and}
  \bibinfo{author}{\bibfnamefont{D.}~\bibnamefont{Shaw}},
  \bibinfo{journal}{Phys.Rev.} \textbf{\bibinfo{volume}{D82}},
  \bibinfo{pages}{063519} (\bibinfo{year}{2010}{\natexlab{a}}),
  \eprint{1005.3735}.

\bibitem[{\citenamefont{Nicolis et~al.}(2009)\citenamefont{Nicolis, Rattazzi,
  and Trincherini}}]{Nicolis:2008in}
\bibinfo{author}{\bibfnamefont{A.}~\bibnamefont{Nicolis}},
  \bibinfo{author}{\bibfnamefont{R.}~\bibnamefont{Rattazzi}}, \bibnamefont{and}
  \bibinfo{author}{\bibfnamefont{E.}~\bibnamefont{Trincherini}},
  \bibinfo{journal}{Phys.Rev.} \textbf{\bibinfo{volume}{D79}},
  \bibinfo{pages}{064036} (\bibinfo{year}{2009}), \eprint{0811.2197}.

\bibitem[{\citenamefont{Pietroni}(2005)}]{Pietroni:2005pv}
\bibinfo{author}{\bibfnamefont{M.}~\bibnamefont{Pietroni}},
  \bibinfo{journal}{Phys.Rev.} \textbf{\bibinfo{volume}{D72}},
  \bibinfo{pages}{043535} (\bibinfo{year}{2005}), \eprint{astro-ph/0505615}.

\bibitem[{\citenamefont{Olive and Pospelov}(2008)}]{Olive:2007aj}
\bibinfo{author}{\bibfnamefont{K.~A.} \bibnamefont{Olive}} \bibnamefont{and}
  \bibinfo{author}{\bibfnamefont{M.}~\bibnamefont{Pospelov}},
  \bibinfo{journal}{Phys.Rev.} \textbf{\bibinfo{volume}{D77}},
  \bibinfo{pages}{043524} (\bibinfo{year}{2008}), \eprint{0709.3825}.

\bibitem[{\citenamefont{Hinterbichler and Khoury}(2010)}]{Hinterbichler:2010es}
\bibinfo{author}{\bibfnamefont{K.}~\bibnamefont{Hinterbichler}}
  \bibnamefont{and} \bibinfo{author}{\bibfnamefont{J.}~\bibnamefont{Khoury}},
  \bibinfo{journal}{Phys.Rev.Lett.} \textbf{\bibinfo{volume}{104}},
  \bibinfo{pages}{231301} (\bibinfo{year}{2010}), \eprint{1001.4525}.

\bibitem[{\citenamefont{Brax et~al.}(2012)\citenamefont{Brax, Davis, Li, and
  Winther}}]{Brax:2012gr}
\bibinfo{author}{\bibfnamefont{P.}~\bibnamefont{Brax}},
  \bibinfo{author}{\bibfnamefont{A.-C.} \bibnamefont{Davis}},
  \bibinfo{author}{\bibfnamefont{B.}~\bibnamefont{Li}}, \bibnamefont{and}
  \bibinfo{author}{\bibfnamefont{H.~A.} \bibnamefont{Winther}}
  (\bibinfo{year}{2012}), \eprint{1203.4812}.

\bibitem[{\citenamefont{Brax et~al.}(2006)\citenamefont{Brax, van~de Bruck,
  Davis, and Green}}]{Brax:2005ew}
\bibinfo{author}{\bibfnamefont{P.}~\bibnamefont{Brax}},
  \bibinfo{author}{\bibfnamefont{C.}~\bibnamefont{van~de Bruck}},
  \bibinfo{author}{\bibfnamefont{A.-C.} \bibnamefont{Davis}}, \bibnamefont{and}
  \bibinfo{author}{\bibfnamefont{A.~M.} \bibnamefont{Green}},
  \bibinfo{journal}{Phys.Lett.} \textbf{\bibinfo{volume}{B633}},
  \bibinfo{pages}{441} (\bibinfo{year}{2006}), \eprint{astro-ph/0509878}.

\bibitem[{\citenamefont{Borisov et~al.}(2012)\citenamefont{Borisov, Jain, and
  Zhang}}]{Borisov:2011fu}
\bibinfo{author}{\bibfnamefont{A.}~\bibnamefont{Borisov}},
  \bibinfo{author}{\bibfnamefont{B.}~\bibnamefont{Jain}}, \bibnamefont{and}
  \bibinfo{author}{\bibfnamefont{P.}~\bibnamefont{Zhang}},
  \bibinfo{journal}{Phys.Rev.} \textbf{\bibinfo{volume}{D85}},
  \bibinfo{pages}{063518} (\bibinfo{year}{2012}), \eprint{1102.4839}.

\bibitem[{\citenamefont{Brax et~al.}(2010{\natexlab{b}})\citenamefont{Brax,
  Rosenfeld, and Steer}}]{Brax:2010tj}
\bibinfo{author}{\bibfnamefont{P.}~\bibnamefont{Brax}},
  \bibinfo{author}{\bibfnamefont{R.}~\bibnamefont{Rosenfeld}},
  \bibnamefont{and} \bibinfo{author}{\bibfnamefont{D.}~\bibnamefont{Steer}},
  \bibinfo{journal}{JCAP} \textbf{\bibinfo{volume}{1008}}, \bibinfo{pages}{033}
  (\bibinfo{year}{2010}{\natexlab{b}}), \eprint{1005.2051}.

\bibitem[{\citenamefont{{Li} and {Efstathiou}}(2012)}]{Li2012a}
\bibinfo{author}{\bibfnamefont{B.}~\bibnamefont{{Li}}} \bibnamefont{and}
  \bibinfo{author}{\bibfnamefont{G.}~\bibnamefont{{Efstathiou}}},
  \bibinfo{journal}{\mnras} \textbf{\bibinfo{volume}{421}},
  \bibinfo{pages}{1431} (\bibinfo{year}{2012}), \eprint{1110.6440}.

\bibitem[{\citenamefont{{Bernardeau} and {Brax}}(2011)}]{Bernardeau2011a}
\bibinfo{author}{\bibfnamefont{F.}~\bibnamefont{{Bernardeau}}}
  \bibnamefont{and} \bibinfo{author}{\bibfnamefont{P.}~\bibnamefont{{Brax}}},
  \bibinfo{journal}{\jcap} \textbf{\bibinfo{volume}{6}}, \bibinfo{pages}{19}
  (\bibinfo{year}{2011}), \eprint{1102.1907}.

\bibitem[{\citenamefont{Li and Efstathiou}(2011)}]{Li:2011qda}
\bibinfo{author}{\bibfnamefont{B.}~\bibnamefont{Li}} \bibnamefont{and}
  \bibinfo{author}{\bibfnamefont{G.}~\bibnamefont{Efstathiou}}
  (\bibinfo{year}{2011}), \eprint{1110.6440}.

\bibitem[{\citenamefont{{Schmidt} et~al.}(2009)\citenamefont{{Schmidt}, {Lima},
  {Oyaizu}, and {Hu}}}]{Schmidt2009}
\bibinfo{author}{\bibfnamefont{F.}~\bibnamefont{{Schmidt}}},
  \bibinfo{author}{\bibfnamefont{M.}~\bibnamefont{{Lima}}},
  \bibinfo{author}{\bibfnamefont{H.}~\bibnamefont{{Oyaizu}}}, \bibnamefont{and}
  \bibinfo{author}{\bibfnamefont{W.}~\bibnamefont{{Hu}}},
  \bibinfo{journal}{\prd} \textbf{\bibinfo{volume}{79}}, \bibinfo{eid}{083518}
  (\bibinfo{year}{2009}), \eprint{0812.0545}.

\bibitem[{\citenamefont{{Oyaizu} et~al.}(2008)\citenamefont{{Oyaizu}, {Lima},
  and {Hu}}}]{Oyaizu2008}
\bibinfo{author}{\bibfnamefont{H.}~\bibnamefont{{Oyaizu}}},
  \bibinfo{author}{\bibfnamefont{M.}~\bibnamefont{{Lima}}}, \bibnamefont{and}
  \bibinfo{author}{\bibfnamefont{W.}~\bibnamefont{{Hu}}},
  \bibinfo{journal}{\prd} \textbf{\bibinfo{volume}{78}}, \bibinfo{eid}{123524}
  (\bibinfo{year}{2008}), \eprint{0807.2462}.

\bibitem[{\citenamefont{Gil-Marin et~al.}(2011)\citenamefont{Gil-Marin,
  Schmidt, Hu, Jimenez, and Verde}}]{GilMarin:2011xq}
\bibinfo{author}{\bibfnamefont{H.}~\bibnamefont{Gil-Marin}},
  \bibinfo{author}{\bibfnamefont{F.}~\bibnamefont{Schmidt}},
  \bibinfo{author}{\bibfnamefont{W.}~\bibnamefont{Hu}},
  \bibinfo{author}{\bibfnamefont{R.}~\bibnamefont{Jimenez}}, \bibnamefont{and}
  \bibinfo{author}{\bibfnamefont{L.}~\bibnamefont{Verde}},
  \bibinfo{journal}{JCAP} \textbf{\bibinfo{volume}{1111}}, \bibinfo{pages}{019}
  (\bibinfo{year}{2011}), \eprint{1109.2115}.

\bibitem[{\citenamefont{{Koyama} et~al.}(2009)\citenamefont{{Koyama}, {Taruya},
  and {Hiramatsu}}}]{Koyama2009}
\bibinfo{author}{\bibfnamefont{K.}~\bibnamefont{{Koyama}}},
  \bibinfo{author}{\bibfnamefont{A.}~\bibnamefont{{Taruya}}}, \bibnamefont{and}
  \bibinfo{author}{\bibfnamefont{T.}~\bibnamefont{{Hiramatsu}}},
  \bibinfo{journal}{\prd} \textbf{\bibinfo{volume}{79}}, \bibinfo{eid}{123512}
  (\bibinfo{year}{2009}), \eprint{0902.0618}.

\bibitem[{\citenamefont{Zhao et~al.}(2009)\citenamefont{Zhao, Pogosian,
  Silvestri, and Zylberberg}}]{Zhao:2008bn}
\bibinfo{author}{\bibfnamefont{G.-B.} \bibnamefont{Zhao}},
  \bibinfo{author}{\bibfnamefont{L.}~\bibnamefont{Pogosian}},
  \bibinfo{author}{\bibfnamefont{A.}~\bibnamefont{Silvestri}},
  \bibnamefont{and}
  \bibinfo{author}{\bibfnamefont{J.}~\bibnamefont{Zylberberg}},
  \bibinfo{journal}{Phys.Rev.} \textbf{\bibinfo{volume}{D79}},
  \bibinfo{pages}{083513} (\bibinfo{year}{2009}), \eprint{0809.3791}.

\bibitem[{\citenamefont{Baker et~al.}(2011)\citenamefont{Baker, Ferreira,
  Skordis, and Zuntz}}]{Baker:2011jy}
\bibinfo{author}{\bibfnamefont{T.}~\bibnamefont{Baker}},
  \bibinfo{author}{\bibfnamefont{P.~G.} \bibnamefont{Ferreira}},
  \bibinfo{author}{\bibfnamefont{C.}~\bibnamefont{Skordis}}, \bibnamefont{and}
  \bibinfo{author}{\bibfnamefont{J.}~\bibnamefont{Zuntz}},
  \bibinfo{journal}{Phys.Rev.} \textbf{\bibinfo{volume}{D84}},
  \bibinfo{pages}{124018} (\bibinfo{year}{2011}), \eprint{1107.0491}.

\bibitem[{\citenamefont{Hu and Sawicki}(2007)}]{Hu:2007nk}
\bibinfo{author}{\bibfnamefont{W.}~\bibnamefont{Hu}} \bibnamefont{and}
  \bibinfo{author}{\bibfnamefont{I.}~\bibnamefont{Sawicki}},
  \bibinfo{journal}{Phys.Rev.} \textbf{\bibinfo{volume}{D76}},
  \bibinfo{pages}{064004} (\bibinfo{year}{2007}), \eprint{0705.1158}.

\bibitem[{\citenamefont{Brax et~al.}(2011{\natexlab{b}})\citenamefont{Brax,
  Davis, and Li}}]{Brax:2011aw}
\bibinfo{author}{\bibfnamefont{P.}~\bibnamefont{Brax}},
  \bibinfo{author}{\bibfnamefont{A.-C.} \bibnamefont{Davis}}, \bibnamefont{and}
  \bibinfo{author}{\bibfnamefont{B.}~\bibnamefont{Li}}
  (\bibinfo{year}{2011}{\natexlab{b}}), \eprint{1111.6613}.

\bibitem[{\citenamefont{{Crocce} and
  {Scoccimarro}}(2006{\natexlab{a}})}]{Crocce2006a}
\bibinfo{author}{\bibfnamefont{M.}~\bibnamefont{{Crocce}}} \bibnamefont{and}
  \bibinfo{author}{\bibfnamefont{R.}~\bibnamefont{{Scoccimarro}}},
  \bibinfo{journal}{\prd} \textbf{\bibinfo{volume}{73}},
  \bibinfo{pages}{063519} (\bibinfo{year}{2006}{\natexlab{a}}),
  \eprint{arXiv:astro-ph/0509418}.

\bibitem[{\citenamefont{{Valageas}}(2007{\natexlab{a}})}]{Valageas2007}
\bibinfo{author}{\bibfnamefont{P.}~\bibnamefont{{Valageas}}},
  \bibinfo{journal}{\aap} \textbf{\bibinfo{volume}{465}}, \bibinfo{pages}{725}
  (\bibinfo{year}{2007}{\natexlab{a}}), \eprint{arXiv:astro-ph/0611849}.

\bibitem[{\citenamefont{{Taruya} and {Hiramatsu}}(2008)}]{Taruya2008}
\bibinfo{author}{\bibfnamefont{A.}~\bibnamefont{{Taruya}}} \bibnamefont{and}
  \bibinfo{author}{\bibfnamefont{T.}~\bibnamefont{{Hiramatsu}}},
  \bibinfo{journal}{\apj} \textbf{\bibinfo{volume}{674}}, \bibinfo{pages}{617}
  (\bibinfo{year}{2008}), \eprint{0708.1367}.

\bibitem[{\citenamefont{{Pietroni}}(2008)}]{Pietroni2008}
\bibinfo{author}{\bibfnamefont{M.}~\bibnamefont{{Pietroni}}},
  \bibinfo{journal}{\jcap} \textbf{\bibinfo{volume}{10}}, \bibinfo{pages}{36}
  (\bibinfo{year}{2008}), \eprint{0806.0971}.

\bibitem[{\citenamefont{{Valageas}}(2007{\natexlab{b}})}]{Valageas2007a}
\bibinfo{author}{\bibfnamefont{P.}~\bibnamefont{{Valageas}}},
  \bibinfo{journal}{\aap} \textbf{\bibinfo{volume}{476}}, \bibinfo{pages}{31}
  (\bibinfo{year}{2007}{\natexlab{b}}), \eprint{0706.2593}.

\bibitem[{\citenamefont{{Valageas} and
  {Nishimichi}}(2011{\natexlab{a}})}]{Valageas2011d}
\bibinfo{author}{\bibfnamefont{P.}~\bibnamefont{{Valageas}}} \bibnamefont{and}
  \bibinfo{author}{\bibfnamefont{T.}~\bibnamefont{{Nishimichi}}},
  \bibinfo{journal}{\aap} \textbf{\bibinfo{volume}{527}}, \bibinfo{pages}{A87+}
  (\bibinfo{year}{2011}{\natexlab{a}}), \eprint{1009.0597}.

\bibitem[{\citenamefont{{Bernardeau} et~al.}(2002)\citenamefont{{Bernardeau},
  {Colombi}, {Gazta{\~n}aga}, and {Scoccimarro}}}]{Bernardeau2002}
\bibinfo{author}{\bibfnamefont{F.}~\bibnamefont{{Bernardeau}}},
  \bibinfo{author}{\bibfnamefont{S.}~\bibnamefont{{Colombi}}},
  \bibinfo{author}{\bibfnamefont{E.}~\bibnamefont{{Gazta{\~n}aga}}},
  \bibnamefont{and}
  \bibinfo{author}{\bibfnamefont{R.}~\bibnamefont{{Scoccimarro}}},
  \bibinfo{journal}{\physrep} \textbf{\bibinfo{volume}{367}},
  \bibinfo{pages}{1} (\bibinfo{year}{2002}), \eprint{arXiv:astro-ph/0112551}.

\bibitem[{\citenamefont{{Laureijs} et~al.}(2011)\citenamefont{{Laureijs},
  {Amiaux}, {Arduini}, {Augu{\`e}res}, {Brinchmann}, {Cole}, {Cropper},
  {Dabin}, {Duvet}, {Ealet} et~al.}}]{Laureijs2011}
\bibinfo{author}{\bibfnamefont{R.}~\bibnamefont{{Laureijs}}},
  \bibinfo{author}{\bibfnamefont{J.}~\bibnamefont{{Amiaux}}},
  \bibinfo{author}{\bibfnamefont{S.}~\bibnamefont{{Arduini}}},
  \bibinfo{author}{\bibfnamefont{J.~.} \bibnamefont{{Augu{\`e}res}}},
  \bibinfo{author}{\bibfnamefont{J.}~\bibnamefont{{Brinchmann}}},
  \bibinfo{author}{\bibfnamefont{R.}~\bibnamefont{{Cole}}},
  \bibinfo{author}{\bibfnamefont{M.}~\bibnamefont{{Cropper}}},
  \bibinfo{author}{\bibfnamefont{C.}~\bibnamefont{{Dabin}}},
  \bibinfo{author}{\bibfnamefont{L.}~\bibnamefont{{Duvet}}},
  \bibinfo{author}{\bibfnamefont{A.}~\bibnamefont{{Ealet}}},
  \bibnamefont{et~al.}, \bibinfo{journal}{arXiv:1110.3193L}
  (\bibinfo{year}{2011}), \eprint{1110.3193}.

\bibitem[{\citenamefont{{Valageas}}(2008)}]{Valageas2008}
\bibinfo{author}{\bibfnamefont{P.}~\bibnamefont{{Valageas}}},
  \bibinfo{journal}{\aap} \textbf{\bibinfo{volume}{484}}, \bibinfo{pages}{79}
  (\bibinfo{year}{2008}), \eprint{0711.3407}.

\bibitem[{\citenamefont{{Goroff} et~al.}(1986)\citenamefont{{Goroff},
  {Grinstein}, {Rey}, and {Wise}}}]{Goroff1986}
\bibinfo{author}{\bibfnamefont{M.~H.} \bibnamefont{{Goroff}}},
  \bibinfo{author}{\bibfnamefont{B.}~\bibnamefont{{Grinstein}}},
  \bibinfo{author}{\bibfnamefont{S.-J.} \bibnamefont{{Rey}}}, \bibnamefont{and}
  \bibinfo{author}{\bibfnamefont{M.~B.} \bibnamefont{{Wise}}},
  \bibinfo{journal}{\apj} \textbf{\bibinfo{volume}{311}}, \bibinfo{pages}{6}
  (\bibinfo{year}{1986}).

\bibitem[{\citenamefont{{Peebles}}(1980)}]{Peebles1980}
\bibinfo{author}{\bibfnamefont{P.~J.~E.} \bibnamefont{{Peebles}}},
  \emph{\bibinfo{title}{{The large-scale structure of the universe}}}
  (\bibinfo{publisher}{Princeton University Press, Princeton, N.J., USA},
  \bibinfo{year}{1980}).

\bibitem[{\citenamefont{{Martin} et~al.}(1973)\citenamefont{{Martin}, {Siggia},
  and {Rose}}}]{Martin1973}
\bibinfo{author}{\bibfnamefont{P.~C.} \bibnamefont{{Martin}}},
  \bibinfo{author}{\bibfnamefont{E.~D.} \bibnamefont{{Siggia}}},
  \bibnamefont{and} \bibinfo{author}{\bibfnamefont{H.~A.}
  \bibnamefont{{Rose}}}, \bibinfo{journal}{\pra} \textbf{\bibinfo{volume}{8}},
  \bibinfo{pages}{423} (\bibinfo{year}{1973}).

\bibitem[{\citenamefont{{Phythian}}(1977)}]{Phythian1977}
\bibinfo{author}{\bibfnamefont{R.}~\bibnamefont{{Phythian}}},
  \bibinfo{journal}{Journal of Physics A Mathematical General}
  \textbf{\bibinfo{volume}{10}}, \bibinfo{pages}{777} (\bibinfo{year}{1977}).

\bibitem[{\citenamefont{{Valageas} and
  {Nishimichi}}(2011{\natexlab{b}})}]{Valageas2011e}
\bibinfo{author}{\bibfnamefont{P.}~\bibnamefont{{Valageas}}} \bibnamefont{and}
  \bibinfo{author}{\bibfnamefont{T.}~\bibnamefont{{Nishimichi}}},
  \bibinfo{journal}{\aap} \textbf{\bibinfo{volume}{532}}, \bibinfo{pages}{A4+}
  (\bibinfo{year}{2011}{\natexlab{b}}), \eprint{1102.0641}.

\bibitem[{\citenamefont{{Valageas}}(2004)}]{Valageas2004}
\bibinfo{author}{\bibfnamefont{P.}~\bibnamefont{{Valageas}}},
  \bibinfo{journal}{\aap} \textbf{\bibinfo{volume}{421}}, \bibinfo{pages}{23}
  (\bibinfo{year}{2004}), \eprint{arXiv:astro-ph/0307008}.

\bibitem[{\citenamefont{Zinn-Justin}(1989)}]{Zinn-Justin1989}
\bibinfo{author}{\bibfnamefont{J.}~\bibnamefont{Zinn-Justin}},
  \emph{\bibinfo{title}{Quantum field theory and critical phenomena}}
  (\bibinfo{publisher}{Oxford: Clarendon Press}, \bibinfo{year}{1989}).

\bibitem[{\citenamefont{{Crocce} and
  {Scoccimarro}}(2006{\natexlab{b}})}]{Crocce2006b}
\bibinfo{author}{\bibfnamefont{M.}~\bibnamefont{{Crocce}}} \bibnamefont{and}
  \bibinfo{author}{\bibfnamefont{R.}~\bibnamefont{{Scoccimarro}}},
  \bibinfo{journal}{\prd} \textbf{\bibinfo{volume}{73}},
  \bibinfo{pages}{063520} (\bibinfo{year}{2006}{\natexlab{b}}),
  \eprint{arXiv:astro-ph/0509419}.

\bibitem[{\citenamefont{{Bernardeau} and {Valageas}}(2010)}]{Bernardeau2010b}
\bibinfo{author}{\bibfnamefont{F.}~\bibnamefont{{Bernardeau}}}
  \bibnamefont{and}
  \bibinfo{author}{\bibfnamefont{P.}~\bibnamefont{{Valageas}}},
  \bibinfo{journal}{\prd} \textbf{\bibinfo{volume}{81}},
  \bibinfo{pages}{043516} (\bibinfo{year}{2010}), \eprint{0912.0356}.

\bibitem[{\citenamefont{{Bernardeau} and {Valageas}}(2012)}]{Bernardeau2012}
\bibinfo{author}{\bibfnamefont{F.}~\bibnamefont{{Bernardeau}}}
  \bibnamefont{and}
  \bibinfo{author}{\bibfnamefont{P.}~\bibnamefont{{Valageas}}},
  \bibinfo{journal}{\prd} \textbf{\bibinfo{volume}{85}}, \bibinfo{eid}{023516}
  (\bibinfo{year}{2012}), \eprint{1109.4223}.

\bibitem[{\citenamefont{{Bernardeau} et~al.}(2008)\citenamefont{{Bernardeau},
  {Crocce}, and {Scoccimarro}}}]{Bernardeau2008}
\bibinfo{author}{\bibfnamefont{F.}~\bibnamefont{{Bernardeau}}},
  \bibinfo{author}{\bibfnamefont{M.}~\bibnamefont{{Crocce}}}, \bibnamefont{and}
  \bibinfo{author}{\bibfnamefont{R.}~\bibnamefont{{Scoccimarro}}},
  \bibinfo{journal}{\prd} \textbf{\bibinfo{volume}{78}},
  \bibinfo{pages}{103521} (\bibinfo{year}{2008}), \eprint{0806.2334}.

\bibitem[{\citenamefont{{Bernardeau} et~al.}(2012)\citenamefont{{Bernardeau},
  {van de Rijt}, and {Vernizzi}}}]{Bernardeau2012a}
\bibinfo{author}{\bibfnamefont{F.}~\bibnamefont{{Bernardeau}}},
  \bibinfo{author}{\bibfnamefont{N.}~\bibnamefont{{van de Rijt}}},
  \bibnamefont{and}
  \bibinfo{author}{\bibfnamefont{F.}~\bibnamefont{{Vernizzi}}},
  \bibinfo{journal}{\prd} \textbf{\bibinfo{volume}{85}}, \bibinfo{eid}{063509}
  (\bibinfo{year}{2012}), \eprint{1109.3400}.

\bibitem[{\citenamefont{{Anselmi} and {Pietroni}}(2012)}]{Anselmi2012}
\bibinfo{author}{\bibfnamefont{S.}~\bibnamefont{{Anselmi}}} \bibnamefont{and}
  \bibinfo{author}{\bibfnamefont{M.}~\bibnamefont{{Pietroni}}},
  \bibinfo{journal}{ArXiv e-prints}  (\bibinfo{year}{2012}),
  \eprint{1205.2235}.

\bibitem[{\citenamefont{{Matarrese} and {Pietroni}}(2008)}]{Matarrese2008}
\bibinfo{author}{\bibfnamefont{S.}~\bibnamefont{{Matarrese}}} \bibnamefont{and}
  \bibinfo{author}{\bibfnamefont{M.}~\bibnamefont{{Pietroni}}},
  \bibinfo{journal}{Modern Physics Letters A} \textbf{\bibinfo{volume}{23}},
  \bibinfo{pages}{25} (\bibinfo{year}{2008}), \eprint{arXiv:astro-ph/0702653}.

\bibitem[{\citenamefont{{Matsubara}}(2008)}]{Matsubara2008}
\bibinfo{author}{\bibfnamefont{T.}~\bibnamefont{{Matsubara}}},
  \bibinfo{journal}{\prd} \textbf{\bibinfo{volume}{77}},
  \bibinfo{pages}{063530} (\bibinfo{year}{2008}), \eprint{0711.2521}.

\bibitem[{\citenamefont{{Sefusatti} et~al.}(2006)\citenamefont{{Sefusatti},
  {Crocce}, {Pueblas}, and {Scoccimarro}}}]{Sefusatti2006}
\bibinfo{author}{\bibfnamefont{E.}~\bibnamefont{{Sefusatti}}},
  \bibinfo{author}{\bibfnamefont{M.}~\bibnamefont{{Crocce}}},
  \bibinfo{author}{\bibfnamefont{S.}~\bibnamefont{{Pueblas}}},
  \bibnamefont{and}
  \bibinfo{author}{\bibfnamefont{R.}~\bibnamefont{{Scoccimarro}}},
  \bibinfo{journal}{\prd} \textbf{\bibinfo{volume}{74}},
  \bibinfo{pages}{023522} (\bibinfo{year}{2006}),
  \eprint{arXiv:astro-ph/0604505}.

\bibitem[{\citenamefont{{Eisenstein} et~al.}(1999)\citenamefont{{Eisenstein},
  {Hu}, and {Tegmark}}}]{Eisenstein1999}
\bibinfo{author}{\bibfnamefont{D.~J.} \bibnamefont{{Eisenstein}}},
  \bibinfo{author}{\bibfnamefont{W.}~\bibnamefont{{Hu}}}, \bibnamefont{and}
  \bibinfo{author}{\bibfnamefont{M.}~\bibnamefont{{Tegmark}}},
  \bibinfo{journal}{\apj} \textbf{\bibinfo{volume}{518}}, \bibinfo{pages}{2}
  (\bibinfo{year}{1999}), \eprint{arXiv:astro-ph/9807130}.

\bibitem[{\citenamefont{{Komatsu} et~al.}(2009)\citenamefont{{Komatsu},
  {Dunkley}, {Nolta}, {Bennett}, {Gold}, {Hinshaw}, {Jarosik}, {Larson},
  {Limon}, {Page} et~al.}}]{Komatsu2009}
\bibinfo{author}{\bibfnamefont{E.}~\bibnamefont{{Komatsu}}},
  \bibinfo{author}{\bibfnamefont{J.}~\bibnamefont{{Dunkley}}},
  \bibinfo{author}{\bibfnamefont{M.~R.} \bibnamefont{{Nolta}}},
  \bibinfo{author}{\bibfnamefont{C.~L.} \bibnamefont{{Bennett}}},
  \bibinfo{author}{\bibfnamefont{B.}~\bibnamefont{{Gold}}},
  \bibinfo{author}{\bibfnamefont{G.}~\bibnamefont{{Hinshaw}}},
  \bibinfo{author}{\bibfnamefont{N.}~\bibnamefont{{Jarosik}}},
  \bibinfo{author}{\bibfnamefont{D.}~\bibnamefont{{Larson}}},
  \bibinfo{author}{\bibfnamefont{M.}~\bibnamefont{{Limon}}},
  \bibinfo{author}{\bibfnamefont{L.}~\bibnamefont{{Page}}},
  \bibnamefont{et~al.}, \bibinfo{journal}{\apjs}
  \textbf{\bibinfo{volume}{180}}, \bibinfo{pages}{330} (\bibinfo{year}{2009}),
  \eprint{0803.0547}.

\bibitem[{\citenamefont{{Lewis} et~al.}(2000)\citenamefont{{Lewis},
  {Challinor}, and {Lasenby}}}]{Lewis2000}
\bibinfo{author}{\bibfnamefont{A.}~\bibnamefont{{Lewis}}},
  \bibinfo{author}{\bibfnamefont{A.}~\bibnamefont{{Challinor}}},
  \bibnamefont{and}
  \bibinfo{author}{\bibfnamefont{A.}~\bibnamefont{{Lasenby}}},
  \bibinfo{journal}{\apj} \textbf{\bibinfo{volume}{538}}, \bibinfo{pages}{473}
  (\bibinfo{year}{2000}), \eprint{arXiv:astro-ph/9911177}.

\bibitem[{\citenamefont{{Eisenstein} et~al.}(2005)\citenamefont{{Eisenstein},
  {Zehavi}, {Hogg}, {Scoccimarro}, {Blanton}, {Nichol}, {Scranton}, {Seo},
  {Tegmark}, {Zheng} et~al.}}]{Eisenstein2005}
\bibinfo{author}{\bibfnamefont{D.~J.} \bibnamefont{{Eisenstein}}},
  \bibinfo{author}{\bibfnamefont{I.}~\bibnamefont{{Zehavi}}},
  \bibinfo{author}{\bibfnamefont{D.~W.} \bibnamefont{{Hogg}}},
  \bibinfo{author}{\bibfnamefont{R.}~\bibnamefont{{Scoccimarro}}},
  \bibinfo{author}{\bibfnamefont{M.~R.} \bibnamefont{{Blanton}}},
  \bibinfo{author}{\bibfnamefont{R.~C.} \bibnamefont{{Nichol}}},
  \bibinfo{author}{\bibfnamefont{R.}~\bibnamefont{{Scranton}}},
  \bibinfo{author}{\bibfnamefont{H.-J.} \bibnamefont{{Seo}}},
  \bibinfo{author}{\bibfnamefont{M.}~\bibnamefont{{Tegmark}}},
  \bibinfo{author}{\bibfnamefont{Z.}~\bibnamefont{{Zheng}}},
  \bibnamefont{et~al.}, \bibinfo{journal}{\apj} \textbf{\bibinfo{volume}{633}},
  \bibinfo{pages}{560} (\bibinfo{year}{2005}), \eprint{arXiv:astro-ph/0501171}.

\bibitem[{\citenamefont{{Valageas}}(2010)}]{Valageas2010}
\bibinfo{author}{\bibfnamefont{P.}~\bibnamefont{{Valageas}}},
  \bibinfo{journal}{\aap} \textbf{\bibinfo{volume}{514}}, \bibinfo{pages}{A46+}
  (\bibinfo{year}{2010}), \eprint{0906.1042}.

\bibitem[{\citenamefont{{Wang} and {Steinhardt}}(1998)}]{Wang1998}
\bibinfo{author}{\bibfnamefont{L.}~\bibnamefont{{Wang}}} \bibnamefont{and}
  \bibinfo{author}{\bibfnamefont{P.~J.} \bibnamefont{{Steinhardt}}},
  \bibinfo{journal}{\apj} \textbf{\bibinfo{volume}{508}}, \bibinfo{pages}{483}
  (\bibinfo{year}{1998}), \eprint{arXiv:astro-ph/9804015}.

\bibitem[{\citenamefont{{Bernardeau}}(1994{\natexlab{a}})}]{Bernardeau1994a}
\bibinfo{author}{\bibfnamefont{F.}~\bibnamefont{{Bernardeau}}},
  \bibinfo{journal}{\apj} \textbf{\bibinfo{volume}{427}}, \bibinfo{pages}{51}
  (\bibinfo{year}{1994}{\natexlab{a}}), \eprint{arXiv:astro-ph/9311066}.

\bibitem[{\citenamefont{{Valageas}}(2002{\natexlab{a}})}]{Valageas2002b}
\bibinfo{author}{\bibfnamefont{P.}~\bibnamefont{{Valageas}}},
  \bibinfo{journal}{\aap} \textbf{\bibinfo{volume}{382}}, \bibinfo{pages}{450}
  (\bibinfo{year}{2002}{\natexlab{a}}), \eprint{arXiv:astro-ph/0107333}.

\bibitem[{\citenamefont{{Valageas}}(2009)}]{Valageas2009}
\bibinfo{author}{\bibfnamefont{P.}~\bibnamefont{{Valageas}}},
  \bibinfo{journal}{\aap} \textbf{\bibinfo{volume}{508}}, \bibinfo{pages}{93}
  (\bibinfo{year}{2009}), \eprint{0905.2277}.

\bibitem[{\citenamefont{{Valageas}}(2002{\natexlab{b}})}]{Valageas2002}
\bibinfo{author}{\bibfnamefont{P.}~\bibnamefont{{Valageas}}},
  \bibinfo{journal}{\aap} \textbf{\bibinfo{volume}{382}}, \bibinfo{pages}{412}
  (\bibinfo{year}{2002}{\natexlab{b}}), \eprint{arXiv:astro-ph/0107126}.

\bibitem[{\citenamefont{{Bernardeau}}(1994{\natexlab{b}})}]{Bernardeau1994}
\bibinfo{author}{\bibfnamefont{F.}~\bibnamefont{{Bernardeau}}},
  \bibinfo{journal}{\aap} \textbf{\bibinfo{volume}{291}}, \bibinfo{pages}{697}
  (\bibinfo{year}{1994}{\natexlab{b}}), \eprint{arXiv:astro-ph/9403020}.

\bibitem[{\citenamefont{{Hellwing} and {Juszkiewicz}}(2009)}]{Hellwing2009}
\bibinfo{author}{\bibfnamefont{W.~A.} \bibnamefont{{Hellwing}}}
  \bibnamefont{and}
  \bibinfo{author}{\bibfnamefont{R.}~\bibnamefont{{Juszkiewicz}}},
  \bibinfo{journal}{\prd} \textbf{\bibinfo{volume}{80}}, \bibinfo{eid}{083522}
  (\bibinfo{year}{2009}), \eprint{0809.1976}.

\bibitem[{\citenamefont{{Li} et~al.}(2012)\citenamefont{{Li}, {Zhao}, and
  {Koyama}}}]{Li2012}
\bibinfo{author}{\bibfnamefont{B.}~\bibnamefont{{Li}}},
  \bibinfo{author}{\bibfnamefont{G.-B.} \bibnamefont{{Zhao}}},
  \bibnamefont{and} \bibinfo{author}{\bibfnamefont{K.}~\bibnamefont{{Koyama}}},
  \bibinfo{journal}{\mnras} \textbf{\bibinfo{volume}{421}},
  \bibinfo{pages}{3481} (\bibinfo{year}{2012}), \eprint{1111.2602}.

\bibitem[{\citenamefont{{Press} and {Schechter}}(1974)}]{Press1974}
\bibinfo{author}{\bibfnamefont{W.~H.} \bibnamefont{{Press}}} \bibnamefont{and}
  \bibinfo{author}{\bibfnamefont{P.}~\bibnamefont{{Schechter}}},
  \bibinfo{journal}{\apj} \textbf{\bibinfo{volume}{187}}, \bibinfo{pages}{425}
  (\bibinfo{year}{1974}).

\bibitem[{\citenamefont{{Li} and {Lam}}(2012)}]{Li2012b}
\bibinfo{author}{\bibfnamefont{B.}~\bibnamefont{{Li}}} \bibnamefont{and}
  \bibinfo{author}{\bibfnamefont{T.~Y.} \bibnamefont{{Lam}}},
  \bibinfo{journal}{ArXiv e-prints}  (\bibinfo{year}{2012}),
  \eprint{1205.0058}.

\bibitem[{\citenamefont{{Lam} and {Li}}(2012)}]{Lam2012}
\bibinfo{author}{\bibfnamefont{T.~Y.} \bibnamefont{{Lam}}} \bibnamefont{and}
  \bibinfo{author}{\bibfnamefont{B.}~\bibnamefont{{Li}}},
  \bibinfo{journal}{ArXiv e-prints}  (\bibinfo{year}{2012}),
  \eprint{1205.0059}.

\bibitem[{\citenamefont{{Li} and {Hu}}(2011)}]{Li2011}
\bibinfo{author}{\bibfnamefont{Y.}~\bibnamefont{{Li}}} \bibnamefont{and}
  \bibinfo{author}{\bibfnamefont{W.}~\bibnamefont{{Hu}}},
  \bibinfo{journal}{\prd} \textbf{\bibinfo{volume}{84}}, \bibinfo{eid}{084033}
  (\bibinfo{year}{2011}), \eprint{1107.5120}.

\bibitem[{\citenamefont{{Ferraro} et~al.}(2011)\citenamefont{{Ferraro},
  {Schmidt}, and {Hu}}}]{Ferraro2011}
\bibinfo{author}{\bibfnamefont{S.}~\bibnamefont{{Ferraro}}},
  \bibinfo{author}{\bibfnamefont{F.}~\bibnamefont{{Schmidt}}},
  \bibnamefont{and} \bibinfo{author}{\bibfnamefont{W.}~\bibnamefont{{Hu}}},
  \bibinfo{journal}{\prd} \textbf{\bibinfo{volume}{83}}, \bibinfo{eid}{063503}
  (\bibinfo{year}{2011}), \eprint{1011.0992}.

\bibitem[{\citenamefont{{Cooray} and {Sheth}}(2002)}]{Cooray2002}
\bibinfo{author}{\bibfnamefont{A.}~\bibnamefont{{Cooray}}} \bibnamefont{and}
  \bibinfo{author}{\bibfnamefont{R.}~\bibnamefont{{Sheth}}},
  \bibinfo{journal}{\physrep} \textbf{\bibinfo{volume}{372}},
  \bibinfo{pages}{1} (\bibinfo{year}{2002}), \eprint{astro-ph/0206508}.

\bibitem[{\citenamefont{{Navarro} et~al.}(1997)\citenamefont{{Navarro},
  {Frenk}, and {White}}}]{Navarro1997}
\bibinfo{author}{\bibfnamefont{J.~F.} \bibnamefont{{Navarro}}},
  \bibinfo{author}{\bibfnamefont{C.~S.} \bibnamefont{{Frenk}}},
  \bibnamefont{and} \bibinfo{author}{\bibfnamefont{S.~D.~M.}
  \bibnamefont{{White}}}, \bibinfo{journal}{\apj}
  \textbf{\bibinfo{volume}{490}}, \bibinfo{pages}{493} (\bibinfo{year}{1997}),
  \eprint{arXiv:astro-ph/9611107}.

\bibitem[{\citenamefont{{Peebles}}(1974)}]{Peebles1974}
\bibinfo{author}{\bibfnamefont{P.~J.~E.} \bibnamefont{{Peebles}}},
  \bibinfo{journal}{\aap} \textbf{\bibinfo{volume}{32}}, \bibinfo{pages}{391}
  (\bibinfo{year}{1974}).

\bibitem[{\citenamefont{{Smith} et~al.}(2003)\citenamefont{{Smith}, {Peacock},
  {Jenkins}, {White}, {Frenk}, {Pearce}, {Thomas}, {Efstathiou}, and
  {Couchman}}}]{Smith2003}
\bibinfo{author}{\bibfnamefont{R.~E.} \bibnamefont{{Smith}}},
  \bibinfo{author}{\bibfnamefont{J.~A.} \bibnamefont{{Peacock}}},
  \bibinfo{author}{\bibfnamefont{A.}~\bibnamefont{{Jenkins}}},
  \bibinfo{author}{\bibfnamefont{S.~D.~M.} \bibnamefont{{White}}},
  \bibinfo{author}{\bibfnamefont{C.~S.} \bibnamefont{{Frenk}}},
  \bibinfo{author}{\bibfnamefont{F.~R.} \bibnamefont{{Pearce}}},
  \bibinfo{author}{\bibfnamefont{P.~A.} \bibnamefont{{Thomas}}},
  \bibinfo{author}{\bibfnamefont{G.}~\bibnamefont{{Efstathiou}}},
  \bibnamefont{and} \bibinfo{author}{\bibfnamefont{H.~M.~P.}
  \bibnamefont{{Couchman}}}, \bibinfo{journal}{\mnras}
  \textbf{\bibinfo{volume}{341}}, \bibinfo{pages}{1311} (\bibinfo{year}{2003}),
  \eprint{arXiv:astro-ph/0207664}.

\bibitem[{\citenamefont{Zhao et~al.}(2011)\citenamefont{Zhao, Li, and
  Koyama}}]{Zhao:2010qy}
\bibinfo{author}{\bibfnamefont{G.-B.} \bibnamefont{Zhao}},
  \bibinfo{author}{\bibfnamefont{B.}~\bibnamefont{Li}}, \bibnamefont{and}
  \bibinfo{author}{\bibfnamefont{K.}~\bibnamefont{Koyama}},
  \bibinfo{journal}{Phys.Rev.} \textbf{\bibinfo{volume}{D83}},
  \bibinfo{pages}{044007} (\bibinfo{year}{2011}), \eprint{1011.1257}.

\bibitem[{\citenamefont{{Spergel} et~al.}(2007)\citenamefont{{Spergel}, {Bean},
  {Dor{\'e}}, {Nolta}, {Bennett}, {Dunkley}, {Hinshaw}, {Jarosik}, {Komatsu},
  {Page} et~al.}}]{Spergel2007}
\bibinfo{author}{\bibfnamefont{D.~N.} \bibnamefont{{Spergel}}},
  \bibinfo{author}{\bibfnamefont{R.}~\bibnamefont{{Bean}}},
  \bibinfo{author}{\bibfnamefont{O.}~\bibnamefont{{Dor{\'e}}}},
  \bibinfo{author}{\bibfnamefont{M.~R.} \bibnamefont{{Nolta}}},
  \bibinfo{author}{\bibfnamefont{C.~L.} \bibnamefont{{Bennett}}},
  \bibinfo{author}{\bibfnamefont{J.}~\bibnamefont{{Dunkley}}},
  \bibinfo{author}{\bibfnamefont{G.}~\bibnamefont{{Hinshaw}}},
  \bibinfo{author}{\bibfnamefont{N.}~\bibnamefont{{Jarosik}}},
  \bibinfo{author}{\bibfnamefont{E.}~\bibnamefont{{Komatsu}}},
  \bibinfo{author}{\bibfnamefont{L.}~\bibnamefont{{Page}}},
  \bibnamefont{et~al.}, \bibinfo{journal}{\apjs}
  \textbf{\bibinfo{volume}{170}}, \bibinfo{pages}{377} (\bibinfo{year}{2007}),
  \eprint{arXiv:astro-ph/0603449}.

\end{thebibliography}

\end{document}